\documentclass[useAMS,usenatbib]{mn2e}

\usepackage{amsmath, amssymb}
\usepackage{epsfig}         
\usepackage{graphicx, float}
\usepackage{multirow}

\usepackage{natbib}
\usepackage{aas_macros}
\usepackage{color}

\usepackage{url}
\usepackage[]{hyperref}

\def\Mpc{~{\rm Mpc}}
\def\Mpch{~h^{-1} {\rm Mpc}}
\def\MpchInverse{~h {\rm Mpc}^{-1}}
\def\MpchArea{~(h^{-1} {\rm Mpc})^2}
\def\MpchAreaInverse{~(h^{-1} {\rm Mpc})^{-2}}
\def\MpchVolume{~(h^{-1} {\rm Mpc})^3}
\def\GpchVolume{~(h^{-1} {\rm Gpc})^3}

\def\kpch{~h^{-1} {\rm kpc}}
\def\Msun{\rm{M}_{\odot}}
\def\Msolar{~h^{-1} \rm{M}_{\odot}}

\newcommand{\dtfe}{\textsc{DTFE}}
\newcommand{\Nexus}{\textsc{NEXUS}}
\newcommand{\nexus}{\textsc{NEXUS+}}
\newcommand{\logFilter}{Log-Gaussian}
\newcommand{\rockstar}{\textsc{rockstar}}

\newcommand{\MI}{\textsc{MS}}
\newcommand{\MII}{\textsc{MS-II}}

\newcommand{\spider}{NEXUS~}
\newcommand{\Spider}{NEXUS+~}
\newcommand{\spiderDen}{NEXUS\_den~}
\newcommand{\spiderTidal}{NEXUS\_tidal~}

\newcommand{\spiderVeldiv}{NEXUS\_veldiv~}
\newcommand{\spiderVelshear}{NEXUS\_velshear~}

\newcommand{\SpideR}{NEXUS+}
\newcommand{\spiderDeN}{NEXUS\_den}
\newcommand{\spiderTidaL}{NEXUS\_tidal}

\newcommand{\spiderVeldiV}{NEXUS\_veldiv}
\newcommand{\spiderVelsheaR}{NEXUS\_velshear}

\newcommand{\mean}[1]{{\langle}{#1}{\rangle}}
\newcommand{\meanTheta}{\widetilde{\omega}}

\newcommand{\Vector}[1]{\mathbf{#1}}

\newcommand{\gsim}{\raisebox{-0.3ex}{\mbox{$\stackrel{>}{_\sim} \,$}}}
\newcommand{\lsim}{\raisebox{-0.3ex}{\mbox{$\stackrel{<}{_\sim} \,$}}}
\newcommand{\eq}[1]{eq. \eqref{#1}}

\newcommand{\refsec}[1]{\S \ref{#1}}
\newcommand{\refappendix}[1]{appendix \S \ref{#1}}
\newcommand{\reftab}[1]{Table \ref{#1}}
\newcommand{\reffig}[1]{Fig.~\ref{#1}}
\newcommand{\reffigS}[1]{Figs~\ref{#1}}
\newcommand{\reffigs}[2]{Figs~\ref{#1}-\ref{#2}~}
\newcommand{\Reffig}[1]{Fig.~\ref{#1}}

\newcommand{\figDir}{fig_pdf/}

\newcommand{\nexusPaper}{CWJ13}

\newcommand{\MCn}[1]{#1}
\title[Evolution of the cosmic web]
{Evolution of the cosmic web}

\author[Cautun et al.]
{\parbox{\textwidth}{Marius Cautun$^{1,2}$\thanks{E-mail : m.c.cautun@durham.ac.uk},
                                Rien van de Weygaert$^{1}$, 
                                Bernard J. T. Jones$^{1}$ and \\ 
                                Carlos S. Frenk$^{2}$ \vspace{.4cm}} \\
$^1$  Kapteyn Astronomical Institute, University of Groningen, PO Box 800, 9747 AV Groningen, The Netherlands \\
$^2$  Department of Physics, Institute for Computational Cosmology, University of Durham, South Road Durham DH1 3LE, UK \\
}

\begin{document}


\maketitle

\begin{abstract}

The cosmic web is the largest scale manifestation of the anisotropic gravitational collapse of matter. It represents the transitional stage between linear and non-linear structures and contains easily accessible information about the early phases of structure formation processes. Here we investigate the characteristics and the time evolution of morphological components. Our analysis involves the application of the NEXUS Multiscale Morphology Filter technique, predominantly its NEXUS+ version, to high resolution and large volume cosmological simulations. We quantify the cosmic web components in terms of their mass and volume content, their density distribution and halo populations. We employ new analysis techniques to determine the spatial extent of filaments and sheets, like their total length and local width. This analysis identifies clusters and filaments as the most prominent components of the web. In contrast, while voids and sheets take most of the volume, they correspond to underdense environments and are devoid of group-sized and more massive haloes.
At early times the cosmos is dominated by tenuous filaments and sheets, which, during subsequent evolution, merge together, such that the present day web is dominated by fewer, but much more massive, structures. The analysis of the mass transport between environments clearly shows how matter flows from voids into walls, and then via filaments into cluster regions, which form the nodes of the cosmic web. We also study the properties of individual filamentary branches, to find long, almost straight, filaments extending to distances larger than $100\Mpch$. These constitute the bridges between massive clusters, which seem to form along approximatively straight lines.

\end{abstract}



\section{Introduction}
\label{sec:evolution:introduction}
On megaparsec scales the matter distribution of the Universe is not uniform, but it forms an intricate pattern which is known as the \textit{Cosmic Web} \citep{Bond1996,WeyBond08a}. The presence of this cosmic pattern, which can easily be seen in the distribution of galaxies, has been suggested by early attempts to map the Universe \citep{Gregory78,deLapparent1986,Geller1989,Shectman96} and, since then, it has been confirmed many times by present day surveys such as 2dFGRS \citep{Colless03}, SDSS \citep[][]{Tegmark2004} and the 2MASS redshift survey \citep{Huchra05}. The cosmic web consists of the largest non-linear structures in the Universe. The network has the massive galaxy clusters as its centres, which are interconnected through filaments and sheets. While the above components give most of the mass in the pattern, the cosmic web volume is dominated by vast and near empty regions known as voids \citep{Aragon07b,nexus2013}.

The cosmic web can be seen as the most prominent manifestation of the anisotropic nature of gravitational collapse, the motor behind the formation of structure in the cosmos \citep{1980lssu.book.....P}. The complex geometrical patterns that form the cosmic web represent a telling illustration of the wealth of structures that can arise under the influence of gravity. N-body computer simulations have illustrated how the large-scale structure of the cosmos evolves into a pronounced and intricate filigree of filamentary features, dented by dense compact clumps at the nodes of the network \citep[][]{Davis1985,White1987,Jenkins98,Colberg05,millSim,Dolag06}. The cosmic pattern forms a highly interconnected network, with galaxy clusters at the intersection of filaments and filaments at the intersection of walls \citep{Doroshkevich80,Klypin1983,Pauls95,Shapiro83,Sathyaprakash96}. These components show structures and substructures over a wide range of scales and densities, which are a clear manifestation of the hierarchical development of the cosmic web \citep{Sheth04,Shethwey04,Shen06}. These, the interconnected and hierarchical nature, are defining characteristics of the cosmic web that pose great difficulties in describing and identifying the large-scale structure of the universe.

\subsection{The theory of the cosmic web}
Understanding the formation and evolution of the large-scale structures cannot be undertaken without considering the role of the large scale tidal field, which is the major driving force shaping the cosmic web. This was first pointed out by \citet[][see also \citealt{Lin1965,Icke73}]{Zeldovich70} which showed the connection between the tidal shear field and the deformation of a fluid element. Subsequently, the gravitational collapse amplifies any initial anisotropies to give rise to highly asymmetrical structures, exhibiting strong planar or filamentary characteristics. According to the Zel'dovich formalism \citep{Zeldovich70}, the final morphology of a structure depends on the eigenvalues of the deformation tensor. Voids correspond to regions with all negative eigenvalues, while sheets, filaments and clusters correspond to domains with one, two and three positive eigenvalues. In the Zel'dovich approximation anisotropic collapse has a well defined sequence, with regions first contracting to form walls, than filaments and only at the end to fully collapse along each direction \MCn{\citep[][]{Arnold1982,Shandarin1984,Shandzeld89,Gurbatov1989,Hidding2013}}. The same predictions arise in the ellipsoidal collapse model \citep{Icke73,White1979}, which is also very widely used in describing anisotropic gravitational collapse. An integral part of this latter model is the inclusion of the external tidal field, which is needed to obtain realistic results \citep{Eisenstein95,Bond96a,Desjacques2008}. The ellipsoidal collapse model is the basis of many advanced descriptions for the distribution of virialized objects within hierarchical structure formation scenarios \citep{Bond96a,Sheth2001,Shen06}.

\MCn{The observed connectivity between the different morphological components arises naturally within the context of the cosmic web theory of \citet{Bond1996}. Their theory embedded the anisotropic evolution of structures in the cosmic web within the context of the hierarchically evolving mass distribution \citep{Bond96a}.} It highlights that the large scale matter distribution can be inferred by knowing the tidal field at a few relevant locations, usually the density peaks acting as cluster seeds. For example, filaments arise from a quadrupolar matter configuration in the initial density field, specified as a tidal shear constraint \citep{vandeWeygaert1996,WeyBond08a}. Such a quadrupolar distribution inevitably evolves to the canonical cluster-filament-cluster configuration, which forms the basis of the cosmic web. 

As we already saw, many of the cosmic web features have at least an embryonic trace in the primordial density field. This has been used by \citet{Doroshkevich70} to study the distribution of deformation tensor eigenvalues in the initial density field \citep[see also][]{Bardeen1986,Bond96a,Pogosyan98}, which, according to Zel'dovich formalism, are related to later time morphological components. Of special interest are the results of \citet{Pogosyan98} which emphasize that primordial overdense regions most likely evolve into clusters and filaments. In contrasts, underdense regions are more likely to become voids and sheets. While these findings are valid for the linear and mildly non-linear stages of evolution, they suggest that filaments are dominant in overdense domains and walls in underdense ones.

\MCn{The early evolution of the cosmic web can be easily understood from the singularities and caustics of the Zel'dovich formalism \citep[][]{Arnold1982b,Arnold1982,Hidding2013}. The web components arise via the formation of caustics and multistream flows, with the more non-linear regions corresponding to the more evolved environments. This description is very useful in outlining the cosmic web spine, as shown recently by \citet{Hidding2013}. An even better description is given by the adhesion model, which, via the introduction of an artificial viscosity term, mitigates some of the late-time limitations of the Zel'dovich approach \citep{Gurbatov1989,Kofman1990,Kofman1992}. This results in a very useful analytical description of the cosmic web and its skeleton \citep{Hidding2010,Hidding2012,Hidding2013}.}

Only recently the large-scale structures have been studied in the non-linear regime, following the application of new cosmic web identification techniques in N-body simulations. Several studies deserve special attention due to their robust and systematic analysis of the cosmic web components. \cite{Hahn2007a} and \cite{2010MNRAS.408.2163A} investigated the distribution of matter and haloes across environments, and showed the dominant role played by clusters and filaments which contain most of the mass and the majority of massive haloes in the universe. \cite{2010MNRAS.408.2163A} have taken the analysis further and explored the connectivity of the different morphological components, with emphasis on the size and inner structure of clusters and filaments. The evolution of the cosmic web has been probed by \cite{Hahn2007b} in terms of the change in mass content, volume fractions and halo population, to show significant changes in the web across cosmic times. A more focused approach was followed by \cite{Bond10}, which analysed the change in the distribution of filaments and their properties. They found that most of the filaments are already in place since high redshift and that most of the evolution is restricted to changes in filament size.

\subsection{Cosmic web identification}
Describing and identifying the cosmic web network, in both numerical simulation and observations, is no easy task due to the overwhelming complexity of the individual structures, their connectivity and the patterns intrinsic multiscale nature. This is clearly suggested by the large number of different methods that have attempted to do so, starting with the two- and higher-point correlation functions \citep{Peebles1975,1980lssu.book.....P,Peacock1999} and continuing with minimal spanning trees \citep{Barrow85,Graham95,Colberg07}, shape statistics \citep{Babul1992,Luo1995}, Minkowski functionals \citep{Mecke1994,Schmalzing1999}, local topological based measures \citep{Sahni1998,Sathyaprakash1998,Shandarin04} and genus statistics \citep{Gott1986,Hoyle2002c,Hoyle2002b}.

Most of the above methods characterize the large-scale pattern in a global and statistical way, but do not offer an approach that can be used locally for the identification of the cosmic web components. Recently, this has changed, after the introduction of several methods developed for the specific task of segmenting the cosmic web into its components: clusters, filaments, walls and voids. There are a variety of methods that attempt to do so, from filament detection via a generalization of the classical Candy model \citep{Stoica05,Stoica07,Stoica10}, to geometric inference formalisms \citep{Chazal09,Genovese10} and tessellation-based algorithms \citep{Gonzalez09}. 
Morse theory \citep[see][]{Colombi2000} forms the basis of the {\it skeleton analysis} \citep{Novikov06,Sousbie08a} and of its more rigorous and mathematically motivated implementation, the DisPerSE algorithm \citep{Sousbie2011a,Sousbie2011b}. These methods identify morphological features with the maxima and saddle points of the density field; and result in an elegant and mathematically rigorous tool for filament identification. Similar to the Watershed Void Finder \citep{Platen07}, the Spineweb procedure \citep{Aragon10,Aragon10b} is a topological approach that uses the intersection of watershed basins for environment identification.

The morphological methods are another important class of detection techniques. They characterize the cosmic web based on the density field Hessian, the tidal and the velocity shear fields \citep{Aragon07a,Aragon07b,Hahn2007a,Hahn2007b,Forero-Romero09,Wu09,Bond09,Bond10,Hoffman2012,nexus2013}. Especially noteworthy are the ones that follow a multiscale approach and allow for the identification of structures at multiple scales. The Multiscale Morphology Filter \citep[MMF;][]{Aragon07a} and its more refined versions, \Nexus{} and \nexus{} \citep{nexus2013}, are examples of such techniques. They are based on a scale-space approach that detects at the same time cosmic web features present at all smoothing scales. It does so by evaluating the density field Hessian over a range of spatial filter sizes and determining at which scales and locations the various morphological signatures are most prominent.

The dynamics of a system entails valuable complementary information towards the identification of the emerging spatial patterns in the cosmic mass distribution. There have been several attempts in this direction. The application of the Monge-Amp\`ere-Kantorovich reconstruction algorithm \citep{Frisch2002,Brenier2003} to the characterization of the nature of voids \citep{Lavaux2010} is an interesting example. More recently, the full 6D phase space information has been invoked towards recognizing single- and multistream regions. Three groups have independently recognized this and proposed the use towards recognizing cosmic web features \citep{Abel11,Falck2012,Neyrinck12,Neyrinck2012b,Shandarin2012}. Their use of the full phase space information allows for a more robust and dynamically motivated characterization of large-scale structure, though it also makes them difficult to use for the analysis of galaxy redshift surveys.

\subsection{Cosmic environments}
Even after the introduction of these advanced identification techniques, the cosmic web components and its properties have not been studied in detail. In fact, most investigations focused on understanding the dependence of halo properties on cosmic web environment. Such studies have shown that indeed there is a systematic dependence of halo properties, like shape, spin and formation redshift, on the environment in which they are embedded \citep{Aragon07a,Hahn2007a,Hahn2007b,Hahn09}. Moreover, there is a distinct correlation between halo shape and spin orientations and the directions of filaments and walls \citep{Altay2006,Aragon07a,Hahn2007a,Hahn2007b,Paz08,Hahn09,Zhang09,Codis2012,Libeskind2013,Aragon2013}. It has been found not only in simulations, but also in galaxy survey data \citep{Jones10,2012arXiv1207.0068T,Zhang2013}.

Most of the studies dealing with the nature and properties of the cosmic web environments are focused on investigating one component at a time. This has limitations since it does not allow for a robust characterization of all web elements within the same framework and moreover it does not permit an analysis of the connections between different cosmic web components. A lot of interest has been put in the investigation of cluster and supercluster properties, especially their size and morphology, to reveal that such objects are highly clustered and that they have a very anisotropic spatial distribution, favouring filamentary configurations \citep{Basilakos2006,Shaw2006,Wray2006,Costa-Duarte2011,Einasto2011,Liivamagi2012}. Filaments also received their fair share of attention, with numerous studies analysing filament properties as length, cross-section and shape \citep{Colberg07,Sousbie08b,Park2009,Bond10,Murphy2011,Pandey2011,Smith2012,Tempel2013a}. Within this context, the filament - cluster connection plays an important role given that these two components embody the spine of the cosmic web \MCn{\citep{Shandarin1984,Shandzeld89,Bond1996,Pimbblet04,Colberg05,WeyBond08a,Kartaltepe2008,Gonzalez09,Noh2011}}. Cosmic voids also pay a major role, given that most of the volume of the universe is in them, with multiple studies focused on characterizing void size, shape and inner structure \citep{Martel1990,vandeWeygaert1991,Weykamp93,Mathis2002,Gottlober03,Colberg05b,Platen2008,vandeWeygaert2011,Aragon-Calvo2013,Ricciardelli2013}.

\subsection{Outline of this paper}
The aim of this study is to investigate the evolution of the cosmic web and the variation in its properties. We do so in a self-consistent way, by employing the \Nexus{} and \nexus{} \citep[][hereafter \nexusPaper{}]{nexus2013} methods to identify in a scale-free way all the features of the cosmic web. This has two major advantages. First, it allows us to directly compare properties of different environments and to characterize the connectivity between components. Secondly, the multiscale nature of the method is instrumental in the detection of both prominent and tenuous structures, and therefore to facilitate a complete description of the filamentary and wall networks.

In a first part of the paper, we study the global properties of the cosmic web within the context of the \Nexus{} approach. \Nexus{} can employ a wide range of tracer fields for identifying large-scale structures, among which we have the density, tidal, velocity divergence and velocity shear fields. While the prominent features are detected in every tracer field, there is a great deal of difference in the identification of the more tenuous structures, which underlines the challenges faced in the detection of morphological components permeating underdense regions. A second part is dedicated to investigating the growth of the cosmic web, with emphasis on the transport of matter between different morphological components. Our analysis focuses on the properties of anisotropic components, i.e. filaments and walls, given that the evolution of these structures has not been properly investigated until now. Moreover, our study characterizes the properties of individual filamentary branches, focusing on properties like shape, length and mean density. In doing so, we introduce a method which uses the branching points to segment the filamentary network into individual objects.

This paper is organized as follows. In \refsec{sec:evolution:data_analysis} we introduce the numerical simulations and the halo samples that we use for our analysis; this is followed by \refsec{sec:evolution:cosmic_web_methods} that gives an overview of the cosmic web detection methods; \refsec{sec:evolution:environment_characterisation} describes the techniques used to measure the extent and mass distribution of filaments and walls; \refsec{sec:evolution:present_properties} presents the properties of the cosmic web at present time and compares between different morphological identification methods; while the evolution of the cosmic web in terms of mass content, halo populations and spatial extent is presented in \refsec{sec:evolution:evolution}, \refsec{sec:evolution:mass_trasport} and \refsec{sec:evolution:morphology_fractal}; the segmentation of the filamentary network and the properties of its branches are analysed in \refsec{sec:evolution:filament_segmentation} and \refsec{sec:evolution:filaments}; we conclude with \refsec{sec:evolution:conclusions} which summarizes the most important findings.


\section{Numerical simulations}
\label{sec:evolution:data_analysis}
In this study we make use of the two high resolution Millennium simulations\footnote{Data from the Millennium/Millennium-II simulation are available on a relational database accessible from \newline http://galaxy-catalogue.dur.ac.uk:8080/Millennium .} (\MI{}; \citealt{millSim} and \MII{}; \citealt{millSim2}). Both simulations are dark matter (DM) only and make use of $2160^3$ particles to resolve structure formation in the \textit{Wilkinson Microwave Anisotropy Probe} 1 cosmogony \citep{Spergel2003} with the following cosmological parameters: $\Omega_m=0.23$, $\Omega_\Lambda=0.75$, $h=0.73$, $n_s=1$ and $\sigma_8=0.9$.

The \MI{} models cosmic evolution in a periodic volume of length $500\Mpch$ with a mass per particle of $m_p=8.6\times10^8\Msolar$ and a gravitational softening length of $5\kpch$. The large volume of the simulation makes it ideal for studying the large-scale structure of the universe with minimal effects from cosmic variance. \MI{} resolves a huge number of haloes, from masses as large as ${\sim}10^{15}\Msolar$ down to small halo masses of only ${\sim}2\times10^{10}\Msolar$. This allows for a characterization of the connection between large-scale structures and gravitationally bound objects over several orders of magnitude in halo mass.

The \MII{} resolves structure formation in a much smaller box of $100\Mpch$ on a side with a particle mass of $m_p=6.89\times10^6\Msolar$ and force softening of $5\kpch$. While the small volume makes \MII{} prone to significant cosmic variance effects, its higher resolution allows us to investigate the cosmic web up to a higher redshift $z$ than for the \MI{} data. Moreover, at high redshift the cosmic variance effects decrease since the homogeneity scale of the universe is also reduced with respect to $z=0$. We mainly use the \MII{} data for illustrative purposes as well as to test possible resolution effects affecting the detection and properties of cosmic environments.

\subsection{Halo finder}
\label{subsec:evolution:rockstar}
We perform the halo and subhalo identification procedure using the \rockstar{} (Robust Overdensity Calculation using K-Space Topologically Adaptive Refinement) phase-space halo finder \citep{Berhoozi2011}. \rockstar{} starts by selecting potential haloes as Friends-of-Friends (FOF) groups in position-space using a large linking length ($b = 0.28$). This first step is restricted to position-space to optimize the use of computational resources, while the analysis of each subsequent step is done using the full 6D phase-space. Each FOF group from the first step is used to create a  hierarchy of FOF phase-space subgroups by progressively reducing the linking length. The phase-space subgroups are selected using an adaptive phase-space linking length such that each successive subgroup has $70\%$ of the parent's particles. \rockstar{} uses the resulting subgroups as potential halo and subhaloes centres and assigns particles to them based on their phase-space proximity. Once all particles are assigned to haloes and subhaloes, an unbinding procedure is used to keep only the gravitationally bound particles. The final halo centres are computed using a small region around the phase-space density maximum associated with each object. The outer boundaries of the haloes are cut at the point where the enclosed overdensity decreases below $\Delta=200$ times the critical density $\rho_c$. Therefore the halo mass $M_{200}$ and radius $R_{200}$ correspond to a spherical overdensity of $200\rho_c$.

\subsection{Density and velocity divergence fields}
\label{subsec:evolution:density_computation}
The methods employed for the identification of the cosmic web components take as input regularly sampled density and velocity fields. We construct these fields using the Delaunay Tessellation Field Estimator \citep[DTFE; ][]{Schaap2000,2009LNP...665..291V,Cautun2011} method. The DTFE algorithm uses the discrete particles position and velocities to extrapolate volume filling density and velocity divergence fields. We make use of the \dtfe{} method because it does not depend on user defined parameters and it preserves the multi-scale character and geometry of the input particle distribution. These features are crucial ingredients for the detection of the anisotropic components of the cosmic web such as filaments and walls (\nexusPaper{}). For simplicity, we express the density in units of the background average density $\bar{\rho}$ as $1+\delta=\rho/\bar{\rho}$. The velocity divergence is given with respect to the Hubble parameter $H$ as $\theta=\Vector{\nabla}\cdot \Vector{v}/H$.

We compute the \dtfe{} density and velocity divergence fields on a grid with spacing $\Delta x=0.4\Mpch$, such that the \MI{} volume is fully covered by a $1280^3$ grid while the \MII{} volume is represented on a $256^3$ grid. Moreover, for resolution studies we compute $\delta$ and $\theta$ values for the \MII{} on grids with $\Delta x=0.2$ and $0.1\Mpch$.


\section{Cosmic web detection using \Nexus{}}
\label{sec:evolution:cosmic_web_methods}
In this work we employ the \Nexus{} and \nexus{} algorithms (\nexusPaper{}) for the segmentation of the cosmic web into its individual components: clusters, filaments, walls and voids. The methods perform the morphological identification of environments using a scale-space formalism which ensures the detection of structures present at all scales. It allows for a complete and unbiased characterization of the cosmic web components, from the prominent features present in overdense regions to the tenuous networks pervading the cosmic voids. This represents a major advantage when studying both the connectivity between components and the time evolution of the cosmic web.

The \Nexus{} and \nexus{} methods are inspired by scale-space analysis techniques used in the medical imaging field for the detections of nodules and blood vessels \citep{DBLP:conf/miccai/FrangiNVV98,Sato983dmulti-scale,li:2040}. These procedures were first introduced in astronomy by \citet{Aragon07b} as the MMF, which used the density field as the basis for the morphological segmentation of the cosmic web. The MMF formalism constitutes the foundation on which \Nexus{} and \nexus{} were developed with the goal of obtaining a more physically motivated and robust method. While both MMF and \Nexus{} share the same philosophy for environment identification, there are some key differences between the two procedures which can result in distinct outcomes. \Nexus{} extends the MMF formalism to incorporate not only the density field, but also tidal, velocity divergence and velocity shear fields. This offer a consistent and physically motivated framework for the detection of the cosmic web components using the full 6D phase space information. The second substantial difference is due to the criteria used to characterize the cosmic web detection threshold. The MMF method uses the percolation of filaments and walls as the threshold for environment identification. Such an approach is prone to resolution effects and moreover does not give consistent results when comparing the mass and volume fractions in cosmic web environments traced by various fields, like density or tidal fields (\nexusPaper{}). In contrast, the threshold approach adopted by \Nexus{} does not suffer such artefacts.

\begin{table}
    \centering
    \caption{ The cosmic web identification methods employed by this study. The central column gives the input field used by each technique as the starting point of the detection procedure. The right most column gives the smoothing filter used by each method. }
    \label{tab:evolution:method_names}
    \begin{tabular}{lll}
        \hline
        Method name & Tracer field & Filter type \\
        \hline
        \spiderDen     &  Density              &  Gaussian \\
        \spiderTidal   &  Tidal                &  Gaussian \\
        \spiderVeldiv  &  Velocity divergence  &  Gaussian \\
        \spiderVelshear & Velocity shear       &  Gaussian \\
        \nexus         &  Density              &  \logFilter{} \\
        \hline
    \end{tabular}
\end{table}

As we already touched upon, the \Nexus{} method has been generalized to include four different environmental tracers: density, tidal, velocity divergence and velocity shear fields. Each of these fields can be used independently of each other for feature detection, giving rise to the methods summarized in \reftab{tab:evolution:method_names}. The density field is an obvious candidate for environmental detection given the prominence of cluster, filamentary and void features in the matter distribution. Similarly, the tidal field is the driver of anisotropic collapse and therefore plays an essential role in formation and evolution of the cosmic web. The use of velocity data for environmental detection is motivated by the close one-to-one relationship in the linear regime between density and velocity divergence, and between the tidal force which is the source of the velocity shear field. Using velocity information opens the other half of the phase space for environment identification. Given the complementary information supplied by each of the four fields, it is not immediately clear which quantity is best suitable for probing the cosmic web evolution. To overcome this, we compare the morphological components identified by each method in the hope of obtaining a better understanding of the advantages and limitations of each approach.

\begin{figure*}
    \centering
    \includegraphics[width=0.87\linewidth]{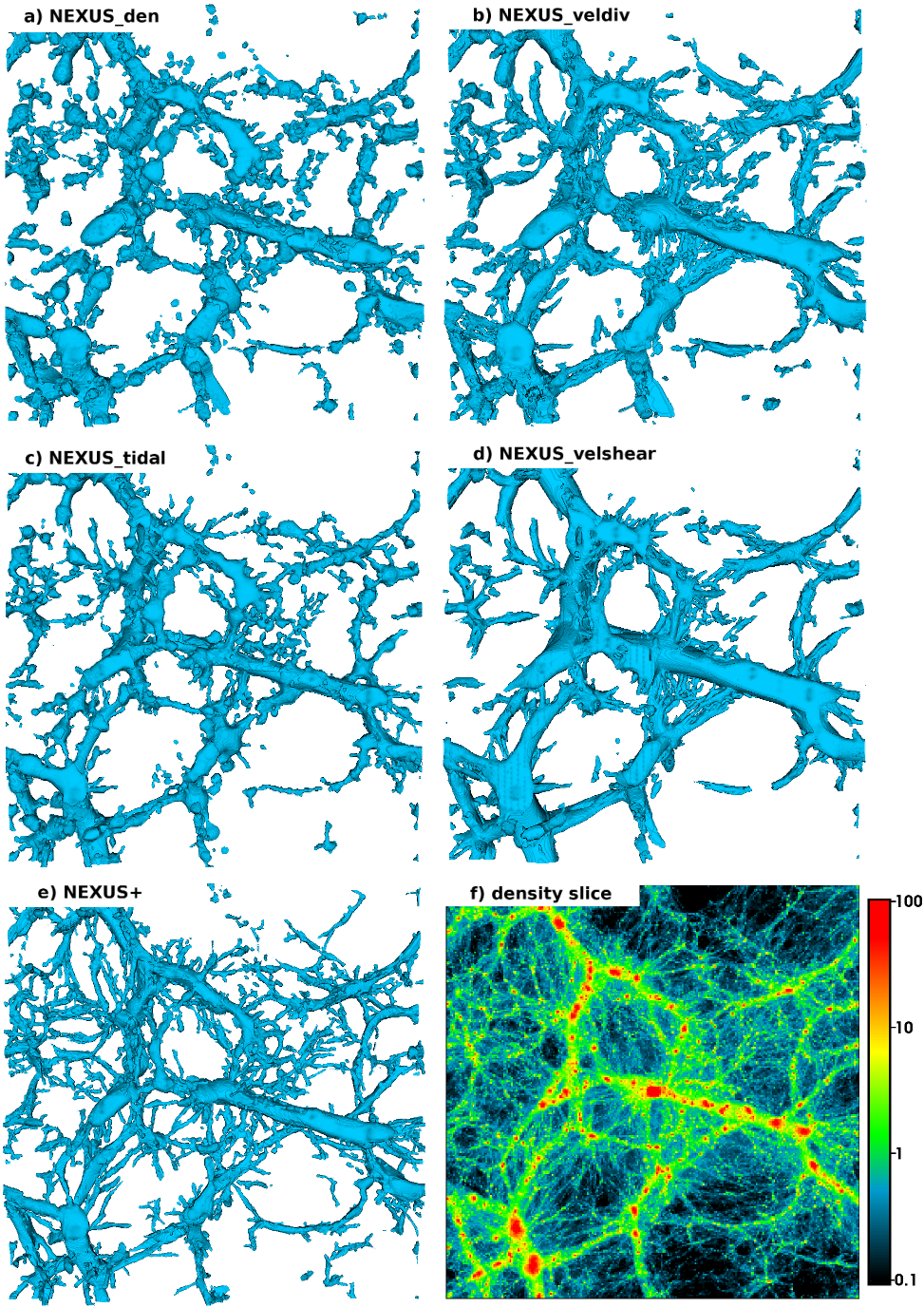} 
    \caption{The filamentary environments in a $100\times100\times10\MpchVolume$ slice centred on the most massive \MII{} halo at present redshift $z=0$. Five of the panels show the filaments detected with: a) \spiderDen, b) \spiderVeldiv,  c) \spiderTidal, d) \spiderVelshear and e) \Spider methods. The sixth panel, f), shows a projection of the density field in the selected volume. The density scale $1+\delta$ is shown on the side of the panel. }
    \label{fig:evolution:env_filaments}
\end{figure*}

The large scale matter distribution is characterized by orders of magnitude difference in the density field between overdense and underdense regions. This variation can be seen also in the size and contrast of filaments and walls, with prominent environments typically found in overdense regions while the underdense regions are dominated by tenuous structures. This poses additional challenges for \Nexus{} which works best when all features have the same contrast (\nexusPaper{}). To cope with these issues, \nexusPaper{} introduced the \nexus{} method which replaces the Gaussian filter employed by \Nexus{} with a \logFilter{} filter. The new filter takes into account the large range of values at which structures can be present and offers a better way of detecting both prominent and tenuous environments. While the new filter is very successful, it involves a logarithmic transform, and therefore can only be applied to positive valued fields. Thus, \nexus{} can only use the density field as cosmic web tracer. Note that the \Nexus{} and \nexus{} algorithms are the same in all respects, except the smoothing filters they use.

\subsection{The \Nexus{} and \nexus{} algorithms}
\label{subsec:evolution:nexus_description}

In the following we briefly summarize the steps that the two algorithms take in the segmentation of the cosmic web. These steps are the same for any of the four input fields used as environmental tracers, but for simplicity we restrict our description to the density field. A more detailed description of the procedures can be found in \nexusPaper{}. The \Nexus{} and \nexus{} methods consist of the following six steps:
\begin{enumerate}
	\item[I)]  Smoothing the input density field with a Gaussian filter of radius $R_n$ in the case of \Nexus{} and a \logFilter{} filter for \nexus{}. It results in a smoothed density in which the dominant features are those with sizes ${\sim}R_n$.
	\item[II)] Computing the eigenvalues for the Hessian matrix of the smoothed density field found in the previous step. The Hessian eigenvalues are sensitive to any morphological features present in the density data.
	\item[III)]Computing the environmental signature at scale $R_n$ using the Hessian eigenvalues. This results in a signature value at each point characterizing how close this region is to an ideal cosmic web node, filament and wall. 
	\item[IV)] Repeating steps \textbf{I)} to  \textbf{III)} for a set of scales $[R_0,R_1,...,R_N]$ with $R_n=2^{n/2}R_0$. For each scale and at each point we obtain a cluster, filament and wall signature.
	\item[V)]  Combining the environmental signature obtained at each scale to obtain a scale independent signature. It results in an environmental signature that characterizes the morphology of each point, independent of smoothing scale $R_n$.
	\item[VI)] Using physical criteria to determine the detection threshold. All points with signature values above the threshold are valid large-scale structures. For cosmic web nodes, the threshold is given by the requirement that most cluster-regions should be virialized. For filaments and walls, the threshold is determined on the basis of the change in filament and wall mass as a function of signature. The peak of the mass variation with signature delineates the most prominent filamentary and wall features of the cosmic web.
\end{enumerate}

The algorithm performs the environment detection by applying the above steps first to clusters, then to filaments and finally to walls. This sequence needs to be followed to make sure that each volume element is assigned only a single environment characteristic. The remaining regions that are not identified as nodes, filaments or sheets, are classified as cosmic voids.

In this work we focus on the characterization of the anisotropic components of the cosmic web, i.e. filaments and walls. Given that differences in the detection of cosmic web nodes can influence the identification of filaments and walls, we chose to perform the cluster identification step using only the \spiderDen{} method. This way, any discrepancies in the identification of the anisotropic components using different procedures are due to the methods themselves and not due to the cluster detection step. A study into the properties of cosmic web nodes and how these vary between different identification methods has been done in \nexusPaper{}.

In practice, the two algorithms are implemented on a grid using the density and velocity divergence fields found by the \dtfe{} method (see \refsec{subsec:evolution:density_computation}). This means that, following the application of the methods, each grid cell is classified as being part of a node, filament, wall or void. The presence of a grid also implies a finite scale given by the grid spacing $\Delta x$ below which we cannot study the cosmic web. This means that we restrict our scale space analysis to features from $R_0=\Delta x$ up to $R_N=8\Mpch$. Larger filter scales do not change our results, while the effect of smaller $\Delta x$ values will be investigated later on.

\subsection{Visual comparison of detection methods}
\label{subsec:evolution:method_comparison}

\begin{table}
    \centering
    \caption[]{The morphological segmentation of the cosmic matter distribution according to the Zel'dovich formalism. The $\lambda_1\ge\lambda_2\ge\lambda_3$ quantities denote the eigenvalues of the deformation tensor. Their sign, positive or negative, determines the morphological characterization. }
    \label{tab:evolution:zeldovich_eigenvalues}
    \begin{tabular}{cccl}
        \hline
        \parbox[][][c]{0.1\textwidth}{\centering $\lambda_1$} & \parbox[][][c]{0.1\textwidth}{\centering $\lambda_2$} & \parbox[][][c]{0.1\textwidth}{\centering $\lambda_3$} & component \\
        \hline
        $+$  & $+$  & $+$   &  cluster  \\
        $+$  & $+$  & $-$   &  filament  \\
        $+$  & $-$  & $-$   &  sheet  \\
        $-$  & $-$  & $-$   &  void  \\
        \hline
    \end{tabular}
\end{table}

Given that we have several approaches for identifying the cosmic web components, it is important to asses the similarities and differences between the outcomes of each method. This is crucial in understanding what are the environments traced by the various fields that we employ: density, tidal, velocity divergence and velocity shear. Having done so, we can decide which method is the most appropriate for following the time evolution of the cosmic web. We already presented in \nexusPaper{} a detailed qualitative comparison of the methods, therefore, in the following, we only summarize some of the results that are of importance for our current study.

To better illustrate our conclusions, we show in \reffig{fig:evolution:env_filaments} the filamentary environments detected by \spiderDeN, \spiderTidaL, \spiderVeldiV, \spiderVelshear and \SpideR. For comparison to the large scale distribution of matter, the lower-right frame shows a projection of the density field in the same volume. The most striking outcome is that the filamentary network is dominated by a few prominent structures with coherent scales of tens of megaparsecs. The prominent filaments are detected by all the methods, though their diameter is dependent on the method used. The \Spider filaments are the thinnest ones, followed by the \spiderDen and \spiderTidal ones. In contrast, the pronounced filaments detected in the velocity fields are the thickest ones. While all the methods detect the most outstanding structures, there are quite some variations when it comes to the more tenuous environments. These are usually located in lower density and sometimes even underdense regions, and therefore have less contrast than the more prominent environments. This makes the detection of tenuous structures much more challenging, which explains the differences that we see between methods. These feeble environments are identified the least by the \spiderTidal and \spiderVelshear approaches, while \Spider finds a much richer network of such structures. It suggests that approaches based on the tidal field \citep{Hahn2007a,Forero-Romero09} or velocity shear field \citep{Hoffman2012} are not very sensitive to the more tenuous structures. Similar differences between methods can be found when analysing the cosmic walls (\nexusPaper{}).

\subsection{The Zel'dovich formalism and \Nexus{} environments}
\label{subsec:evolution:zeldovich}

\begin{figure}
    \centering
    \includegraphics[width=\linewidth]{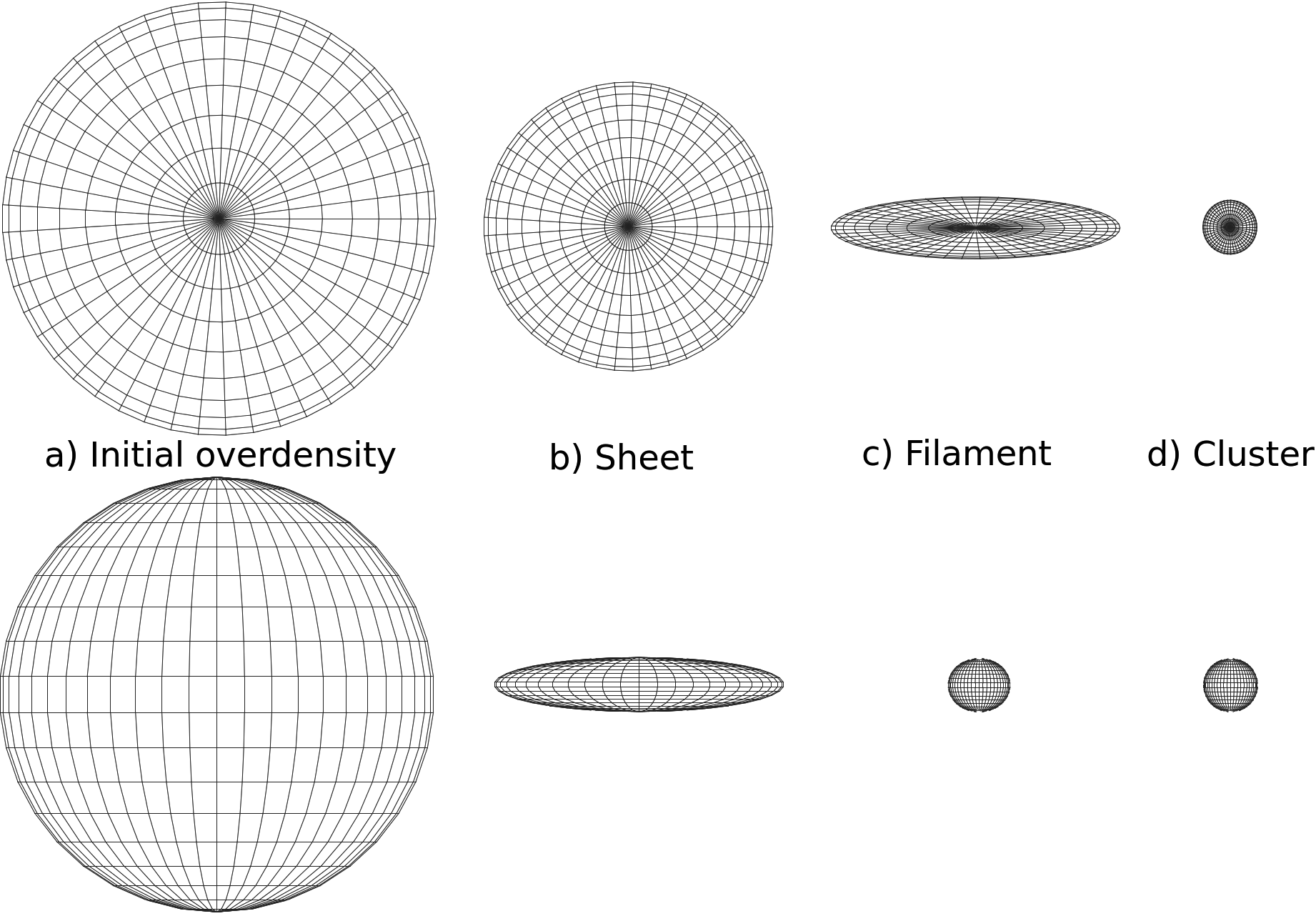} 
    \caption{ The formation of structure according to the Zel'dovich formalism. The sequence starts with the left most panel which shows an ellipsoidal overdensity from two perpendicular angles. The overdensity collapse proceeds most strongly along one axis to form a sheet, followed by the full contraction of the second axis to form a filament. At last, full collapse takes place resulting in a 3D virialized structure. }
    \label{fig:evolution:env_example_zeldovich}
\end{figure}

The Zel'dovich formalism \citep{Zeldovich70} offers a natural way of describing anisotropic collapse and therefore the formation of the cosmic web. It has been found to give a good description of structure formation in the linear and mildly non-linear stages. This suggests that the Zel'dovich formalism can offer a reasonable description of large-scale structures, given that the cosmic web is at the transitional stage between linear primordial and fully non-linear structures. This raises questions about the common points as well as the differences between \Nexus{} and Zel'dovich predictions.

The Zel'dovich formalism offers a first-order Lagrangian approximation to the formation and evolution of cosmic structure. In the Zel'dovich approximation, the motion of a fluid element is determined by the primordial density fluctuations, following a ballistic displacement approach. At some time $t$, the Eulerian position $\Vector{x}(t)$ of the fluid element is given by
\begin{equation}
	\Vector{x}(t) = \Vector{q} + D(t) \; \nabla \psi(\Vector{q}) \;,
	\label{eq:evolution:zeldovich}
\end{equation}
where $\Vector{q}$ is the initial or Lagrangian position of the element. The quantity $D(t)$ denotes the linear growth factor and $\psi$ is the Lagrangian displacement potential \citep{1980lssu.book.....P}. The latter is the primordial linearly extrapolated gravitational potential, up to a constant multiplication factor. Using this prescription, we can describe how an initial mass element $\bar{\rho}d^3\Vector{q}$ gets mapped at a later time $t$ to $\rho(\Vector{x})d^3\Vector{x}$. The mass within the mapped volume is conserved, i.e. $\bar{\rho}d^3\Vector{q} = \rho(\Vector{x})d^3\Vector{x}$, which, after a few algebraic manipulations, leads to
\begin{equation}
	\rho(\Vector{x}) = \frac{\bar{\rho}}{ [1-D\;\lambda_1(\Vector{q})] \; [1-D\;\lambda_2(\Vector{q})] \; [1-D\;\lambda_3(\Vector{q})] } \;.
	\label{eq:evolution:zeldovich_rho}
\end{equation}
Here $\rho(\Vector{x})$ denotes the density at Eulerian position $\Vector{x}$ and $\bar{\rho}$ symbolizes the mean cosmic density. The three $\lambda_1\ge\lambda_2\ge\lambda_3$ quantities denote the eigenvalues of the deformation tensor
\begin{equation}
	\psi_{ij}(\Vector{q}) = \frac{\partial^2\psi(\Vector{q})}{\partial q_i\partial q_j} \;.
\end{equation}

Similarly to the \Nexus{} techniques, the Zel'dovich formalism can be used to identify the cosmic web components. This can be easily appreciated from \eq{eq:evolution:zeldovich_rho}, which describes the evolution of the density at a later time in terms of the primordial matter distribution. The formation of pancakes, filaments and clusters is dictated by the eigenvalues of the deformation tensor, as given in \reftab{tab:evolution:zeldovich_eigenvalues}. For example, clusters form in the regions with three positive eigenvalues. The evolution of these domains is via a well defined sequence as illustrated in \reffig{fig:evolution:env_example_zeldovich}, where we sketch the collapse of an ellipsoidal overdensity. As time evolves, the overdensity contracts along all directions, but most strongly along the direction corresponding to the largest eigenvalue $\lambda_1$. The full collapse along this axis takes place when $1-D(t)\;\lambda_1\rightarrow 0$, resulting in a sheet as shown in panel b). The contraction follows along the second axis to form a filamentary configuration and ends with the collapse along the third direction to form a 3D virialized object. This suggests that one can define a sequence of morphologies, each one associated with a well defined stage of the anisotropic gravitational collapse. As shown in \reffig{fig:evolution:env_example_zeldovich}, these morphologies evolve in time and moreover, at any one epoch, we can find a range of intermediate states.

\begin{figure}
    \centering
    \includegraphics[width=.85\linewidth]{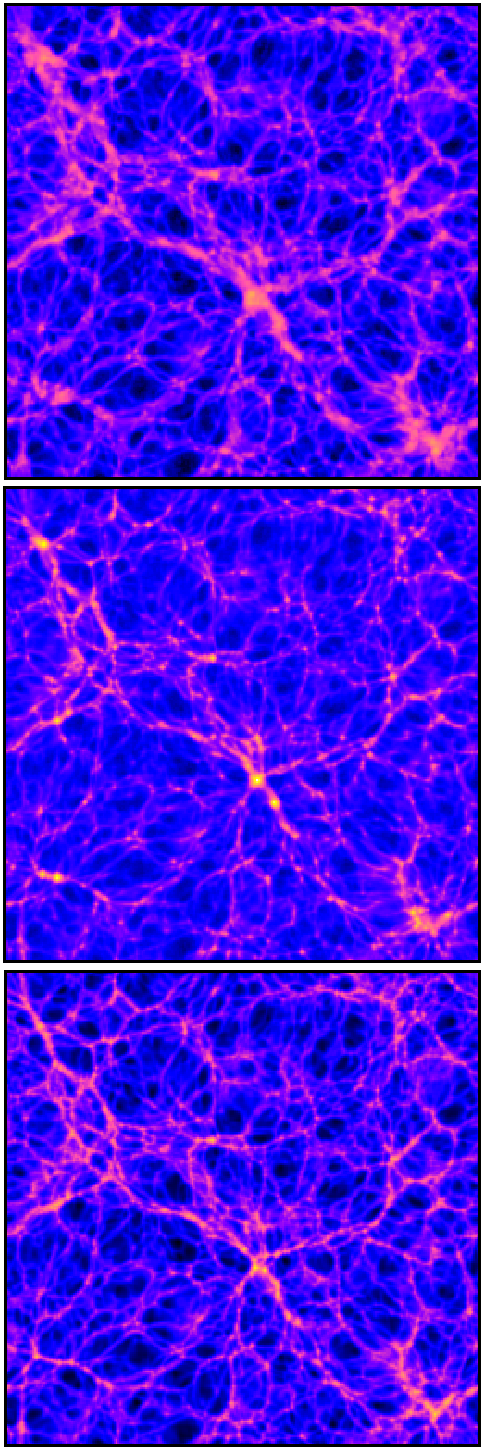}
    \caption{ The large-scale structure of the universe as predicted by the Zel'dovich approximation (top panel), an N-body simulation (centre panel) and the adhesion model (bottom panel). For each case the initial conditions are the same which leads to the formation of the same large scale pattern. Courtesy of \citet{Hidding2010}. }
    \label{fig:evolution:zeldovich_approximation}
\end{figure}

Out of all the different versions of the \Nexus{} technique, \spiderTidal shares the largest number of common points with the Zel'dovich formalism. For example, both approaches use the eigenvalues of the tidal tensor for identifying the cosmic web components. But, most crucially, \spiderTidal uses the tidal tensor computed at the redshift for which we need to identify the different morphological components. In contrast, the Zel'dovich formalism always uses the primordial tidal tensor, neglecting non-linear effects that arise during the subsequent gravitational collapse of matter. Such non-linear effects are important when studying large-scale structures, given that the cosmic web represents the transitional stage between linear structures and fully developed non-linear objects. The eigenvalue threshold used to characterize morphological components represents another crucial difference between the two methods. Within the Zel'dovich approximation, the distinction between positive versus negative eigenvalues is important since they lead to different morphological structures. But using such a criterion for the present time leads to unrealistic structures \citep{Hahn2007a,Forero-Romero09}, which is why \spiderTidal uses a non-zero eigenvalue threshold that varies with redshift, optimized for the detection of the most prominent cosmic web components (\nexusPaper{}).

In spite of these differences, there is a good correspondence between the predictions of the Zel'dovich formalism and the \Nexus{} detections, as seen in \reffig{fig:evolution:zeldovich_approximation}. Except small differences, we find the same large-scale structures in the top and centre panels. Moreover, the figure also illustrates the main limitation of the Zel'dovich approximation, which breaks down when different matter streams cross paths, since then the motion is dominated by the gravitational field of these non-linear structures. This limitation is overcome in the adhesion model via an artificial viscosity term and results in a better description of the later stages of anisotropic collapse \citep{Gurbatov1989,Kofman1990,Kofman1992}. Recently, \citet{Hidding2010} and \citet{Hidding2012,Hidding2013} have shown that adhesion theory is a very useful analytical description of the cosmic web, as seen from the bottom frame of \reffig{fig:evolution:zeldovich_approximation}.

It is important to note that the cosmic web segmentation performed by \Nexus{} or \nexus{} does not always have a one-to-one correspondence with the anisotropic collapse stages predicted by the Zel'dovich and the adhesion formalism. For example, we find many haloes, which are fully collapsed objects, inside filaments and walls. This suggests that the environments we identify characterize the collapse stages on megaparsec scales and not on those of individual haloes. Moreover, it is conceivable that our methods identify filaments and walls that are still in their formation phase, before they fully collapsed along their axes. This is the case since we do not check the virialization state of our detected structures. While this later issue may play some role at high redshift, we suspect that close to the present time it is insignificant. A more thorough analysis of this point is outside the scope of this study and remains to be investigated at another time.

\section{Characterizing the cosmic web environments}
\label{sec:evolution:environment_characterisation}
We are interested in characterizing the properties of morphological components beyond global quantities like mass and volume fractions. Therefore, in the following, we introduce a few new methods to describe the spatial extent and the mass distribution of filaments and walls, at different points along these structures. In particular, we are interested in measuring the length of the filamentary network, as well as the diameter and the linear density of filaments at each position along these objects. Similarly for sheets, we want to measure the total extent of the wall network, as well as the thickness and the surface density of walls. 

\begin{figure*}
    \centering
    \includegraphics[width=\linewidth]{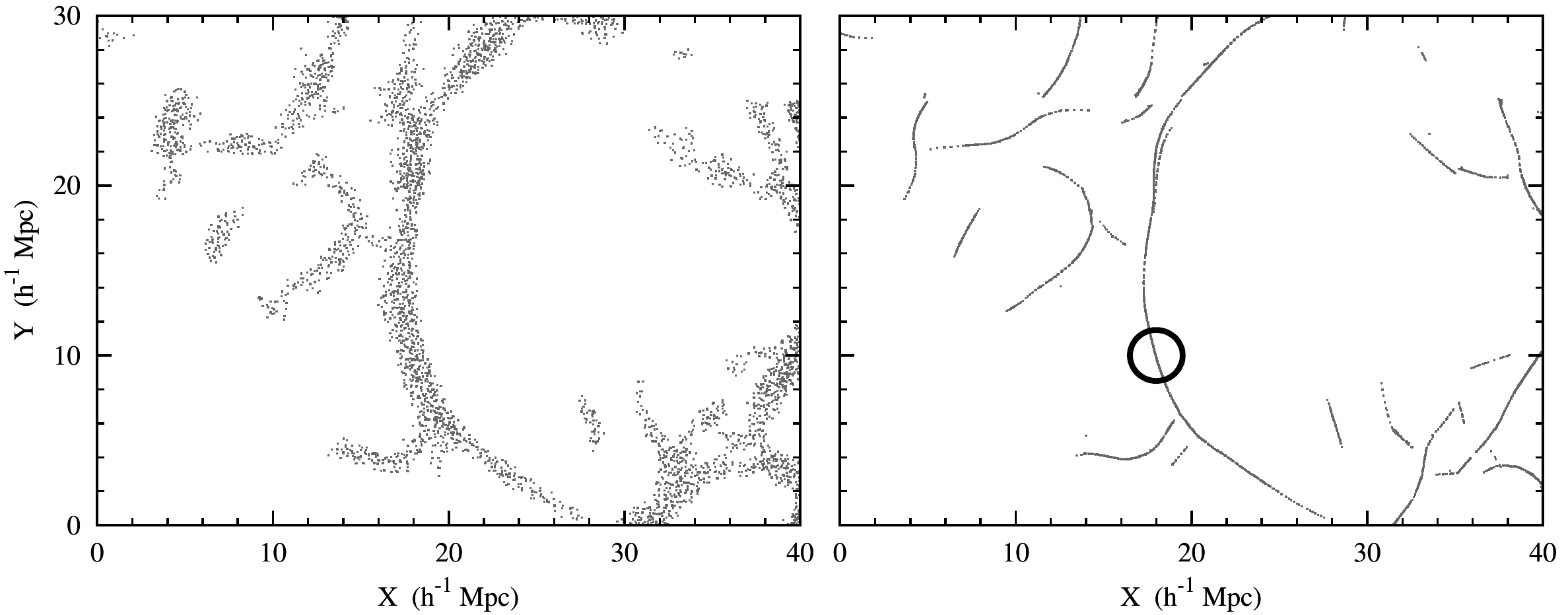}
    \caption{ The compression of the filamentary network towards its central axis. Each point represents a voxel that has been identified as part of a filament. The left frame shows the filamentary network before the contraction. The right panel shows the resulting central axis of the filaments, as sampled by the filament voxels. For more details on the procedure see \refappendix{appendix:evolution:filament_contraction}. The black circle visible in the centre of the right frame shows the filter size used by our approach. The projection shows a $10\Mpch$ thick region of the \MII{}. }
    \label{fig:evolution_fila_contraction_example}
\end{figure*}

\subsection{Compressing filaments and walls }
\label{subsec:evolution:fila_contraction_description}

The first steps in computing the properties of large-scale structures at each point along these objects involves compressing the morphological components to their central axis for filaments and to their central plane for walls. The procedure works by displacing each filament voxel towards the central spine of the object, always moving the voxel perpendicular to the filament orientation. For sheets, every wall voxel is displaced towards the central plane of the structure, always shifting along the normal to the wall plane. The complete description of the compression algorithm is presented in \refappendix{appendix:evolution:filament_contraction}. The compression approach is very useful since, after its application, all the voxels along a filament segment are compressed to a line with the same length (e.g. \reffig{fig:evolution:example_test_filament}). This allows for a simple characterization of the spatial extent and mass distribution of the filament or wall object, as we will see shortly. 

We illustrate the outcome of the compression procedure in \reffig{fig:evolution_fila_contraction_example}, where we show the filaments and their central spine in a small cosmological volume. In the figure, each point represents a voxel identified as being part of a filament, with the two frames showing the distribution of voxels before and after the contraction procedure. The figure clearly shows that the method works very well in compressing the filament network, and even though not shown, it works equally well for wall environments too. Comparing the two panels, we find that the filament spine corresponds very well to the position and orientation of the input filaments. Moreover, any artefacts visible in the figure are mainly due to projection effects and not to failings of the method, as can be seen when inspecting the full 3D data.

\subsection{Computing the filament length, diameter and linear density}
\label{subsec:evolution:fila_properties_2}
In the following we introduce a few more elaborate methods of describing filamentary environments. Concerning their physical extent, filaments can be characterized in terms of their length and their local diameter. The former determines the span of both individual objects as well as that of the entire filamentary network. The latter characterizes the typical width of representative filamentary regions. A complementary approach characterizes the distribution of mass along filaments. This is easily captured by computing the linear mass density, which gives the typical mass of filament segments of unit length.

\subsubsection{Filament length}
\label{subsec:evolution:fila_length_computation}

\begin{figure}
    \centering
    \includegraphics[width=\linewidth]{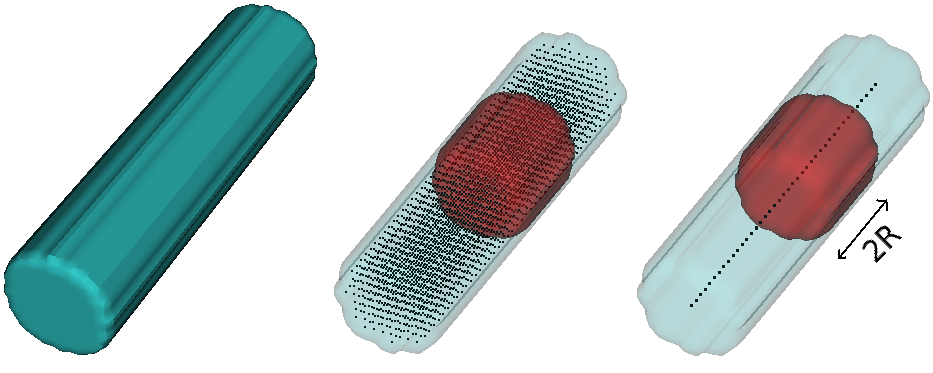}
    \caption{ Illustration of the process used to compute filament length, diameter and linear mass density. We exemplify this for a straight filament with constant diameter (left frame). In \Nexus{}, filaments are sampled at a discrete set of grid points, whose centres are shown as points in the centre panel. Following the compression procedure, the filament grid points are displaced to the central axis of the object (right panel). The dark-red shaded areas show a segment of length $2R$ along the filament. }
    \label{fig:evolution:example_test_filament}
\end{figure}

The compressed networks can be used to compute the length of filaments, and in a very similar way, the area of walls. For simplicity, we illustrate this with the help of \reffig{fig:evolution:example_test_filament} where we show a perfectly straight filament. As in the case of the \Nexus{} filaments, this object is sampled by a set of voxels indicated in the figure by points. Following the compression procedure, the filament voxels are compacted to a line, as shown in the right most panel. For explanatory purposes, we show only a fraction of the points along the spine, to make it clear that the filament axis is sampled by a discrete set of points.   

Given that the compressed filaments are represented by discrete distributions of points, we need to use a filtering procedure to smooth out any shot noise. For this we choose a spherical filter of radius $R$, where the value of $R$ is motivated by two requirements. First, $R$ needs to be significantly larger than the distance between neighbouring points along the central axis of filaments. The limit in this case is the very thin filaments which have only one voxel along their cross-section. For such objects, their spine is sampled at distances equal to the grid spacing $\Delta x$. Secondly, the smoothing scale needs to be much smaller than the typical radius of curvature of filaments, otherwise we underestimate the true length. Given the two complimentary constraints, we compromised on the value $R=2\Mpch$. To put this filter scale within context, we show it as a black solid circle of radius $R$ in the right frame of \reffig{fig:evolution_fila_contraction_example}. We find that the typical curvature radius of filaments is large compared to the filter size and therefore, to a good approximation, the filaments are straight within the smoothing scale $R$. While spurious artefacts may arise from the intersection of two or more filaments, the compression algorithm automatically cuts the branches from the main trunk and therefore this issue is minimal. In fact, most of the filamentary intersections seen in \reffig{fig:evolution_fila_contraction_example} are due to projection effects.

To compute the length of filaments, we proceed by estimating the contribution to the length brought by each voxel of the filament. For this, we place at each point along the filament spine a sphere of radius $R$ and count the number of points $N_\rmn{points}$ encompassed within the sphere. If the sphere intersects the filaments anywhere except at its ends, then the sphere encloses a filament segment of length $2R$. This is the segment enclosed by the dark shaded cylinder shown in \reffig{fig:evolution:example_test_filament}. We assume that the length $2R$ of the segment is distributed uniformly between the enclosed points, therefore, the point at the centre of the sphere has a contribution of
\begin{equation}
	\Delta l = \frac{2R}{N_\rmn{points}}
\end{equation}
to the total filament length. Once we find the $\Delta l$ value associated with each filament voxel, the length of the complete filament is obtained by summing over the contribution of all the points. This method can be used to compute the length of individual filamentary objects (see \refsec{subsec:evolution:filament_individual_extent}) as well as the total length of the filamentary network (see \refsec{subsec:evolution:evolution_filament_length}).

\subsubsection{Filament diameter}
\label{subsec:evolution:fila_diameter_computation}

Given the central axis of filaments, we can compute the width and mass density for each unit of filament length. To better exemplify this, we consider the case of a straight filament with constant diameter $D_\rmn{filament}$, as shown in \reffig{fig:evolution:example_test_filament}. If we take a segment of length $\Delta L = 2R$ along the filament spine, then the number of voxels $N_\rmn{points}$ contained in this segment is given by
\begin{equation}
	N_\rmn{points} = \frac{\pi \; \Delta L \; D_\rmn{filament}^2}{4 \; V_\rmn{voxel}}
	\label{eq:evolution:fila_prop_diameter_1} \;,
\end{equation}
where $V_\rmn{voxel}$ denotes the volume of a voxel. In practice, we know the number of points contained in the filament segment, but not its diameter. Therefore, we can invert the above relation to obtain 
\begin{equation}
	D_\rmn{filament} = \sqrt{ \frac{4 \; V_\rmn{voxel} \; N_\rmn{points}}{\pi \; \Delta L} }
	\label{eq:evolution:fila_prop_diameter_2} \;.
\end{equation}

\begin{figure}
    \centering
    $\begin{array}{cc}
    \includegraphics[width=.17\linewidth]{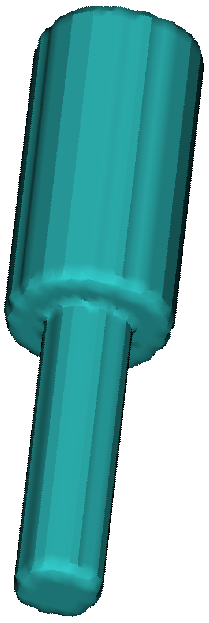}  &
    \includegraphics[width=.79\linewidth,angle=0]{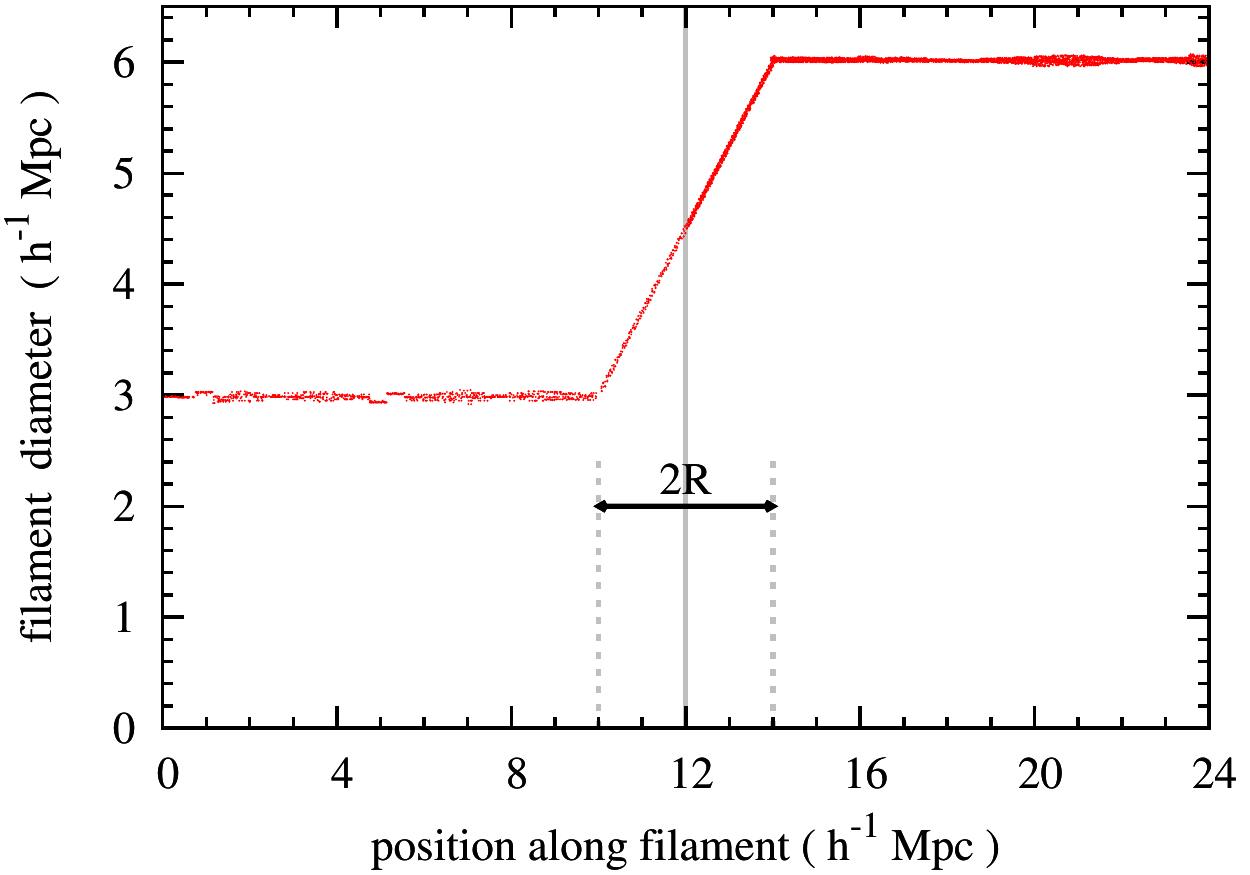}
    \end{array}$
    \caption{ Illustration of what the local filament diameter actually computes. \textit{Left panel:} It shows a straight filament that has two different diameters. Half of it has a $3\Mpch$ diameter, while the other half has a $6\Mpch$ diameter. \textit{Right panel:} It show the local filament diameter measured along segments of length $2R$. The solid vertical line shows the coordinate where the filament diameter changes from $3$ to $6\Mpch$. }
    \label{fig:evolution:example_test_filament_double}
\end{figure}

The above expression measures the local filament diameter along representative stretches of the filament network. In other words, it computes the mean diameter of a filament segment of length $2R$. To better understand this, we exemplify it using a test case. In \reffig{fig:evolution:example_test_filament_double} we show a straight filament that has two different diameters, with a sharp transition from a $3$ to a $6\Mpch$ diameter value. The measurement of the filament diameter as described above is shown in the right frame of the figure. It clearly shows that the computed quantity is the average diameter within a window function of length $2R$ along the filament. The figure also illustrates that there is some scatter in the $D_\rmn{filament}$ measurements. This arises from the fact that the filament is sampled in a discrete manner. Thinner filaments, sampled with fewer voxels along any given segment, will show a larger scatter in the estimated $D_\rmn{filament}$ values.

The computation of the local filament diameter depends on the value of the filament segment $\Delta L$. Small $\Delta L$ values result in noisy estimates due to the discrete sampling of the object. Too long segments result in measuring the average diameter over large filament stretches, washing out some of the local variation in the extent of filaments. Therefore, we select a value $\Delta L=2R=4\Mpch$ which we found to be a good compromise between the two requirements.

\subsubsection{Filament linear mass density}
\label{subsec:evolution:fila_linear_density_computation}
Similar to the way we just computed the local filament diameter, we can also measure the distribution of mass in each representative filament stretch. For an ideal filament with a constant mass distribution along its length, the mass $\Delta M$ contained in a segment $\Delta L$ is given by
\begin{equation}
	\Delta M = \Delta L \; \zeta_\rmn{filament}
	\label{eq:evolution:fila_prop_density_1} \;.
\end{equation}
The filament linear density $\zeta_\rmn{filament}$ represents the mass contained in a unit length of the filament. In general, the mass $\Delta M$ contained in a filament segment can be easily computed by summing the mass contained in each voxel that is part of that filament stretch. Therefore, the linear density can be computed as
\begin{equation}
	\zeta_\rmn{filament} = \frac{\Delta M}{\Delta L}
	\label{eq:evolution:fila_prop_density_2} \;.
\end{equation}
Similarly to the filament diameter, the linear mass density is a local quantity that characterizes representative filament stretches. Therefore, the computation of $\zeta_\rmn{filament}$ has analogous behaviour and properties as the $D_\rmn{filament}$ quantity that we just discussed.

\begin{figure*}
    \centering
    \includegraphics[width=\linewidth]{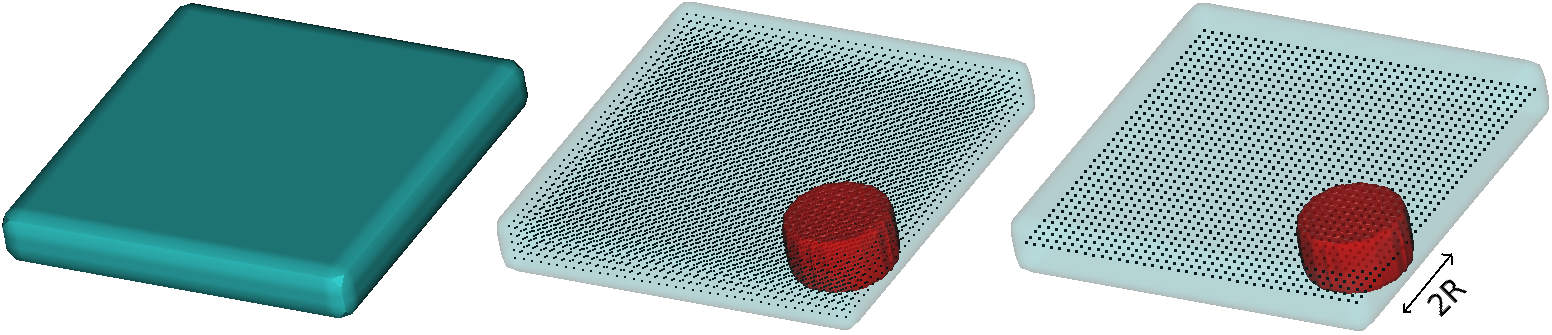}
    \caption{ Illustration of the process used to compute wall area, thickness and surface mass density. We exemplify this for a planar sheet with constant thickness (left frame). In \Nexus{}, walls are sampled at a discrete set of grid points, whose centres are shown as points in the centre panel. Following the compression procedure, the wall grid points are displaced to the central plane of the object (right panel). The dark shaded area shows a circular cross-section of radius $R$ along the plane of the wall. }
    \label{fig:evolution:example_test_wall}
\end{figure*}

\subsection{Computing the wall area, thickness and surface density}
\label{subsec:evolution:wall_properties_2}
Similarly to the way filaments are described by their length, diameter and linear density, sheet environments are characterized by their total area, thickness and surface density. In fact, the methods used to compute wall properties are very much alike to those employed to determine filamentary attributes. Therefore, in the following, we only focus on the few key differences between the two morphological components.

\subsubsection{Wall area}
\label{subsec:evolution:wall_area_computation}

To illustrate the measurement of the area of walls, we present in \reffig{fig:evolution:example_test_wall} the example of a fully planar wall with constant thickness. The centre and right panels show the voxels used to sample this structure as well as the resulting central plane of the object, after applying the wall compression procedure.

Measuring the total area of sheets reduces to finding the contribution of each wall voxel to this quantity. For this, we place at each point along the central plane of the wall a sphere of radius $R$. The points enclosed by this sphere correspond to the wall voxels represented by the darker shaded region in \reffig{fig:evolution:example_test_wall}. Therefore, these $N_\rmn{points}$ correspond to a circular cut of area $\pi R^2$ perpendicular to the wall plane. Assuming that the area of the circular cut is uniformly distributed among all the points enclosed by it, the area contribution associated with a point is given by
\begin{equation}
	\Delta a = \frac{\pi R^2}{N_\rmn{points}}
\end{equation}
The total area of the wall is given by summing over all the contributions of each point that is part of a sheet. Note that computing the wall area is affected by the same issues raised when determining the filament length in \refsec{subsec:evolution:fila_length_computation}.

\subsubsection{Wall thickness}
\label{subsec:evolution:wall_thickness_computation}
To compute the wall thickness $t_\rmn{wall}$, we consider a wall section of area $\Delta A=\pi R^2$. This is shown as the dark region in \reffig{fig:evolution:example_test_wall}. The number of voxels contained in this domain is proportional to the wall thickness $t_\rmn{wall}$ and is given by
\begin{equation}
    N_\rmn{points} = \frac{t_\rmn{wall} \;\Delta A} {V_\rmn{voxel}}
	\label{eq:evolution:wall_prop_thickness_1} \;.
\end{equation}
In practice, we know the number of points contained in the selected region, but not the local thickness of the sheet. Therefore, we can invert the above relation to obtain the local wall thickness as
\begin{equation}
	t_\rmn{wall} = \frac{V_\rmn{voxel} \; N_\rmn{points}}{\Delta A}
	\label{eq:evolution:wall_prop_thickness_2} \;.
\end{equation}
The thickness $t_\rmn{wall}$ defined above is a local quantity that characterizes the mean wall width on representative stretches along these structures. For a sheet of constant thickness, $t_\rmn{wall}$ is the actual width of these objects. In the more realistic case of a varying wall thickness, $t_\rmn{wall}$ gives the mean width within the given smoothing window. For further details see \refsec{subsec:evolution:fila_diameter_computation}.

\subsubsection{Wall surface mass density}
\label{subsec:evolution:wall_density_computation}
The amount of mass per unit area contained in walls is computed in the same way as the sheet thickness. The total mass of a wall circular cut of area $\Delta A=\pi R^2$ is given by
\begin{equation}
	\Delta M = \Delta A \; \sigma_\rmn{wall}
	\label{eq:evolution:wall_density_1} \;,
\end{equation}
where $\sigma_\rmn{wall}$ denotes the local surface mass density of the wall. This can be easily computed by inverting the above equation, to obtain 
\begin{equation}
	\sigma_\rmn{wall} = \frac{\Delta M}{\Delta A}
	\label{eq:evolution:wall_density_2} \;.
\end{equation}
The quantity $\Delta M$ denotes the total mass enclosed in the dark shaded region of \reffig{fig:evolution:example_test_wall}, which is the sum over the mass contained in each voxel enclosed by the selected region. Therefore, $\sigma_\rmn{wall}$ is the mean surface density of a representative stretch of wall area.


\section{Properties of cosmic web environments at $z=0$}
\label{sec:evolution:present_properties}
This section focuses on quantifying the characteristics of the cosmic web environments at the present time of $z=0$. In doing so, we investigate the outcomes of the five methods that we employ for the segmentation of the cosmic web into its components. The goal is to obtain a more quantitative comparison of the methods and therefore a better assessment of their similarities and differences.

\subsection{Mass and volume fractions}
\label{subsec:evolution:mass_fraction}

\begin{figure}
    \centering
    \includegraphics[width=0.75\linewidth]{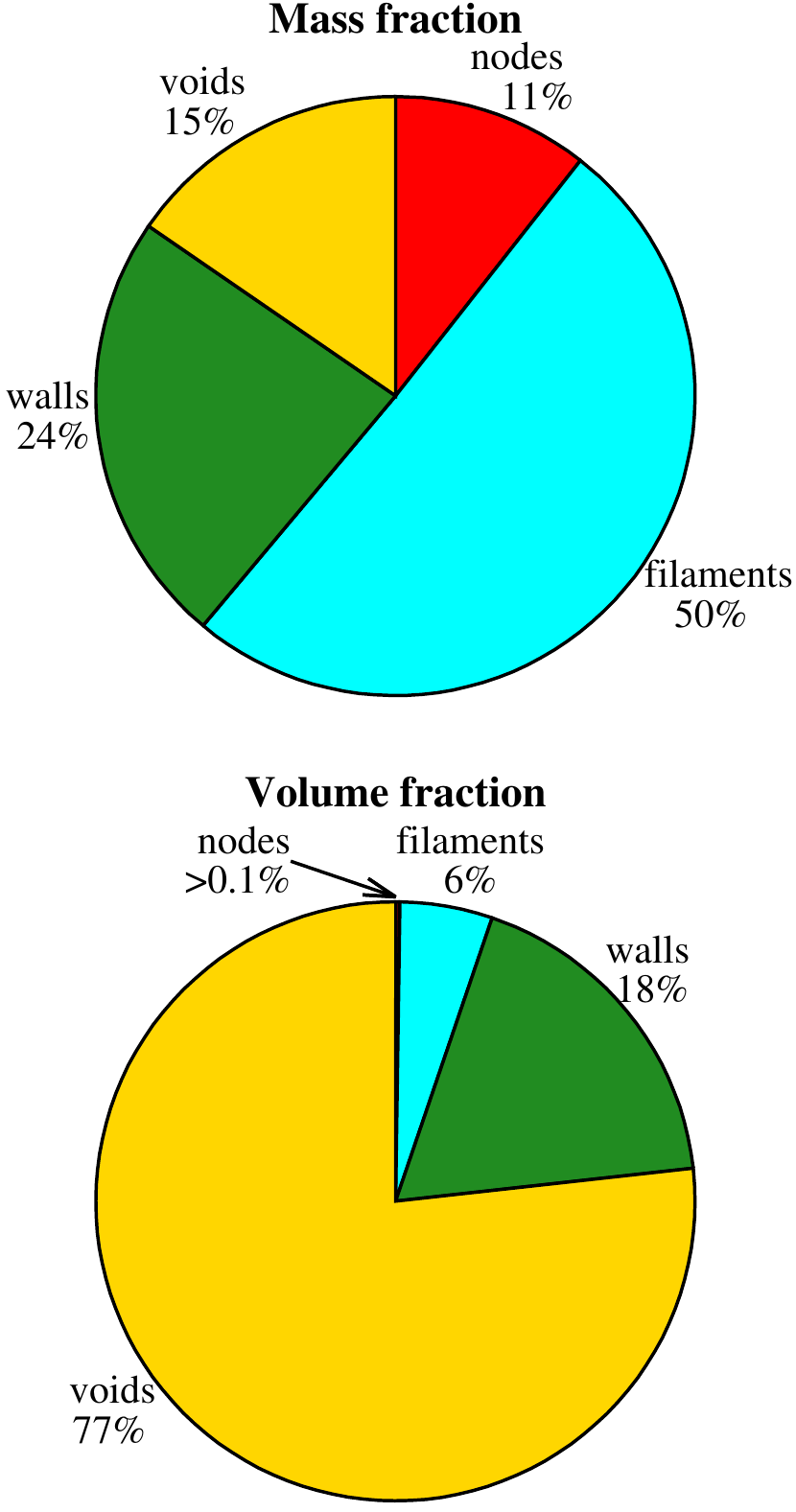}
    \caption{The mass and volume fractions occupied by cosmic web environments detected by the \Spider method. }
    \label{fig:nexus+_fractions}
\end{figure}

One of the easiest way of studying large scale environments involves evaluating global quantities, like the mass and volume content of these structures. Such a determination is shown in \reffig{fig:nexus+_fractions}, which presents the mass and volume fractions occupied by the cosmic web components identified by \SpideR. We find that while the nodes have a significant fraction of the total mass content of the universe, they occupy a negligible volume. Following that, we have the filamentary network which contains, into a relatively small volume, half of the total cosmic matter distribution. The walls have a fair share of both the mass and volume fractions, with a mean density $1+\delta\sim1$. On the other hand, voids take by far the largest volume fraction in the universe, but they only have $\sim15\%$ of the total mass content. This makes voids the most underdense regions, with an average density of $1+\delta=0.2$ which is in very good agreement with the predictions of the excursion set formalism \citep{Shethwey04}.

\begin{figure}
    \centering
    \includegraphics[width=\linewidth]{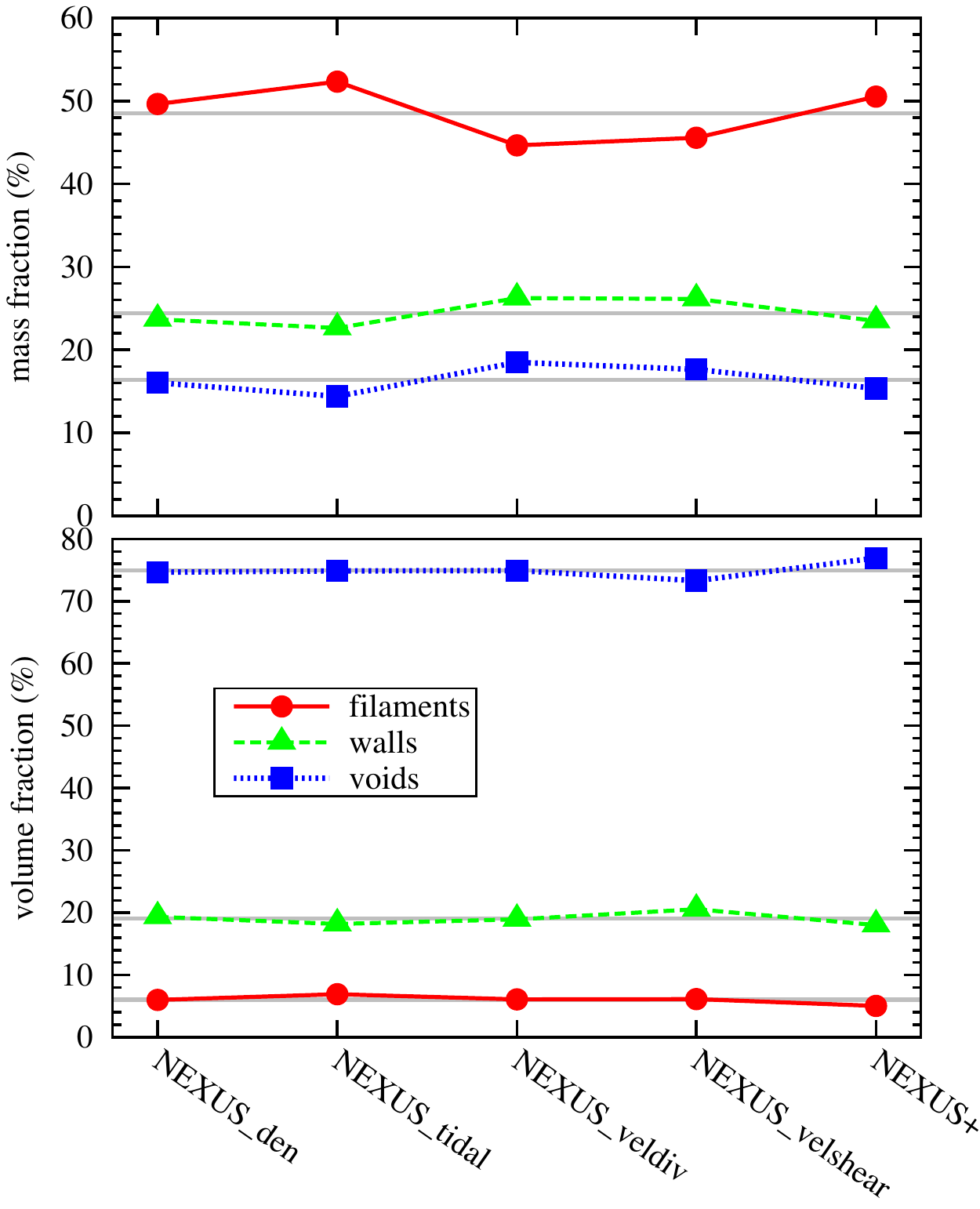}
    \caption{The mass (top panel) and volume (bottom panel) fractions occupied by the cosmic web components as identified by the \spider and \Spider methods. The results are obtained for a redshift of $z=0$. We give the outcome for: filaments (circles with solid line), walls (triangles with dashed line) and voids (squares with dotted line). The horizontal lines for each case show the average result of the five methods. All methods give similar results, with only minor variations. }
    \label{fig:mass_volume_fraction}
\end{figure}

\begin{table*}
    \centering
    \caption{ The mass and volume fractions of each cosmic web component as found by the current work and previous studies. Most results were obtained by analysing N-body simulations, with the exception of the last two rows which present analytical results. \citet{Doroshkevich70} predictions are for a Gaussian random field, while \citet{Shen06} results use the excursion set formalism. }
    \label{tab:evolution_mass_fraction}
    \begin{tabular}{l cccl rccc}
        \hline
        \multirow{2}{*}{Method} & \multicolumn{4}{c}{Mass fraction ($\%$)} & \multicolumn{4}{c}{Volume fraction ($\%$)} \\
          & cluster & filament & wall & void & cluster & filament & wall & void \\
        \hline
        \nexus{} - this work                 & 11 & 50 & 24 & 15 & $<0.1$ &   6 & 18 & 77 \\ [.03cm]
        \cite{2010MNRAS.408.2163A}  & 28 & 39 &  6 & 27 &        0.4 &   9 &   5 & 86 \\ [.03cm]
        \cite{Hahn2007a}                     &  - &  - &  - &  - &               2 & 31 & 54 & 13 \\ [.03cm]
        \cite{Forero-Romero09}           &  9 & 40 & 35 & 16 &           1 & 15 & 42 & 42 \\ [.03cm]
        \cite{Hoffman2012}                 & 17 & 34 & 36 & 13 &        0.4 &   5 & 27 & 68 \\ [.03cm]
        \cite{Shandarin2012}               & 41 & 18 & 17 & 24 &        0.7 &   2 &   5 & 93 \\ [.03cm]
        \hline
        \cite{Doroshkevich70}             &   8 &  42 &   42 &   8 &           8 & 42 & 42 &  8 \\ [.03cm]
        \cite{Shen06}                          & 46 & 26 &  27 &   1 &           - &   - &   - &  - \\
        \hline
    \end{tabular}
\end{table*}

\Reffig{fig:mass_volume_fraction} shows how the mass and volume fractions change when considering the filaments, walls and voids found by the other comic web identification methods. With symbols we show the actual mass and volume fractions while the solid horizontal lines show the average over the five methods, for each environment in question. We find that all the methods return approximatively the same values for the mass and volume fractions, but that there are some differences that are especially visible in the mass fraction result. The methods based on the density field, i.e. \spiderDeN, \spiderTidal and \SpideR, suggest that filaments contain a larger share of the mass than the velocity based approaches. These former methods use the density field as starting point and therefore are more likely to characterize higher density regions as being part of filaments, and therefore assigning a larger mass to filamentary environments. The remaining two environments show an opposite trend, with  walls and voids identified by \spiderVeldiv and \spiderVelshear containing a larger mass fraction. This is an outcome of the smaller mass found in filaments, which implies that there is a larger mass share left to be split between walls and voids. In contrast, the volume fraction shows a much better agreement between different methods. While there are discrepancies between the results of various methods, these are small at ${<}10\%$. Therefore, the mass and volume fraction are robust environmental characteristics that are largely independent of the underlying field used to identify cosmic web features.

Even though our five approaches find consistent mass and volume filling fractions, previous studies on this topic have found a wide range of values \citep{Doroshkevich70,Shen06,Hahn2007a,Forero-Romero09,2010MNRAS.408.2163A,Shandarin2012}. For example, the void volume fraction can vary from ${\sim}10\%$ \citep{Hahn2007a,Forero-Romero09} up to ${\sim}90\%$ \citep{Aragon07b,Shandarin2012}. To underline this point, we summarize in \reftab{tab:evolution_mass_fraction} the mass and volume fraction of each cosmic web component obtained by previous studies. These large discrepancies arise because different studies use various criteria for segmenting the cosmic web, with different precepts giving sometimes very disparate results. Therefore, it is very challenging to compare our results with previous works. This is obvious even when contrasting with the mass and volume fractions found by the MMF algorithm, which is very similar to our \spiderDen method. For instance, \cite{2010MNRAS.408.2163A} find a mass fraction of ${\sim}30\%$ for MMF cluster regions, which is a factor of 3 larger than our result. The disparity is due to \cite{2010MNRAS.408.2163A} identifying as clusters any quasi-spherical objects above the resolution limit of their simulations, while we limit our selection to only the most massive such objects. The discrepancies between methods become even larger when comparing results for filaments and walls, since usually the detection of these components is sensitive to the identification of cluster regions.

In particular, the analysis of the DM phase space sheet illustrated in \citet[][see also \citealt{Abel11,Falck2012}]{Shandarin2012} gives a very natural way of characterizing large-scale structures. Such an approach allows for the identification of single- and multi-stream regions which, according to the Zel'dovich formalism \citep{Zeldovich70}, correspond to different stages of the anisotropic collapse of matter. The single-stream regions correspond to voids, while the multi-stream regions are indicative of walls, filaments and clusters. In particular, the \citet{Shandarin2012} method illustrates the dominance of voids in terms of volume and that ${\lsim}60\%$ of the mass in the universe is in clusters and filaments. These results are in good agreement with the \Nexus{} and MMF findings, suggesting that the morphology of the density field is a good tracer of the anisotropic collapse of matter.

The fact that our five approaches give consistent results suggests that the same cosmic web pattern is imprinted in all the cosmological fields: density, tidal tensor, velocity divergence and velocity shear. Therefore, discrepancies between different studies are mainly due to variations in the criteria used to characterize the cosmic web, and not because of the underlying cosmic field used to trace the large scale environments.

\subsection{Cross-correlation between identification methods}
\label{subsec:evolution:cross_correlation}

\begin{figure*}[t]
    \centering
    \includegraphics[width=.99\linewidth]{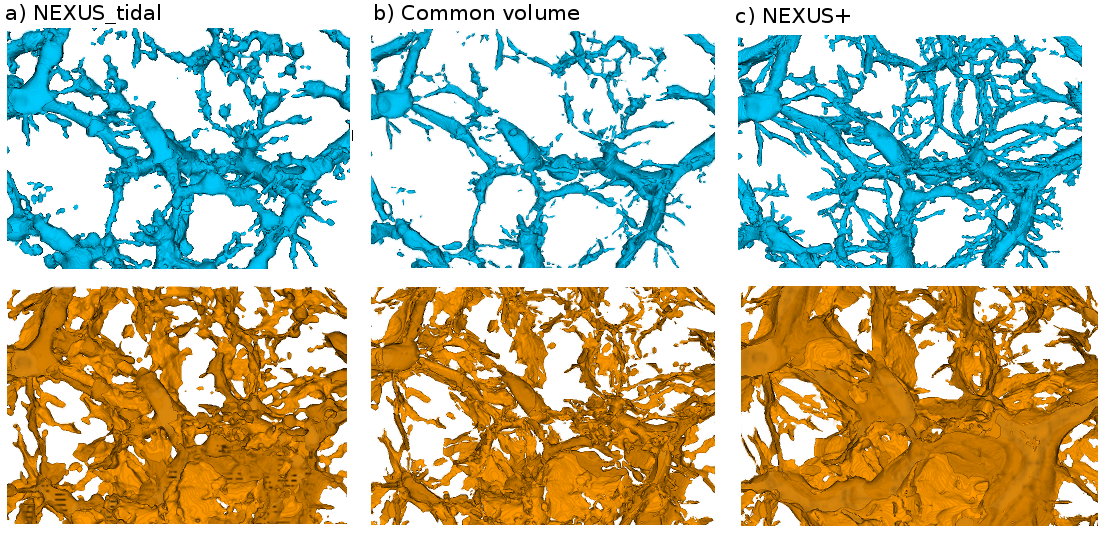}   
    \caption{ Identifying the common regions between filaments (top-row) and walls (bottom-row) detected using two different methods. The left and right columns give the cosmic web components identified by \spiderTidal and \nexus{} respectively. The centre column gives the common volume regions that were identified by both methods as part of filaments and walls respectively. }
    \label{fig:env_fila_intersection}
    \vskip 0.5truecm
    \centering
    \mbox{\hskip -0.truecm\includegraphics[width=1.0\linewidth]{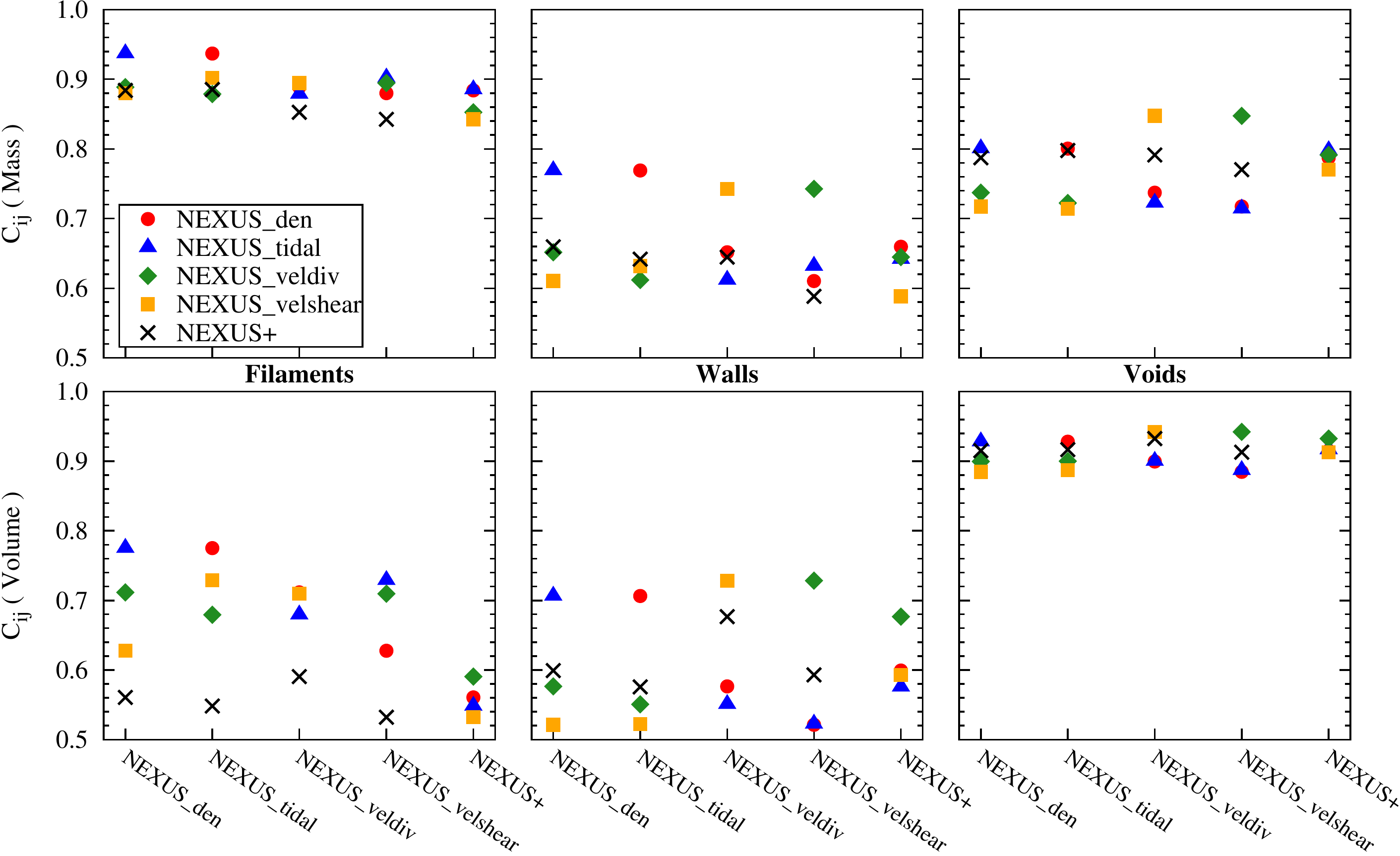}}
    \caption{The mass (upper row) and volume (bottom row) cross-correlation coefficients between environments identified with \spiderDeN, \spiderTidaL, \spiderVeldiV, \spiderVelshear and \SpideR. We show the cross-correlation for filaments (left column), walls (centre column) and voids (right column). For each method, we compute the mass and volume cross-correlation with each of the other four identification methods. }
    \label{fig:cross-correlation}
\end{figure*}

The visual impression given by \reffig{fig:evolution:env_filaments} suggests that there are important differences between the filaments detected with the five approaches, but these variations do not seem to be captured by the global quantities analysed above. To get a better characterization of these discrepancies we resort to cross-correlate the volume and mass contained in environments obtained via different methods. This is illustrated in \reffig{fig:env_fila_intersection} where we show the filaments and walls identified by \spiderTidal and \nexus{}. The middle column shows the volume regions that were identified by both methods as being part of filaments and walls respectively. To quantify the overlap between detection methods, we define the volume cross-correlation coefficient $C_{ij}(V)$ as
\begin{equation}
	C_{ij}(V) = 2\frac{V_{i\cap j}}{V_i+V_j} \;,
\end{equation}
where $V_{i\cap j}$ is the volume of the common regions identified by both methods $i$ and $j$. It corresponds to the volume shown in the middle column of \reffig{fig:env_fila_intersection}. With $V_i$ and $V_j$ we denote the environmental volumes found by procedures $i$ and $j$ respectively. The volume cross-correlation coefficient is computed separately for each cosmic web component and quantifies the percentage of volume that is common to environments detected via different methods. Similarly, the mass cross-correlation coefficient is given by
\begin{equation}
	C_{ij}(M) = 2\frac{M_{i\cap j}}{M_i+M_j} \;,
\end{equation}
where $M_{i\cap j}$ is the mass contained in regions found by both methods as being part of the same environment. It corresponds to the mass enclosed in the volume shown in the middle column of \reffig{fig:env_fila_intersection}. With $M_{i}$ we denote the mass contained in the morphological component detected by method $i$. The $C_{ij}(M)$ quantity characterizes what fraction of the mass is identified as being contained in the same cosmic environment by two different methods. 

The mass and volume cross-correlation coefficients are illustrated in \reffig{fig:cross-correlation}. We present results for filament, wall and void environments for each possible pair of detection methods. In the case of filaments, we find a large value for the mass cross-correlation coefficient which implies that all methods identify roughly the same mass distribution as being contained inside filaments. In contrast, the much lower values of the volume cross-correlation suggest that there is a much larger variation in the spatial regions that are identified as filaments. These two findings indicate that most of the mass in filaments is contained within prominent structures that are identified by all the methods. On top of that, there are additional tenuous filaments, which even though contain only a small share of the total mass in filaments, they do occupy a similar volume with the more pronounced structures. The detection of these tenuous filaments varies greatly between methods which considerably lowers the $C_{ij}(V)$ value. This interpretation is in agreement with the visual impression given by \reffig{fig:env_fila_intersection}.

For cosmic walls we find that both the mass and volume cross-correlation coefficients are quite low, with typical values $\lsim 0.65$. Therefore, wall regions can vary greatly between the outcomes of different methods. This can easily be appreciated from \reffig{fig:env_fila_intersection}, which shows that the overlap between \spiderTidal and \nexus{} walls is not very substantial. There are two sets of approaches that show a closer match between the cosmic sheets they identify: \spiderDen matches better with \spiderTidaL; and \spiderVeldiv matches with \spiderVelsheaR. While these two sets stand out compared with the other pairs, even their cross-correlation coefficients are low. In the case of voids, we find a low $C_{ij}(M)$, but a high $C_{ij}(V)$ value; in contrast with the results for filaments. It suggests that most of the void volume is given by very low density regions that are consistently identified by all the methods as being part of voids. But a significant part of the mass in voids comes from the higher density regions around the void edges, whose inclusion or exclusion from voids varies from method to method.

\subsection{Mass distribution}
\label{subsec:evolution:density}

\begin{figure}
    \centering
    \includegraphics[width=\linewidth]{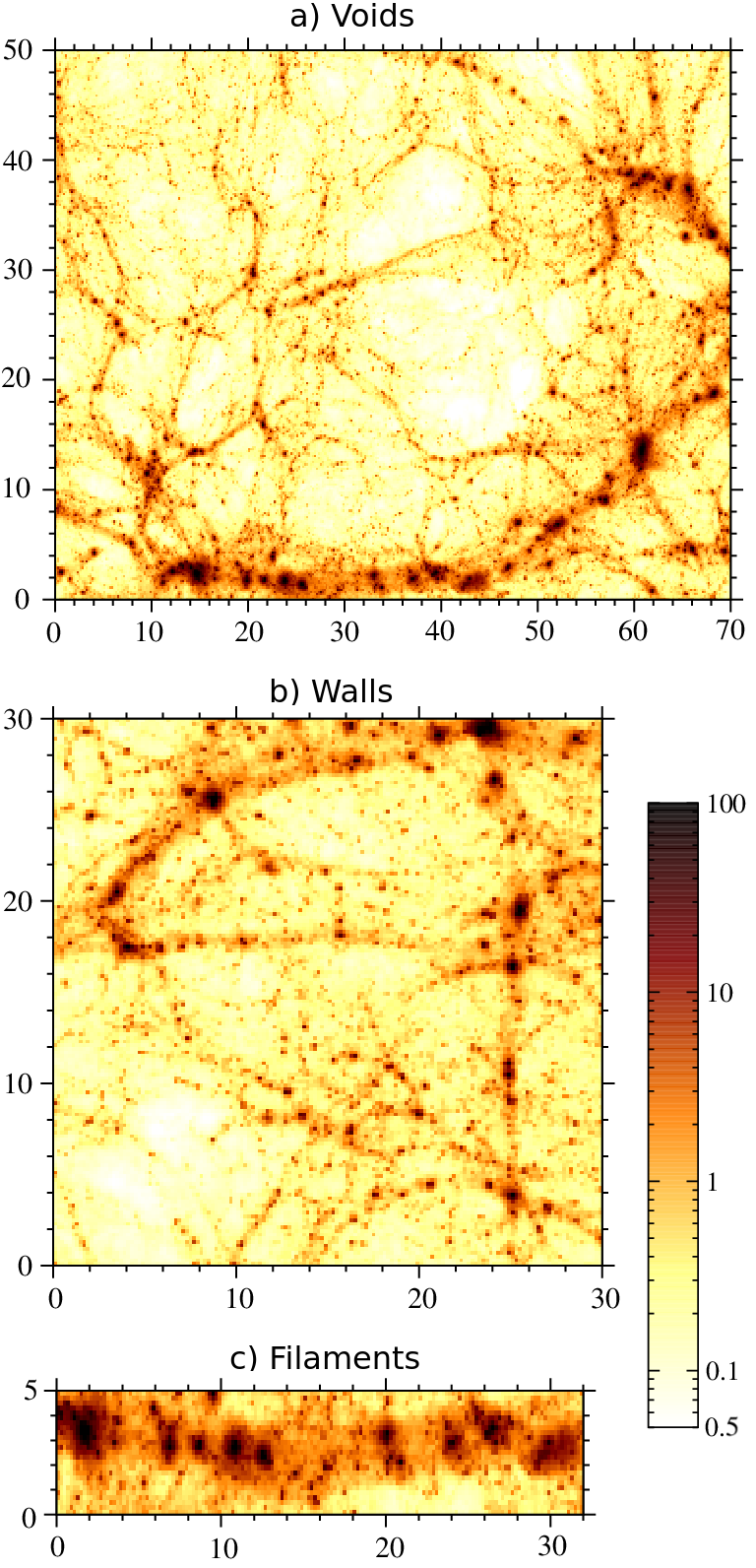} 
    \caption{ The density distribution across a few typical void (top), wall (centre) and filament (bottom) stretches. The density scale is shown in the right-most column, with light and dark patches corresponding to underdense and overdense regions. Note that each panel has a different physical size as indicated by the coordinate ticks. } 
    \label{fig:evolution:env_density_maps}
\end{figure}

Global quantities like mass and volume fraction offer only a very basic overview of the attributes of morphological components. In fact, the properties of large-scale structures can differ from object to object, and they can even vary across the same structure. This can be easily appreciated from \reffig{fig:evolution:env_filaments} which clearly shows the extent of variation in filament widths between different filamentary branches. The same is true regarding the distribution of matter across environments, with a large degree of diversity between different objects as well as across the same structure. This is illustrated in \reffig{fig:evolution:env_density_maps}, where we show the density distribution in a few representative stretches of void, wall and filamentary regions. Panel a) shows a large void bounded by prominent walls in the top and right-hand side. The inside of the underdense region is not smooth, but in fact shows a significant variation in density from point to point. A similar diversity is observed in the sheet shown in panel b) or in the filament given in frame c). 

A major feature of the cosmic web is its hierarchical nature, which can be easily assessed from \reffig{fig:evolution:env_density_maps}. This is especially indicative when focusing on the void region shown in panel a). The insides of the void have a large amount of substructure, with some corresponding to tenuous walls, which delimit smaller sub-void regions. This is a clear manifestation of the hierarchical distribution of voids, with smaller voids enclosed within larger underdense regions \citep{Weykamp93,Shethwey04,Platen07,Aragon10b,Aragon-Calvo2013,Rieder2013}. The same hierarchical nature is present in the distribution of walls and filaments, as can be seen in \reffigS{fig:evolution:env_filaments} and \ref{fig:env_fila_intersection}. Thick filaments and walls branch into thinner structures that pervade most of the cosmos, even in very underdense regions \citep[][]{Weykamp93,Gottlober03,Platen07,Aragon-Calvo2013}. The great deal of structures and substructures present over a wide range of scales and densities is a clear manifestation of the hierarchical development of the cosmic web \citep{Sheth04,Shethwey04,Shen06}.

\begin{figure}
    \centering
    \includegraphics[width=\linewidth]{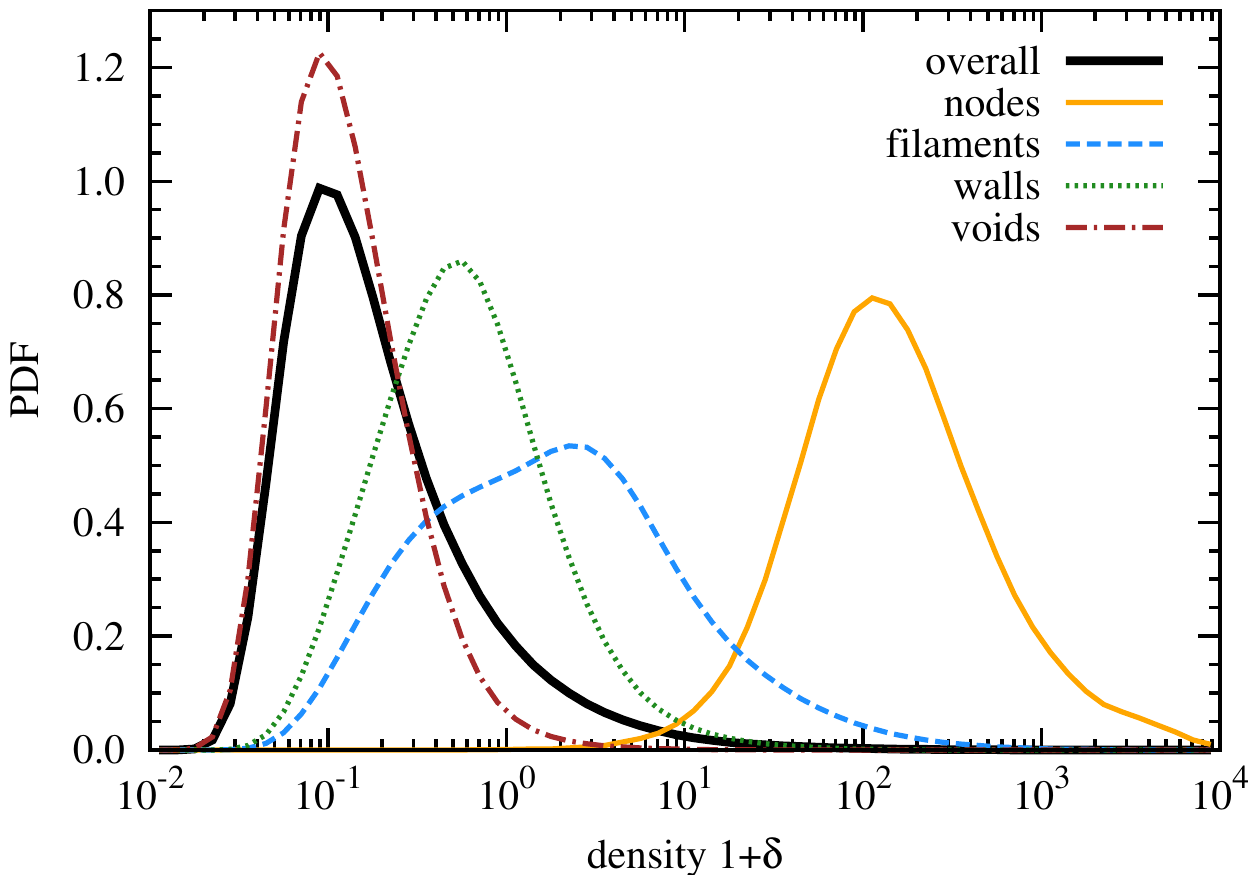}
    \caption{The density PDF in each environment of the cosmic web as detected by \SpideR. The histogram was obtained using the \dtfe{} density field on a regular grid with spacing $\Delta x=0.4\Mpch$. No additional smoothing was used. }
    \label{fig:density-histogram}
    \centering
    \includegraphics[width=\linewidth]{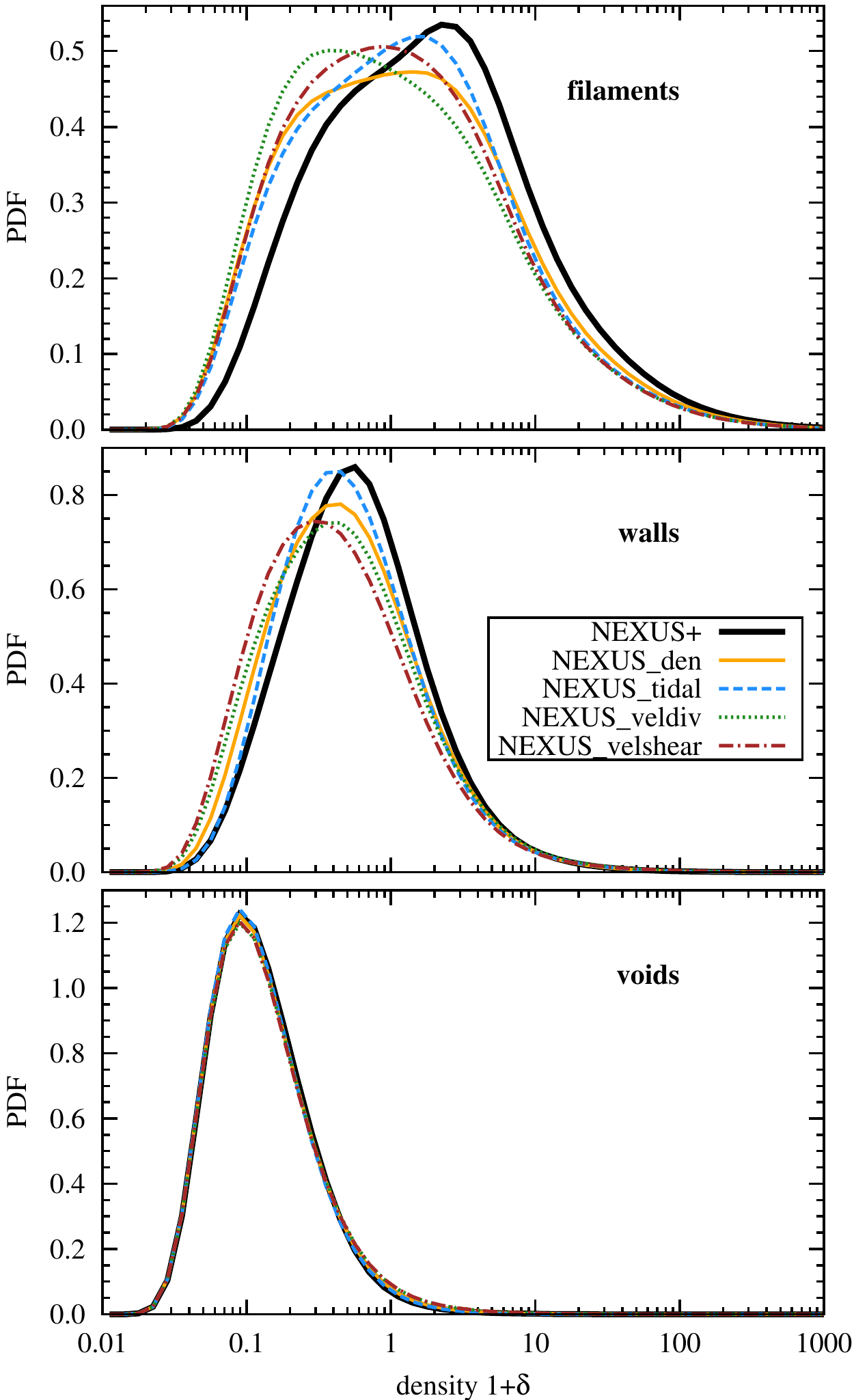}
    \caption{A comparison of the density PDF when using environments identified by \spiderDeN, \spiderTidaL, \spiderVeldiV, \spiderVelshear and \SpideR. The three panels show: filaments (top-right), walls (bottom-left) and voids (bottom-right). The histogram was obtained using the \dtfe{} density field on a regular grid with spacing $\Delta x=0.4\Mpch$. No additional smoothing was used. }
    \label{fig:density-histogram_comparison}
\end{figure}

The analysis of the density distribution represents the simplest way of characterizing the variation of the matter content across environments. This is even more interesting, given that previous studies have used density thresholds as a simple method of identifying clusters and filaments \citep{Eke96,Shandarin04,Dolag06}. Studying the typical densities of morphological components allows us to both asses how successful such methods are and also to get a better understanding of large-scale structures. \Reffig{fig:density-histogram} shows the probability distribution function (PDF) of the density field segregated according to cosmic environments. We use the \dtfe{} density extrapolated to a regular grid with spacing $\Delta x=0.4\Mpch$, with no additional smoothing. While the actual density values depend on the smoothing scale, the qualitative conclusions that we arrive to are largely independent of the filter scale. 

The figure clearly shows that various environments are characterized by different density values. The node regions have typically the highest density, with values $\gsim100$. The filaments also represent overdense environments, though to a lesser degree than clusters. Following that, we have the walls for which the density PDF peaks just below an overdensity of $1$. And finally, the voids have significantly lower density values, with the distribution reaching a maximum at $1+\delta\sim0.1$. Given the large width of the distributions, we find significant overlaps between the density PDF of different environments. This means that a simple density threshold is not sufficient in identifying the cosmic web components.

Our results on the density segregation are in agreement with the previous findings of \citet{Hahn2007a} and \citet{2010MNRAS.408.2163A}. Both studies showed that cluster regions are the most dense, followed by filaments and walls, while voids are very underdense. They found, similarly to us, that each morphological component occupies a large range of densities and that there is significant overlap in density values between different environments. A more quantitative comparison with these studies is difficult, given that each one uses a different smoothing scale. This shifts and distorts the shape of the density PDF, and therefore does not allow for even simple comparisons like contrasting the position of the peaks.

\begin{figure*}
    \centering
    \includegraphics[width=.7\linewidth]{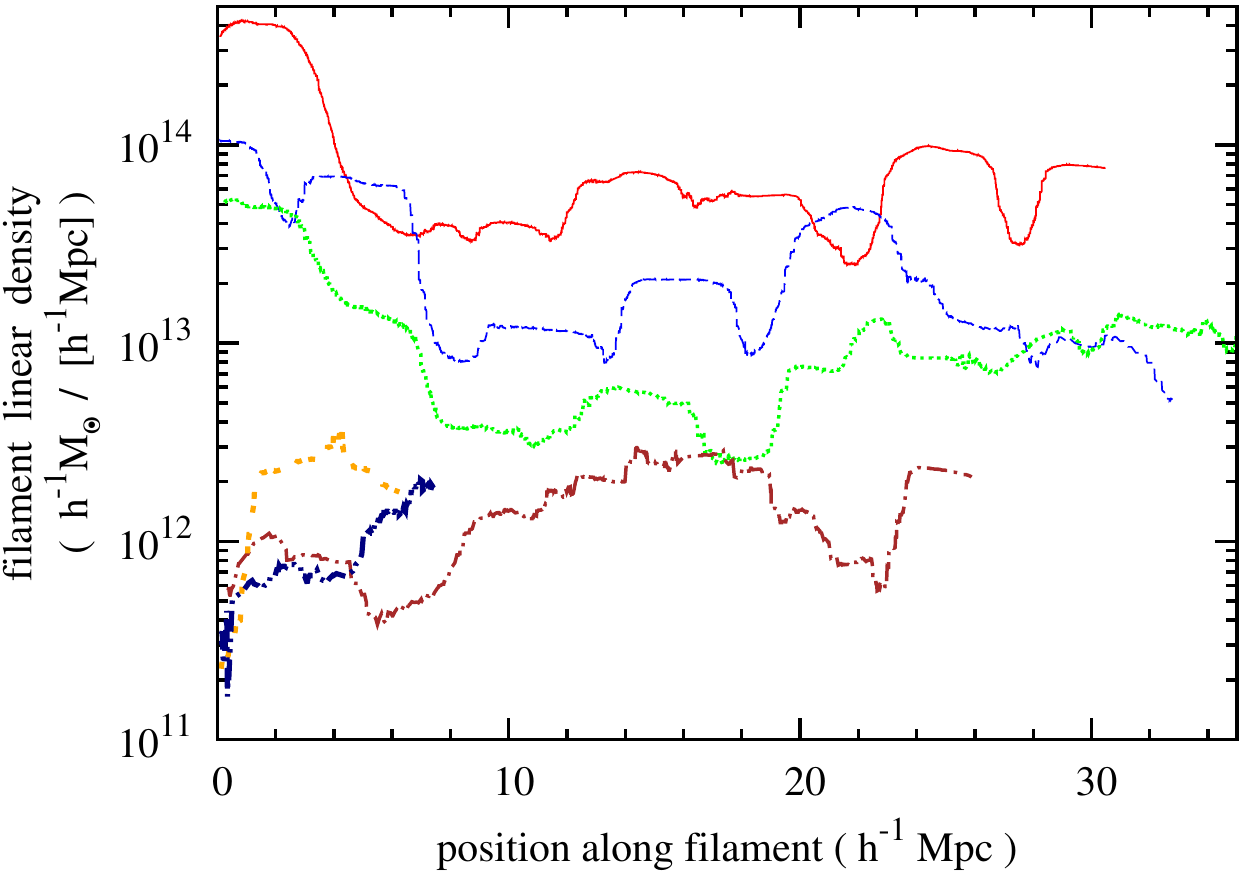}  
    \caption{ The linear mass density across a few representative filamentary branches. We exemplify this for six structures, with the three massive ones corresponding to filaments between clusters. The remaining three examples, with smaller $\zeta_\rmn{filament}$ values are objects found in underdense regions. }
    \label{fig:evolution:test_fila_density}
\end{figure*}

Of particular interest is the study of \citet{Pogosyan98}, which predicted the density span of morphological environments within the Zel'dovich formalism. \citet{Pogosyan98} emphasized that primordial overdense regions most likely evolve into clusters and filaments, while underdense regions are more likely to become voids and sheets. These predictions, while limited to the linear and mildly non-linear stages of evolution, are in good qualitative agreement with our results. In particular, clusters and voids are mostly limited to overdense and underdense regions respectively. The other two components seem to span both overdense and underdense volume, though filaments are more likely to be found at higher density while walls in underdense regions, as pointed out by \citet{Pogosyan98}.

The density PDF can also be used to get a better understanding of the various environmental detection methods that we use. For doing so, we present in \reffig{fig:density-histogram_comparison} the density distribution in filaments, walls and voids identified by each of \spider and \Spider approaches. We find that all the methods give very consistent results, though there are some small differences. For example, \Spider identifies filaments and walls that have slightly higher density values than the other methods. In contrast, the velocity based methods find filaments and walls that have lower density values, but only to a very minor extent. 

Studying the matter content of representative filament and wall stretches represents another, more interesting, way of describing the mass distribution across the cosmic web. It can be done by employing the linear and surface mass density concepts defined in \refsec{subsec:evolution:fila_properties_2} and \refsec{subsec:evolution:wall_properties_2}, which characterize the local mass content of filament and wall segments. To better illustrate these quantities, \reffig{fig:evolution:test_fila_density} shows the linear density $\zeta_\rmn{filament}$ across a few filamentary branches. We find that $\zeta_\rmn{filament}$ shows a strong variation not only between different filaments, but also along the same structure. The large $\zeta$ variations seen along the same filament are due to the massive haloes, which contain a significant fraction of the mass. Given that $\zeta_\rmn{filament}$ is an average quantity along filament segments of length $2R=4\Mpch$, the presence of such massive haloes is shown as a top-hat like profile of width $2R$. This is clearly visible for most of the examples shown in the figure.

\begin{figure}
    \centering
    \includegraphics[width=\linewidth]{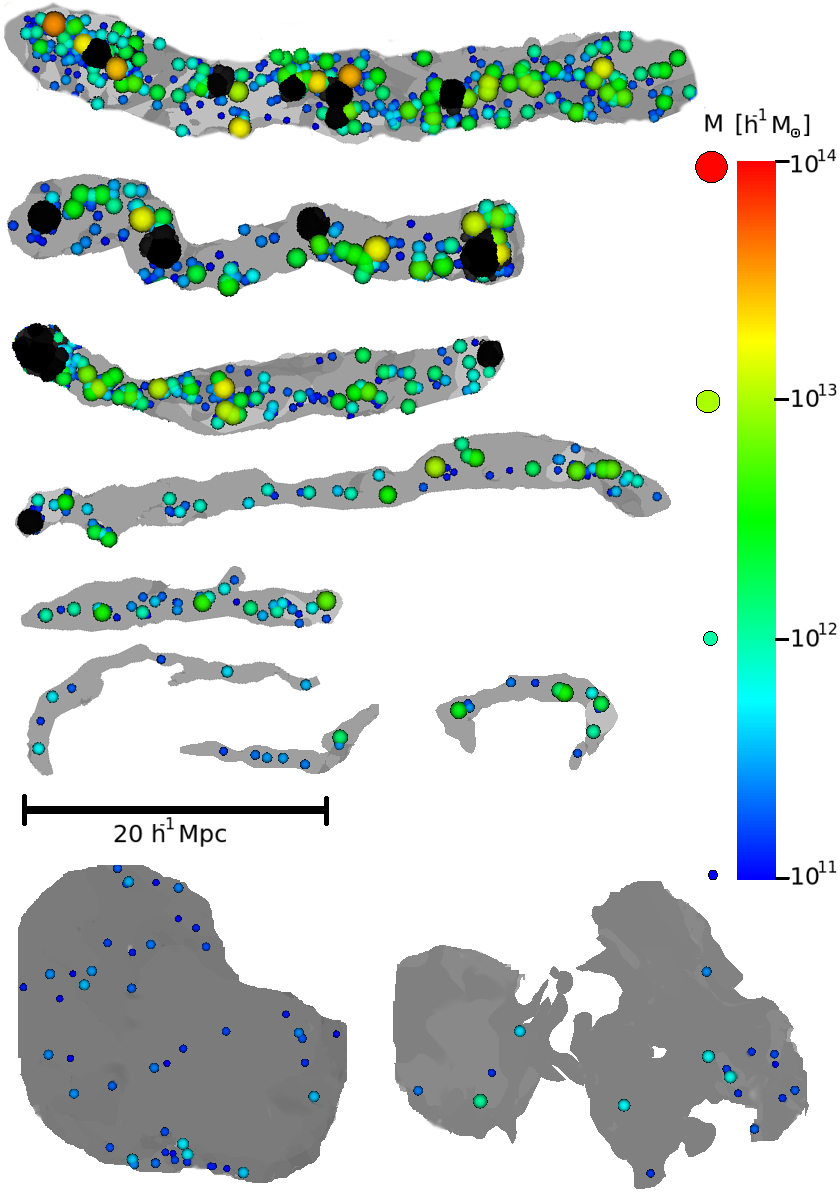} 
    \caption{ The population of haloes in a few typical filaments (top rows) and walls (bottom-most row). Haloes are shown via points whose size and color depends on the halo mass as shown in the right-hand legend. For filaments, the black points show haloes found in cluster regions. The length scale of the objects is shown via the horizontal bar on top of the last row of graphs. }
    \label{fig:evolution:env_example_haloes}
\end{figure}

\Reffig{fig:evolution:test_fila_density} illustrates another crucial find. Prominent filaments extending between clusters have a higher mass per unit length than filaments found in other regions. This is clearly seen in \reffig{fig:evolution:test_fila_density}, where the solid and dashed curves are examples of prominent filaments while the three objects with the lowest $\zeta_\rmn{filament}$ are found in underdense regions. It raises two important observations. First, structure formation theory predicts that filaments arise from a primordial quadrupolar mass distribution which gives rise to the canonical cluster-filament-cluster configuration \citep{Bond1996,vandeWeygaert1996,WeyBond08a}. This prediction agrees with the configuration of prominent filaments, but not with that of the more tenuous objects. It possibly suggests that structures located in underdense regions correspond to very weak primordial quadrupolar distributions or just to chance alignments. Secondly, more massive filaments have many more haloes, especially higher mass ones, than their tenuous counterparts (see \reffig{fig:evolution:env_example_haloes} for examples of halo population across different filaments). Hence, the filaments connecting clusters are much richer in galaxies and therefore more easily detectable in galaxy redshift surveys \citep{Drinkwater2004,Pimbblet04,Kartaltepe2008,Gonzalez09}.

\begin{figure}
    \centering
    \includegraphics[width=\linewidth]{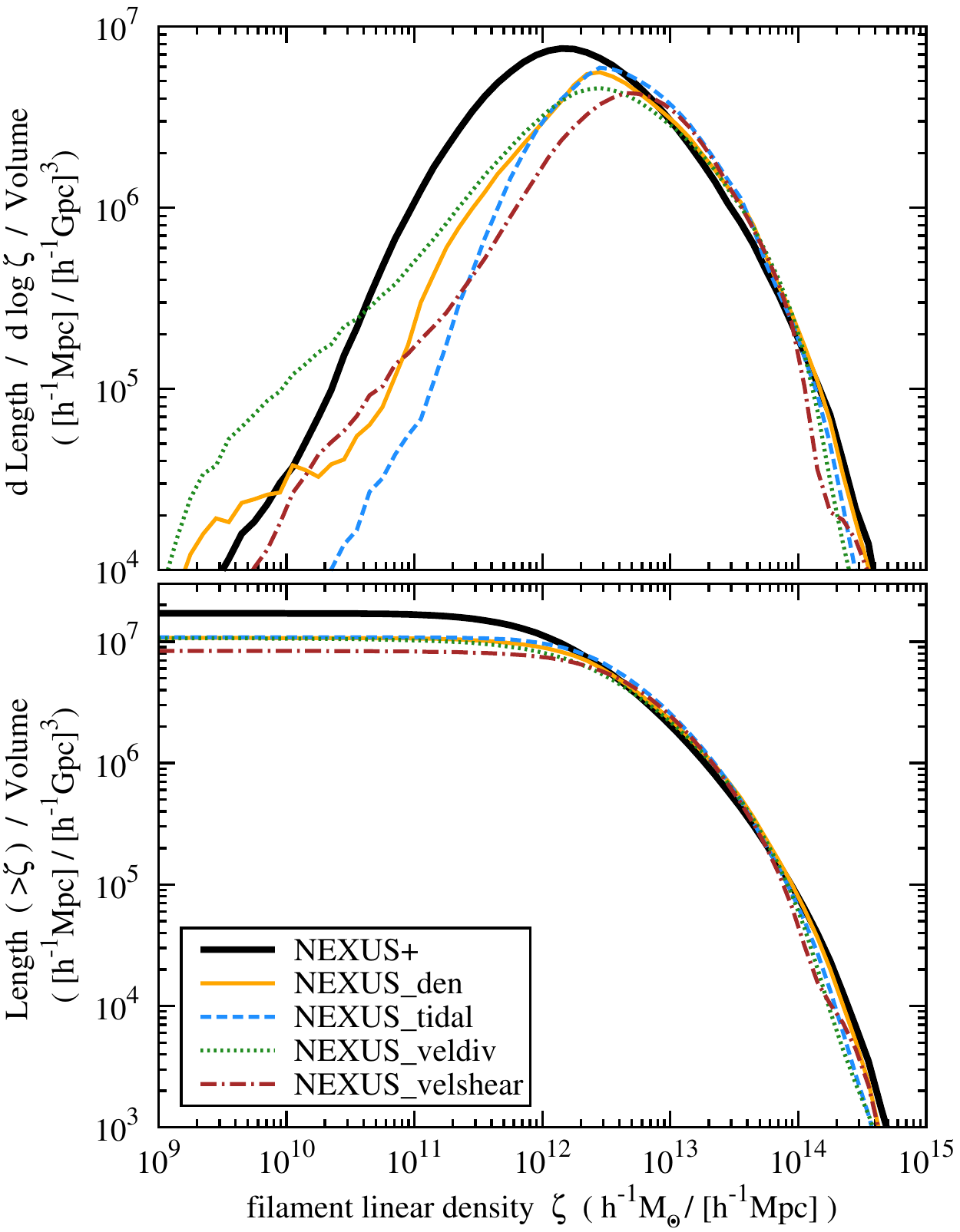}
    \caption{ The distribution of mass across filament environments identified by \spiderDeN, \spiderTidaL, \spiderVeldiV, \spiderVelshear and \SpideR. It gives the total length of filaments that have a certain linear mass density $\zeta_\rmn{filament}$ (top panel) and the cumulative length above $\zeta_\rmn{filament}$ (bottom panel). }
    \label{fig:evolution:fila_linear_density}
\end{figure}
\begin{figure}
    \centering
    \includegraphics[width=\linewidth]{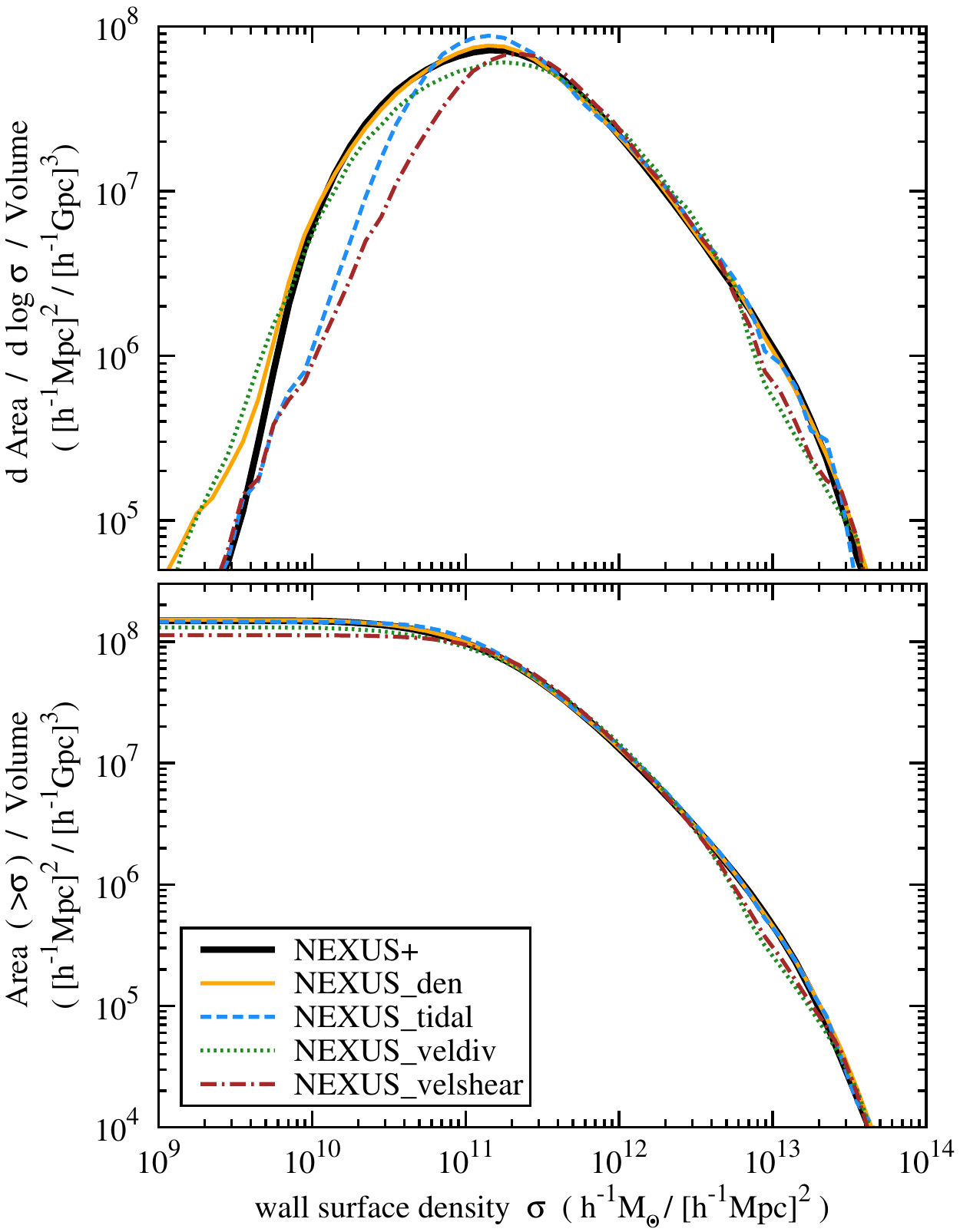}
    \caption{ The distribution of mass across wall environments identified by \spiderDeN, \spiderTidaL, \spiderVeldiV, \spiderVelshear and \SpideR. It gives the total area of walls that have a certain wall surface mass density $\sigma_\rmn{wall}$ (top panel) and the cumulative area above $\sigma_\rmn{wall}$ (bottom panel).  }
    \label{fig:evolution:wall_surface_density}
\end{figure}


The large variation in filament linear density between different objects as well as along the same filament raises questions about what is the typical $\zeta_\rmn{filament}$. To further quantify this, we count the total length of filament segments with a fixed $\zeta_\rmn{filament}$ value. This is shown in \reffig{fig:evolution:fila_linear_density}, where we plot the length of the filamentary network as a function of linear density. First, we focus on the thick black curve which presents the \nexus{} results. These filaments have a wide range in $\zeta_\rmn{filament}$, from very tenuous filaments in voids which barely have a ${\sim}10^{10}\Msun/\Mpc$, to very massive filamentary segments with masses similar to those of cluster haloes. It shows how diverse the filamentary environments are, from very crowded to very sparse regions. Their common link is a highly anisotropic matter distribution with a distinctive overdensity along one direction. 

The very low mass per unit length of the more tenuous environments hints to the observational challenge of detecting theses structures in galaxy redshift survey, given that these systems are most probably inhabited by low luminosity galaxies far apart. This can be easily appreciated from \reffig{fig:evolution:env_example_haloes}, which in the bottom part of the left panel shows a few typical void filaments. Such structures are only sparsely inhabited by $10^{12}\Msun$ and lower mass haloes, which implies that even though present, such tenuous objects are not conspicuous features in the spatial distribution of galaxies. The configuration of three aligned galaxies inside a void found by \citet[][]{Beygu2013} is probably an example of such a thin void filament \citep{Rieder2013}. Within this context, it is interesting to compare with the filaments detected by \citet{Bond10} in the SDSS data. While using a different detection method that is most sensitive to the prominent filaments, \citet{Bond10} found a total filament length that is a factor of ${\sim}10$ smaller than our result. If we assume that they identified only the most massive such structures, it suggests that only filaments with $\zeta_\rmn{filament}\gsim5\times10^{12}\Msun/\Mpc$ can be easily detected in galaxy redshift data. Therefore, while theoretical models predict a wealth of filamentary structures, from very high to low mass ones, most of this filamentary network seems to be outside the detection limit of current galaxy surveys. This is because most of the network consists of tenuous structures that are only sparsely sampled by low mass haloes, and equivalently by low brightness galaxies.

We already saw that various cosmic web identification methods result in different morphological components. It suggests that we need to further investigate if the $\zeta_\rmn{filament}$ findings are sensitive to the environmental detection method. This is analysed in the left panel of \reffig{fig:evolution:fila_linear_density}. As expected, we find that the total length associated with massive filaments, i.e. $\zeta_\rmn{filament}\gsim 10^{13}\Msun/\Mpc$, is the same across all the five methods analysed here. In contrast, the extent of the more tenuous structures is highly sensitive to the identification technique, with \nexus{} identifying the largest amount of such objects. These quantitative findings agree with the visual impression given by \reffig{fig:evolution:env_filaments}, with prominent filaments detected by all the methods, while the thin ones show a large degree of variation. In particular, it reiterates the impression that \nexus{} is especially suited for the identification of both prominent and tenuous structures, and therefore to characterize the range of environments in both overdense and underdense regions.

\begin{figure}
    \centering
    \mbox{\hskip -1.0truecm\includegraphics[width=1.12\linewidth]{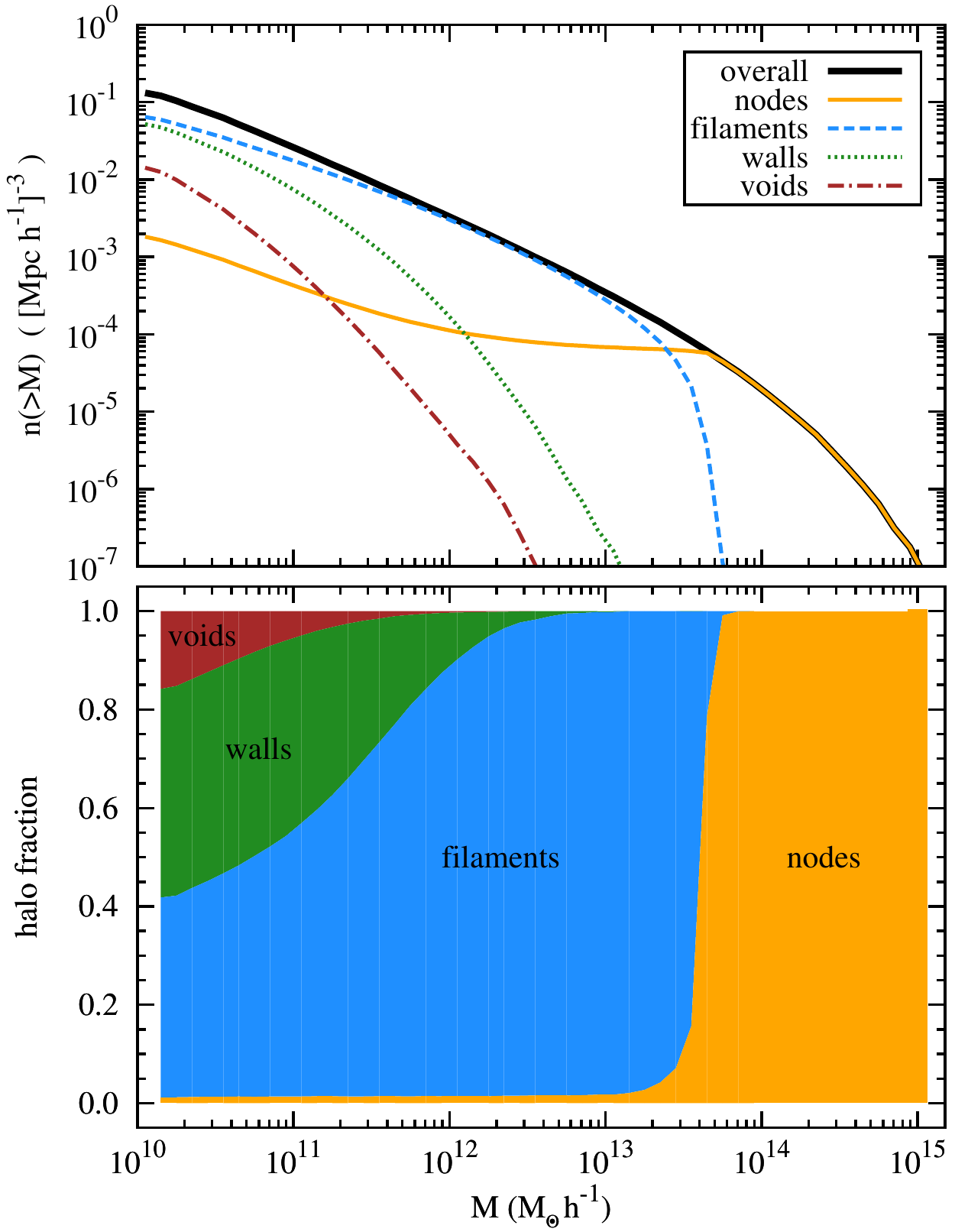}}
    \caption{The upper panel shows the cumulative halo mass function split according to the \nexus{} environment in which the halo resides. The bottom frame shows the fraction of haloes in each environment as a function of halo mass. }
    \label{fig:evolution:halo_mass_function}
\end{figure}

Equivalent to the way we characterized filaments in terms of their linear density, the mass distribution across sheet environments is described by the surface density of walls $\sigma_\rmn{wall}$. This is shown in \reffig{fig:evolution:wall_surface_density}, where we plot the total area of walls as a function of $\sigma_\rmn{wall}$. Focusing on the black thick curve, which gives the \nexus{} results, we find that the surface density of sheets extends over many orders of magnitude, also showing the diversity of wall environments. We find that most sheet sections have $\sigma_\rmn{wall}<10^{12}\Msolar/\MpchAreaInverse$, which implies that these regions are typically populated with haloes smaller than our own galaxy. This view is strengthened by the right panel of \reffig{fig:evolution:env_example_haloes}, which shows the distribution of haloes across a few typical walls. Sheet haloes are typically both low mass and sparsely distributed. Therefore, most wall sections are populated with sparsely distributed lower brightness galaxies, which makes the detection of cosmic sheets very challenging, especially in galaxy surveys. 

Moreover, the low contrast of walls is probably behind the large discrepancy in sheet identification between different methods (e.g. \reffig{fig:cross-correlation}). It reiterates the findings of previous studies according to which walls are challenging to identify because of their reduced contrast with respect to the background and due to their planar nature \citep[e.g.][]{Aragon07,2010MNRAS.408.2163A}. While there is a large discrepancy in the detection of sheets, the mass distribution of the resulting structures seems to be quite similar. This can be seen from the right panel of \reffig{fig:evolution:fila_linear_density}, where we show the area of the wall network at constant $\sigma_\rmn{wall}$ as identified by various methods. Similar to filaments, the extent of the massive walls is the same for all the detection techniques, with differences restricted to less massive structures with $\sigma_\rmn{wall}\lsim10^{12}\Msolar/\MpchAreaInverse$.

\subsection{Halo distribution}
\label{subsec:evolution:halo}

Haloes play a prominent role within the current theories of structure formation, given that the central parts of these objects are the sites of galaxy formation. Therefore, the differences in the halo population across the cosmic web components are indicative of variations with large scale environment in the population of galaxies and their properties. 

We exemplify the relation between haloes and environment by showing in \reffig{fig:evolution:env_example_haloes} the spatial distribution of haloes across a few representative filament and wall stretches. The figure clearly exemplifies that haloes are well segregated across environments, with the most massive such objects living in cluster and prominent filaments. Walls typically host $10^{12}\Msolar$ and lower mass objects, while void regions are populated with even lower mass haloes \citep[][]{Hahn2007a,Hahn2007b,Aragon07}. 

\begin{figure}
    \centering
    \includegraphics[width=\linewidth]{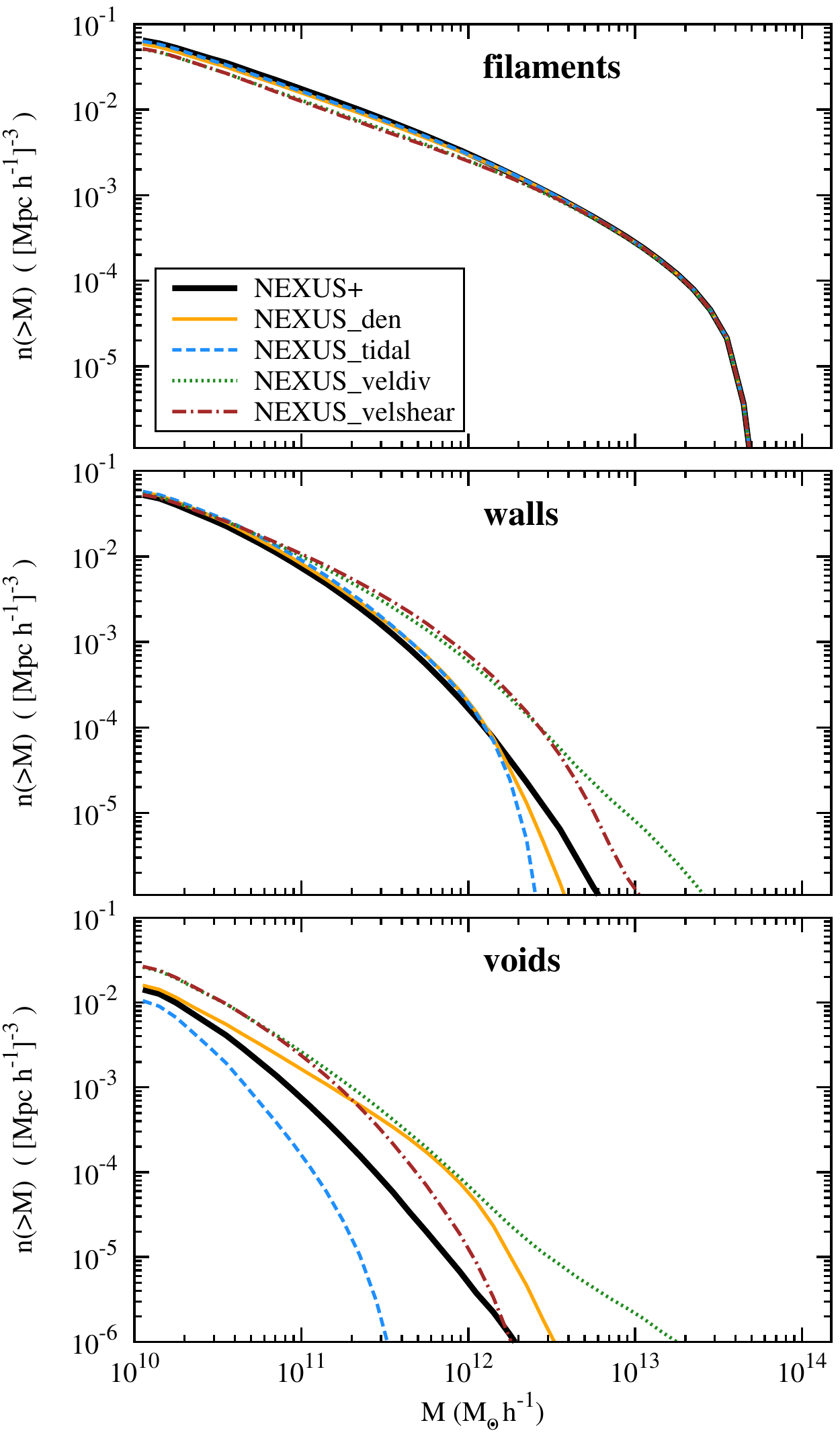}
    \caption{ A comparison of the halo mass function when using environments identified by \spiderDeN, \spiderTidaL, \spiderVeldiV, \spiderVelshear and \SpideR. The three panels show: filaments (top-right), walls (bottom-left) and voids (bottom-right). }
    \label{fig:evolution:halo_mass_function_comparison}
\end{figure}

Even more interesting, the halo population does not only vary between environments, but also between structures with similar morphological features. This is seen in the left panel of \reffig{fig:evolution:env_example_haloes}, where we show several filamentary branches; the top ones correspond to prominent structures while the bottom ones represent void filaments. We observe a clear trend of the halo distribution with filament properties. Thicker filaments, which are typically outstretched between cluster pairs, are populated with more massive haloes which are also more tightly packed together. In contrast, haloes in tenuous filaments are typically low mass, similar to the ones in walls, and are widely spaced apart. The major implications of these findings were discussed in \refsec{subsec:evolution:density}.

The variation of the halo population with environment is quantified in \reffig{fig:evolution:halo_mass_function}. It shows the halo mass function segregated according to the morphological component in which the haloes reside. At the higher mass end, we find that the most massive $\gsim 5\times 10^{13}\Msolar$ haloes are exclusively located in cluster regions. Less massive objects are typically found in filaments, with haloes in sheets and voids representing a significant share of the halo population only at very low masses.

In particular, less than $10\%$ of haloes more massive than $10^{12}\Msolar$ are found in sheets which implies that very few luminous galaxies are found in walls. Less massive objects are also rarely found in walls, with fewer than $20\%$ of $10^{11}\Msolar$ haloes residing in this environment. It suggests that most of the galaxies that are easily observable in typical galaxy redshift surveys (e.g. 2dFGRS \citealt{Colless03}, SDSS \citealt{Tegmark2004}) are found in filament and cluster regions, with only a small fraction of them in walls. For voids, we find an even starker contrast, with at most $5\%$ of $10^{11}\Msolar$ and higher mass haloes located in this environment. It explains why redshift surveys find large regions of the Universe almost or completely devoid of galaxies. These findings have important consequences for the identification of large-scale structure in galaxy surveys. The rarity of galaxies in walls suggest that these environments may be best detected in terms of the matter distribution surrounding them, and not necessary due to the luminous objects they contain \citep{Trujillo2006,Aragon07}. Similarly, voids are best identified as regions nearly, but not completely, devoid of galaxies.

In the three panels of \reffig{fig:evolution:halo_mass_function_comparison} we explore how much the environmental halo mass function changes when varying the cosmic web identification method. In the case of filaments, all methods return very similar results, with minor differences only at the low mass end. In contrast, for walls and voids, there is a much larger discrepancy between the results of different identification procedures. In particular, we find a large difference in the number of massive objects residing in sheets and voids. The velocity based methods, \spiderVeldiv and \spiderVelsheaR, return a significantly larger number of massive wall and void haloes than the rest of the density based methods. While this variation is large, the disparity is mainly restricted to the mass range where sheet and void haloes are only a minor fraction of the total population of same mass objects. 
The halo mass function discrepancy between methods is easily understood when realizing that sheet and void haloes are the ones that were not identified as part of cluster or filament environments. Any small differences in the detection of clusters and filaments get amplified in the population of walls and void haloes, since these latter haloes are only a small fraction of the overall population.


\begin{figure*}
    \centering
    \includegraphics[width=.9\linewidth]{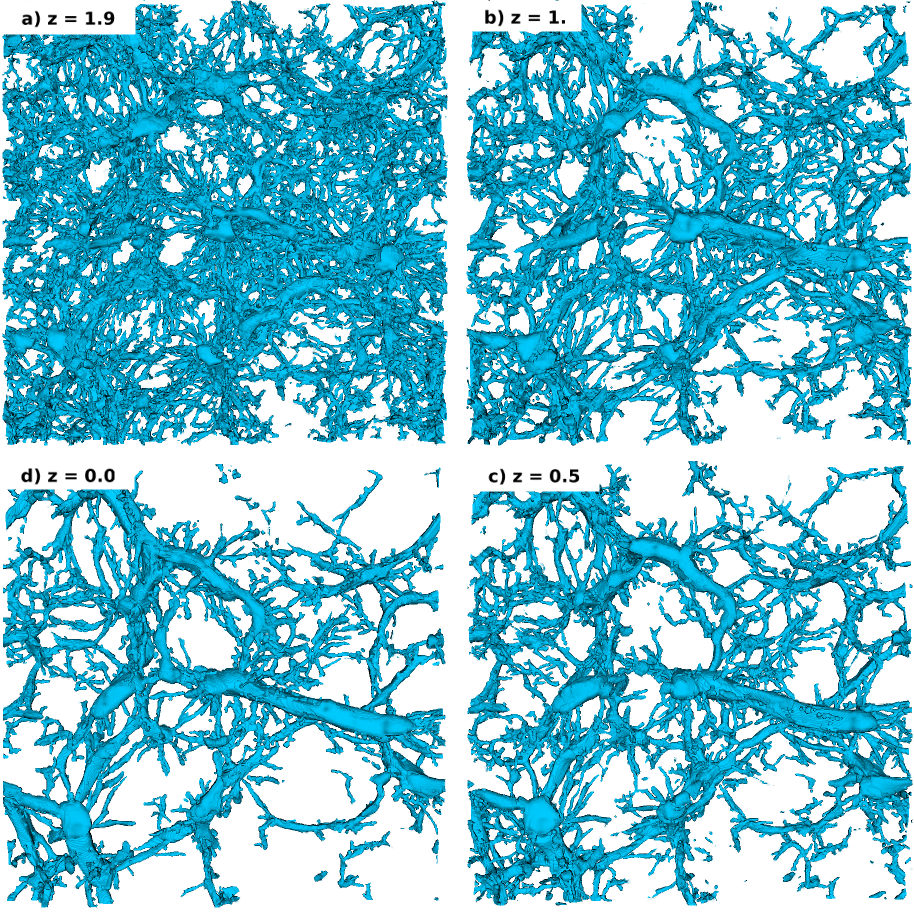}
    \caption{The time evolution of filamentary environments in a $100\times100\times10\MpchVolume$ slice centred on the most massive halo in the \MII{}. The panels show in clockwise direction, starting with the upper left corner, the filaments at a redshift of: a) $z=1.9$, b) $z=1.0$, c) $z=0.5$ and d) $z=0.0$. }
    \label{fig:env_fila_evolution}
\end{figure*}

\begin{figure*}
    \centering
    \includegraphics[width=.9\linewidth]{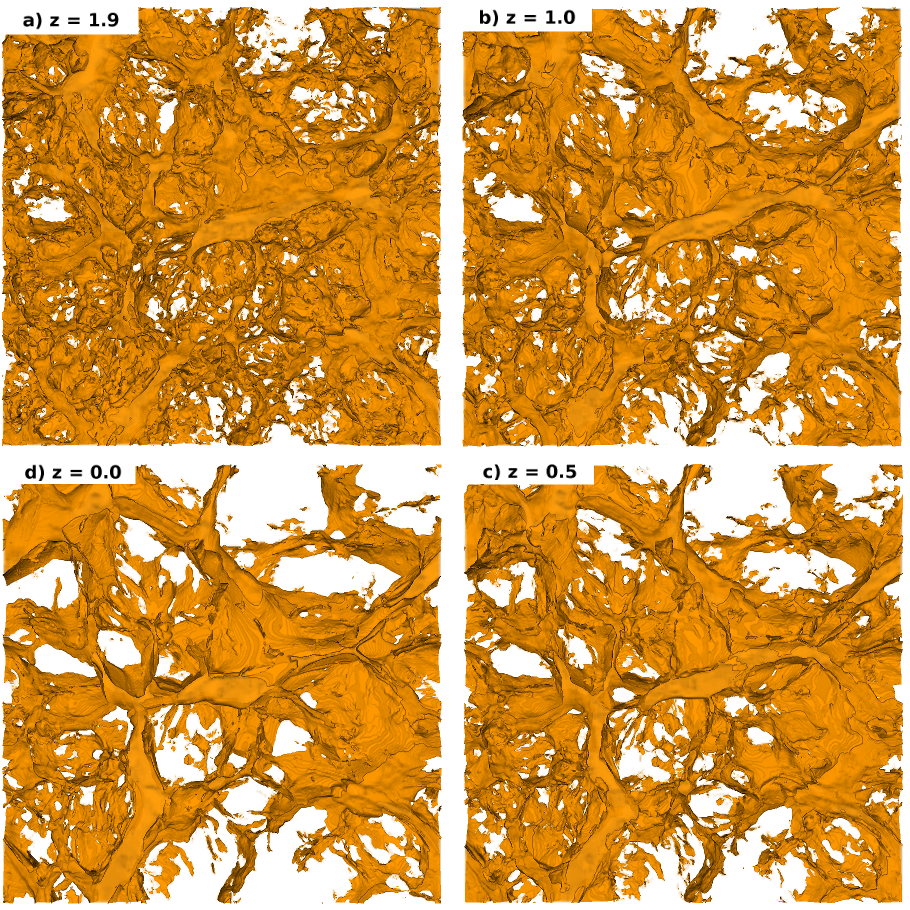}
    \caption{The time evolution of cosmic sheets in a $100\times100\times5\MpchVolume$ slice centred on the most massive halo in the \MII{}. The panels show in clockwise direction, starting with the upper left corner, the wall environments at a redshift of: a) $z=1.9$, b) $z=1.0$, c) $z=0.5$ and d) $z=0.0$. }
    \label{fig:env_wall_evolution}
\end{figure*}

\section{Evolution of mass distribution across cosmic web environments}
\label{sec:evolution:evolution}

This section is the first dedicated to investigating the evolution of the cosmic web from an early time to the present epoch. The main goal is to understand what shapes the large-scale structures that we find at the present time, from the very massive clusters connected by prominent filaments, to the huge voids that dominate galaxy redshift surveys. We approach this task from two directions. We first analyse how global cosmic web properties, like mass and volume fractions, change with redshift. Secondly, we investigate how the mass distribution across individual web elements changes from early times till present.

In the previous section we analysed several methods for identifying the cosmic web components. We found that while there are a lot of similarities between the environments detected by the five approaches, there are also some discrepancies. These variations are consistent along all methods and there is no approach that stands out as significantly better than the rest. Given these findings, in the rest of this work we restrict our investigation to \Spider environments. We selected this technique since we found it to be better at capturing the tenuous environments of underdense regions.

Before proceeding to quantify the development of the cosmic web, we illustrate in \reffig{fig:env_fila_evolution} the evolution of filamentary environments starting with a redshift of $z=1.9$ down to the present time. At early times, the filaments form a complex network that pervades most of the cosmic volume, with the exception of the most underdense regions. The typical non-linear scale at $z=2$ is significantly smaller than the $10\Mpch$ thickness of the slice, which makes it difficult to visually distinguish individual structures. Nonetheless, we can still make a few general observations. While the filamentary network has a few thick structures, it is dominated by small scale filaments. These thin filaments seem to be packed much more tightly close to prominent structures, suggesting that overdense regions have a higher richness of filaments. In the next frame, at $z=1$, we already find that most of the tenuous structures have disappeared and that we can more easily see the pronounced filaments. These prominent structures are also identified in panel a), but they are surrounded by a multitude of thinner objects that obscure their presence. Going forward in time, to $z=0.5$ and $0$, we find that the evolution of the cosmic web significantly slows down, with only minor changes after $z=0.5$. Most of the variations after $z=1$ are restricted to the population of thin filaments. The most marked effect is the emptying of the underdense regions, with large voids empty of filaments clearly visible at the present time.

It is interesting to observe that most of the prominent filaments found at present can already be seen at high $z$ \citep[this observation has previously been pointed out by][]{Bond10}. Compared to $z=1$, which offers a better illustration, we find that these massive filaments show little evolution in shape and size. In fact, most of the change in the filamentary network is restricted to the more tenuous filaments, whose disappearance is driven by merging with the more pronounced structures.

The evolution of walls is very similar to that of the filamentary network, as can be seen from \reffig{fig:env_wall_evolution} which shows the cosmic sheets in a $5\Mpch$ slice through the \MII{} volume. At early times there are a large number of walls that split the volume into numerous small voids. As time increases, the tenuous sheets disappear and leave behind a network of prominent structures. These are the cosmic walls that we see today and which segment the \MII{} volume into several large voids. Similar to filaments, we can make two main observations. First, the time evolution of sheets is most evident when analysing the most feeble structures, with variations significantly slowing down after $z=0.5$. And secondly, the pronounced walls are already in place since very early redshift and they show little evolution since then.

\subsection{Mass and volume fraction}
\label{subsec:evolution:evolution_mass_fraction}

\begin{figure}
    \centering
    \includegraphics[width=\linewidth]{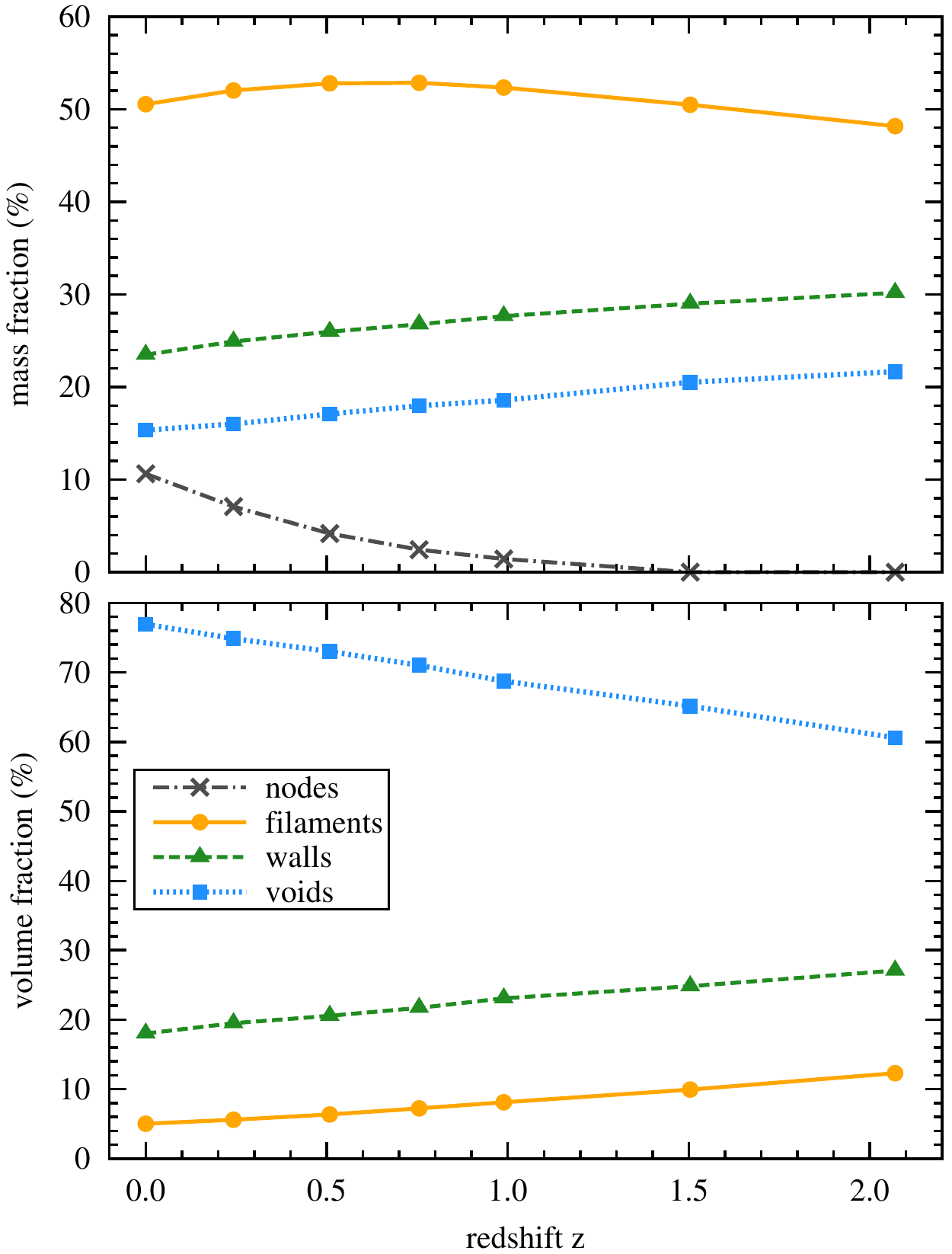}
    \caption{The time evolution of the mass (top panel) and volume (bottom panel) filling fractions for: nodes(crosses), filaments (solid with circles), walls (dashed with triangles) and voids (dotted with squares). The effect of cosmic variance on the mass and volume fraction is smaller than the size of the symbols and it is not shown. } 
    \label{fig:evolution_mass_volume_fraction}
\end{figure}

The simplest way to characterize the cosmic web evolution is to track the mass and volume content of each of its components. This is shown in \reffig{fig:evolution_mass_volume_fraction}, with the mass fraction in the left panel and the volume fraction in the right frame. We find that cluster environments start to contain a significant fraction of matter only at late times, after $z=1$. Even though they appear late on the cosmic stage, their influence grows rapidly such that at the present time they contain $10\%$ of the mass. The filaments have a more complex evolution, with an initial increase in mass until around $z\sim0.5$, after which we find a slight decrease. The reduction in mass is due to the formation of the cosmic web nodes that accumulate a considerable share of mass, predominantly from filaments, as showed later in \refsec{sec:evolution:mass_trasport}. In terms of volume fraction, filaments show a factor of $2$ decrease from $z=2$ to present. This means that the same mass fraction gets accumulated into fewer, but more massive filaments.

The cosmic sheets are described by a decreasing mass and volume content. Compared to early redshift, at present time the walls contain ${\sim}20\%$ less mass and volume. The void environments show a similar decrease in mass fraction, but show an opposite trend in volume fraction. This suggests that voids do not only increase by merging with other voids as seen in \reffig{fig:env_wall_evolution}, but also by taking over regions that were previously identified as walls and possibly filaments.

\begin{figure}
    \centering
    \includegraphics[width=\linewidth]{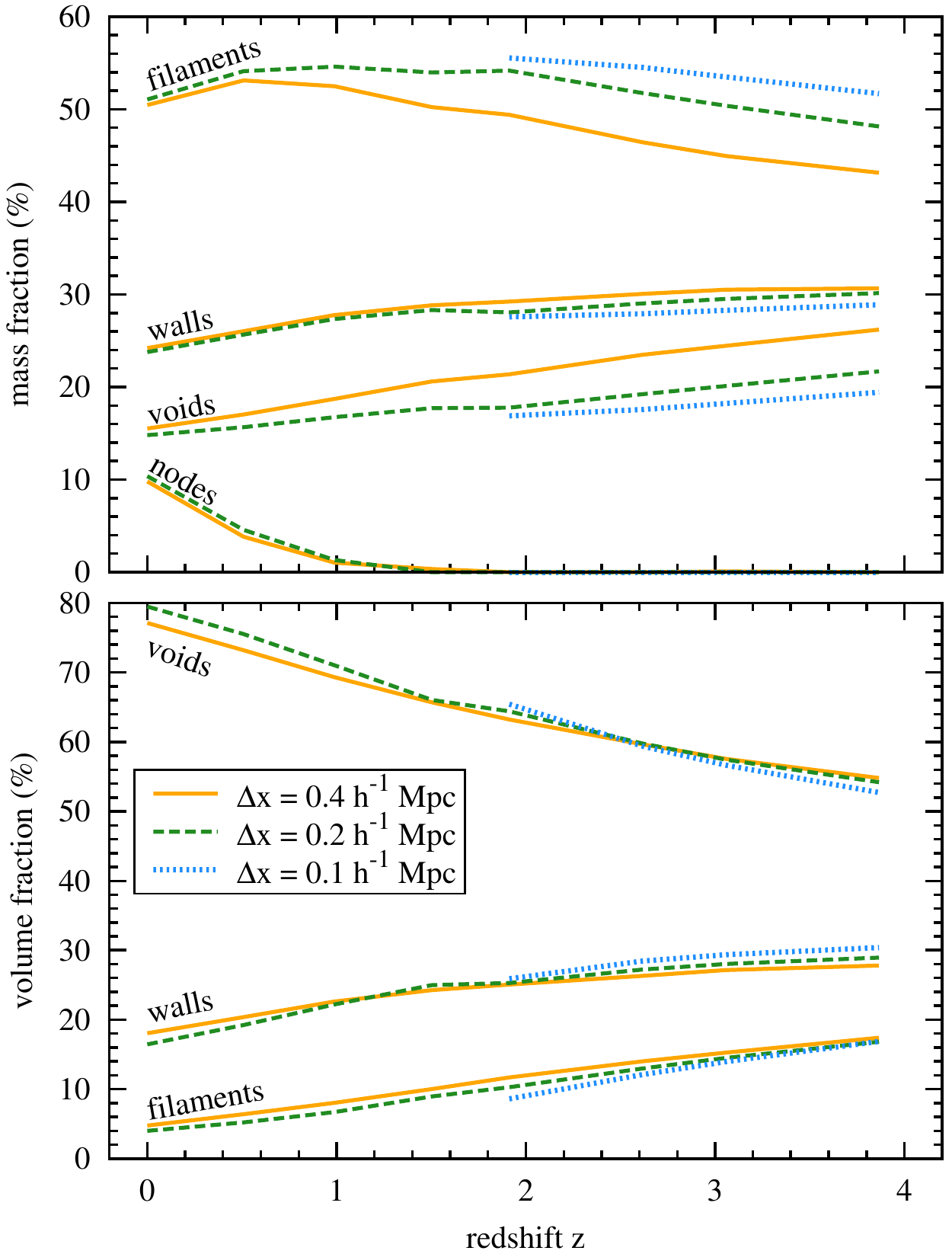}
    \caption{The variation of node, filament, wall and void properties with scale and redshift. We present the mass and volume fraction found for the \MII{} data using three different scales: $\Delta x=0.4\Mpch$ (solid line), $\Delta x=0.2\Mpch$ (dashed line) and $\Delta x=0.1\Mpch$ (dotted line). For each case $\Delta x$ gives the grid spacing as well as the smallest smoothing scale used to identify environments. Using smaller $\Delta x$ values allows for the detection of thinner and more tenuous large-scale structures. }
    \label{fig:evolution_resolution_study}
\end{figure}

These results paint an interesting picture, with a universe that evolves to be dominated by voids in terms of volume and by very dense regions, i.e. clusters and filaments, in terms of mass. This is in accordance with the standard picture of anisotropic collapse, which predicts that at later times more of the matter content of the universe is inside collapsed objects \MCn{\citep[][]{Zeldovich70,Icke73,Shandarin1984,Shandzeld89,Gurbatov1989,Pogosyan98,Shen06,WeyBond08a,Desjacques2008,Rossi2012,Rossi2013}}. Our findings also show the dominant role played by filaments, which not only at present, but also at high redshift, contain the largest share of mass. This offers a natural explanation of why filaments are such a prominent feature of the galaxy distribution \citep{deLapparent1986,Drinkwater2004,Pimbblet04,Porter2005,Einasto2011d}.

To understand the evolution of the environmental mass fraction we need to study how matter flows across the same and also between different morphological components. This is investigated in great details in \refsec{sec:evolution:mass_trasport}, where we quantify the mass transport across different environments. Without going into details, we note that the time variation of the mass fraction seen in \reffig{fig:evolution_mass_volume_fraction} is consistent with the large scale flow of matter as given by the velocity field. The matter flows out of voids towards sheets, inside walls it streams towards the filaments that surround these planar structures, while the matter inside filaments moves towards cluster regions \MCn{\citep{Shandarin1984,Shandzeld89,vanHaarlem1993,WeyBond08a}}. This is exemplified in \reffig{fig:evolution:env_velocity_flow} which shows the large scale velocity field across a few typical void, sheet and filament stretches. According to this picture, voids always loose mass while clusters always become more massive. Walls and filaments have a more complex, time dependent behaviour, since these environments have both an inflow and outflow of matter. For example, filaments gain mass from wall regions while at the same time matter stream out of them towards clusters. Therefore, the sheets and filaments can switch from gaining to losing mass, depending on the balance of inflow versus outflow. This is in good agreement with the quantitative results seen in \reffig{fig:evolution_mass_volume_fraction}. Especially noteworthy is the change in filament mass fraction, with filaments growing in mass until $z{\sim}0.7$, while decreasing afterwards. It implies that at later times more of the filament mass flows in cosmic web nodes than it arrives from sheets.

\reffigS{fig:env_fila_evolution} and \ref{fig:env_wall_evolution} show that at high redshift the cosmic web components are dominated by thin structures and only at later times the prominent configurations become prevailing. The preponderance of the thin structures raises questions about what is the smallest filtering scale needed to identify most of these narrow elements. We further investigate this in \reffig{fig:evolution_resolution_study} where we present the mass and volume fraction in cosmic web environments as a function of the smallest scale $\Delta x$ used for their identification. At present we find that decreasing the scale below the $\Delta x = 0.4\Mpch$ that we used for the previous results does not change our results significantly. This implies that a value of $\Delta x = 0.4\Mpch$ is sufficient in identifying most of the environments at $z=0$. Going to a higher redshift, we find that using a finer grid and smaller scales results in different mass fractions in filaments and voids. In contrast, the volume fractions seem less sensitive to scale, the same holds true for the wall and node mass fraction. While these global quantities do not change with scale, there exist differences in the detection of individual environments between various scales. These discrepancies are mostly restricted to thin structures, below the minimum scale used by the identification procedure.

\Reffig{fig:evolution_resolution_study} suggests that studying the cosmic web components at high redshift necessitates the use of smaller scales and finer grids to better capture the thin environments. While this is computationally feasible for the small \MII{} volume, it becomes a very memory and time intensive task to do a similar analysis for the much larger \MI{} volume. Moreover, going to smaller scales can be counter-productive from an observational point of view. While there are considerably more thin structures at high redshift, these are very difficult to probe observationally. Typical observations at high $z\gsim1$ are limited to the most massive objects, e.g. luminous galaxies in redshift surveys \citep[e.g.]{Tegmark2004} and prominent filaments in gravitational lensing measurements \citep[e.g.]{Kartaltepe2008}, and therefore cannot easily detect the tenuous structures.

For the rest of this study, we limit our \MI{} analysis to the $z\le2$ regime and use filtering scales of $\Delta x = 0.4\Mpch$. This is best thought as characterizing the large-scale structures that are accessible at scales of ${\sim}0.4\Mpch$ or larger. As we already argued, only such structures are observationally accessible at present or in the near future. Moreover, from a theoretical perspective, such a choice gives for the redshift range $z\le2$ similar results to having smaller $\Delta x$ values, with the differences restricted to the population of thin structures.


\begin{figure}
    \centering
    \includegraphics[width=\linewidth]{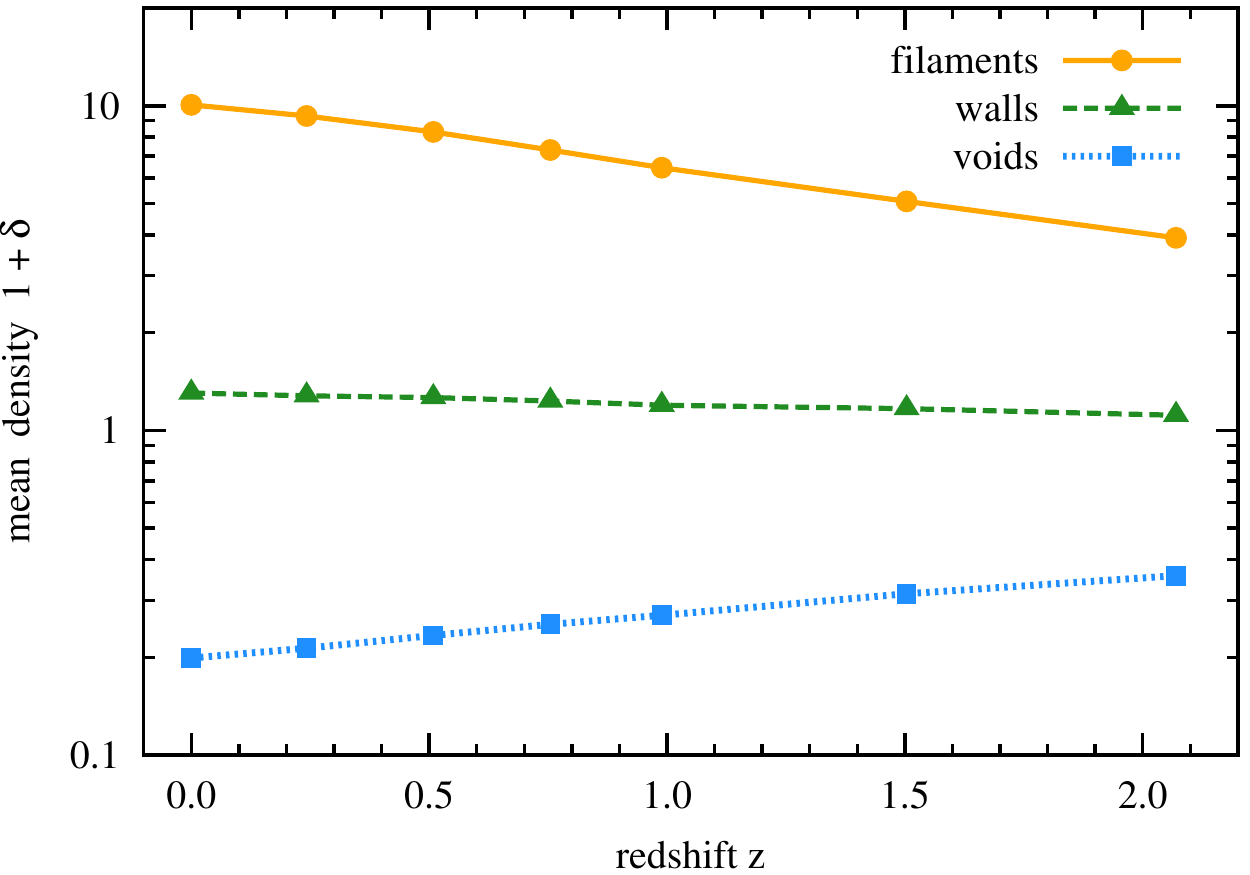}
    \caption{The time evolution of the mean density for: filaments (solid with circles), walls (dashed with triangles) and voids (dotted with squares). The density $1+\delta$ is expressed in units of the mean background density at each redshift. } 
    \label{fig:evolution_mean_density}
\end{figure}

\begin{figure*}
    \centering
    \includegraphics[width=.99\linewidth]{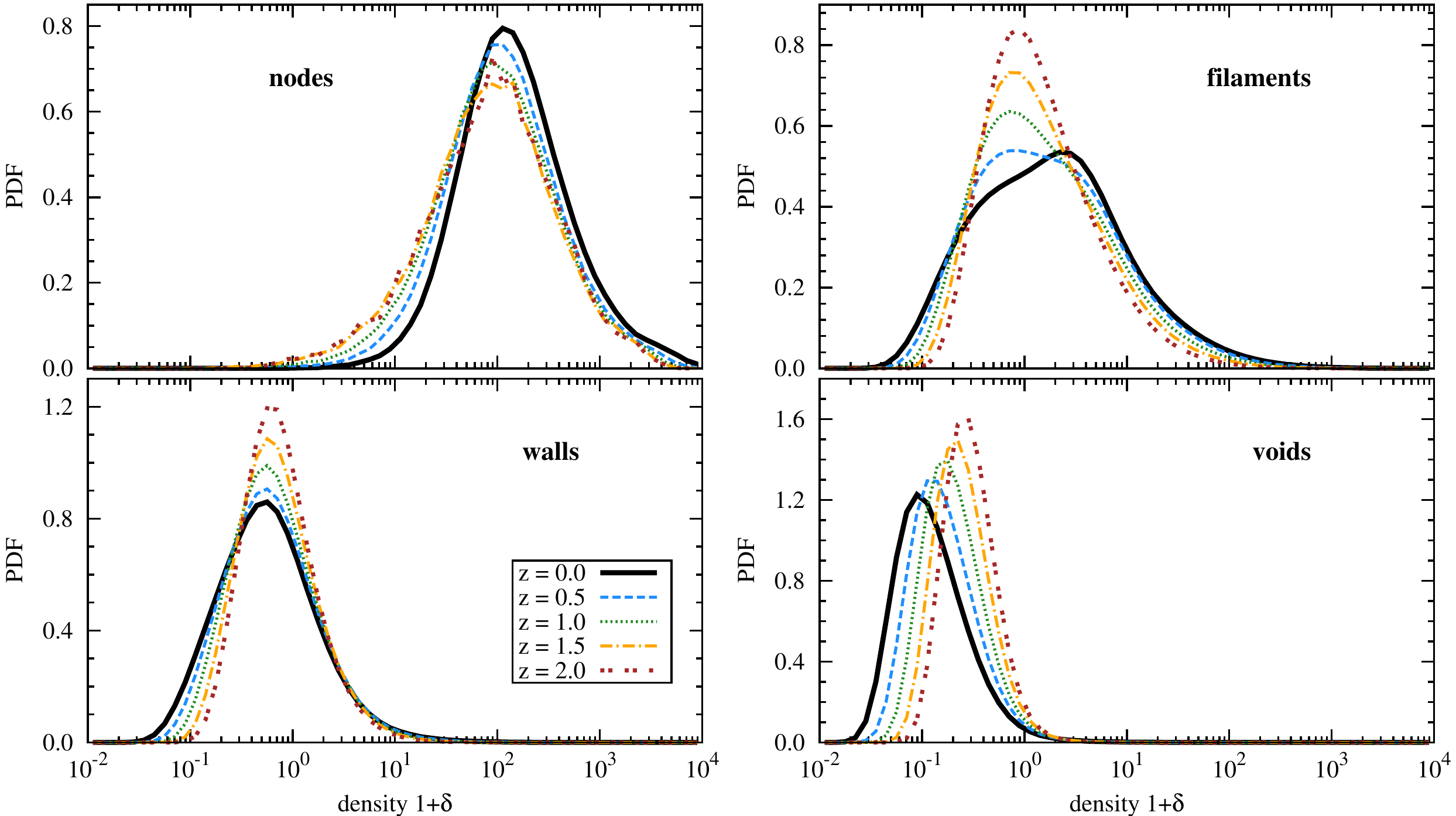}
    \caption{ The evolution with redshift of the density PDF for each cosmic web environment: nodes (top-left), filaments (top-right), walls (bottom-left) and voids (bottom-right). The density $1+\delta$ is expressed in units of the mean background density at each redshift. The histogram was obtained using the \dtfe{} density field on a regular grid with spacing $\Delta x=0.4\Mpch$. No additional smoothing was used. }
    \label{fig:evolution_density_pdf}
\end{figure*}

\subsection{Density distribution}
\label{subsec:evolution:evolution_density_distribution}

Given that both the mass and volume contained within cosmic web components depends on redshift, it is natural to further investigate the time variation of the density in each environment. \Reffig{fig:evolution_mean_density} shows the mean density in filaments, walls and voids as a function of redshift. The average density in filaments increases rapidly with time such that it doubles from $z=2$ to present. This is a result of the mass in the filamentary network getting concentrated in a few massive structures, as can clearly be seen from \reffig{fig:env_fila_evolution}. Walls show a more quiet evolution, with the mean density of this environment very close to the average background density. Contrary to filaments, voids show a considerable decrease in density, from $1+\delta\sim 0.4$ at $z=2$ down to $1+\delta\sim 0.2$ at present.

We further explore the evolution of density in \reffig{fig:evolution_density_pdf} where we show the density PDF of each cosmic web component at several redshifts. The density PDF shows a significant change with time which gives further insight on the evolution of large scale environments. Before proceeding to analyse every panel individually, it is important to note that the area under each curve is constant and equal to unity. This typically means that curves with higher peak values have a smaller width, spanning a narrower range of densities. The change in the height of the PDF curves shows that at high redshift each environment, except clusters, is characterized by a tighter density range which becomes more extended at present time. The main factor contributing to this evolution is the increase in difference between underdense and overdense regions as the universe gets older. Therefore, at later times there is a larger density range that environments can occupy.

In the case of cluster regions, the density PDF is shifted towards lower values at high $z$ and also shows a slightly wider range. Given that we identify the cosmic web nodes as the regions with an enclosed mean density equal to the virial density, most of the evolution is accounted by the increase of the virial density with time. In fact, if we were to rescale the density axis according to the value of the virial ratio, most of the variation would disappear, with only some evolution at the low density tail of the distribution. In the case of filaments, we find a more complex behaviour. Compared to the high redshift results, at present filaments have a more extended density range at both tails of the distribution. It implies a dual nature to filamentary evolution, with some regions becoming more dense, while at the same time there are other filamentary regions that become more underdense. This, together with \reffig{fig:evolution_mean_density}, paints a picture of filamentary environments that while becoming more massive at later times, they also become more mass segregated, with a higher contrast between high and low density regions inside filaments. 

In contrast, walls and voids show a much simpler progression. In the case of walls, while the high density tail does not change significantly, there are notably more underdense regions. Given that the mean wall density is almost constant with time, it suggests that most of the mass content of sheets is located in a few very massive regions, but small in volume. Many of the remaining wall regions have low to very low densities, with no massive haloes in them. This explains why cosmic sheets are so difficult to identify, in both simulations and observations. The remaining component, voids, shows a clear shift of the density PDF towards lower $1+\delta$ values at present time, which suggests a significant emptying of void regions.

\subsection{Linear and surface density distribution}
\label{subsec:evolution:evolution_linear_density}

\begin{figure}
    \centering
    \includegraphics[width=\linewidth]{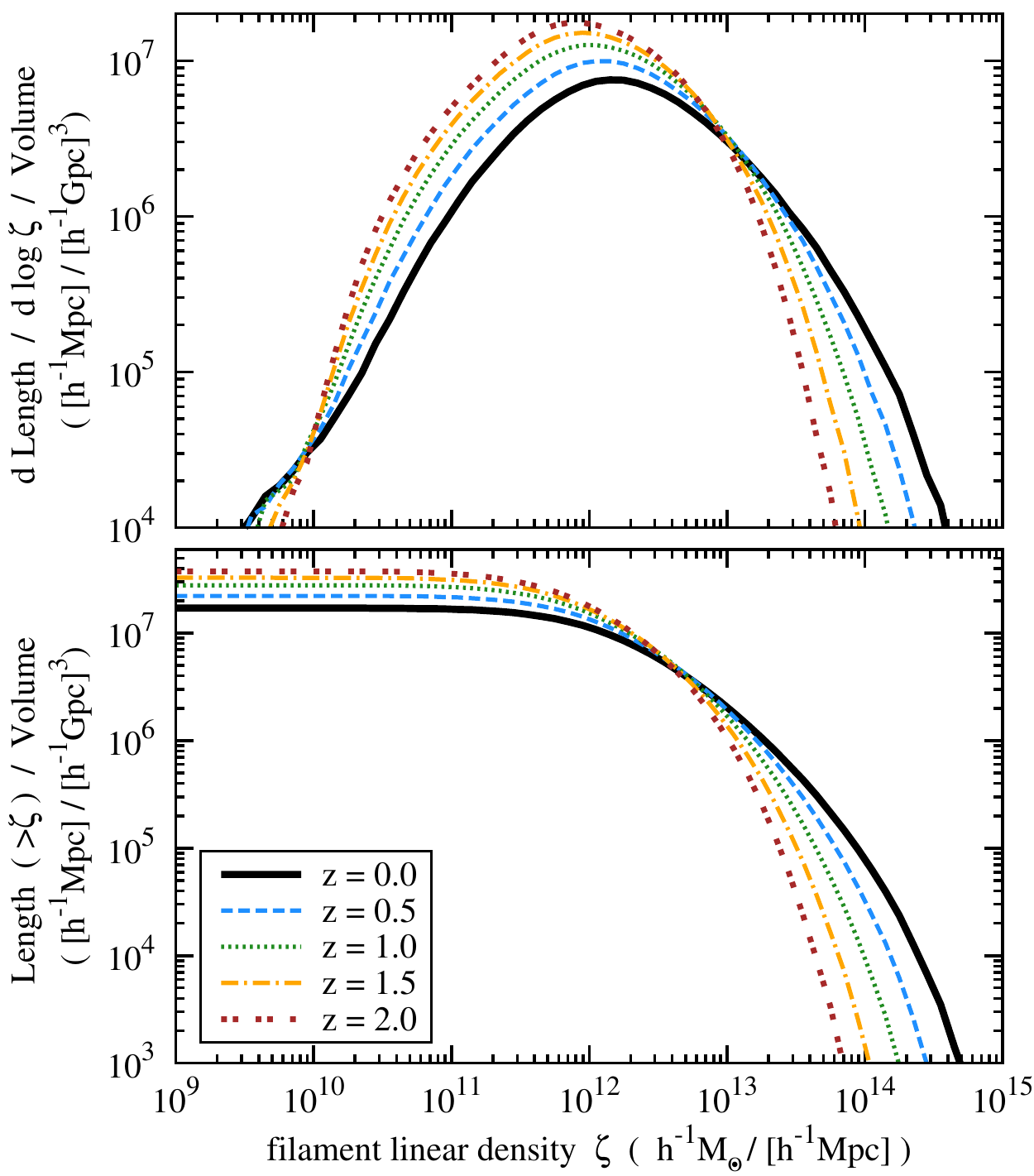}
    \caption{ The length of filaments per unit volume as a function of the filament linear mass density and redshift. It gives both the differential length (top panel) as well as the cumulative one (bottom panel). }
    \label{fig:evolution_env_mass}
\end{figure}

Another way of characterizing the mass distribution across filaments and sheets is in terms of their linear/surface densities, as we already did in \refsec{subsec:evolution:density}. We characterize the time evolution of the linear density $\zeta_\rmn{filament}$ of filaments in the left panel of \reffig{fig:evolution_env_mass}, which shows the length of the filamentary network at various linear densities and different redshifts. The figure shows how typical filament segments evolve to be much more massive at present time. This change can be seen in the shift of the distribution peak towards higher $\zeta_\rmn{filament}$ values, with late time filaments significantly more massive. Of especial interest is the considerable increase in the number of massive filamentary segments, which shows the tendency to accumulate mass in just a few structures. These correspond to filaments around cluster and group massed haloes, which increase tremendously in length since $z=2$. In contrast, the extent of less massive filaments decreases considerably towards present time.

\begin{figure}
    \centering
    \includegraphics[width=\linewidth]{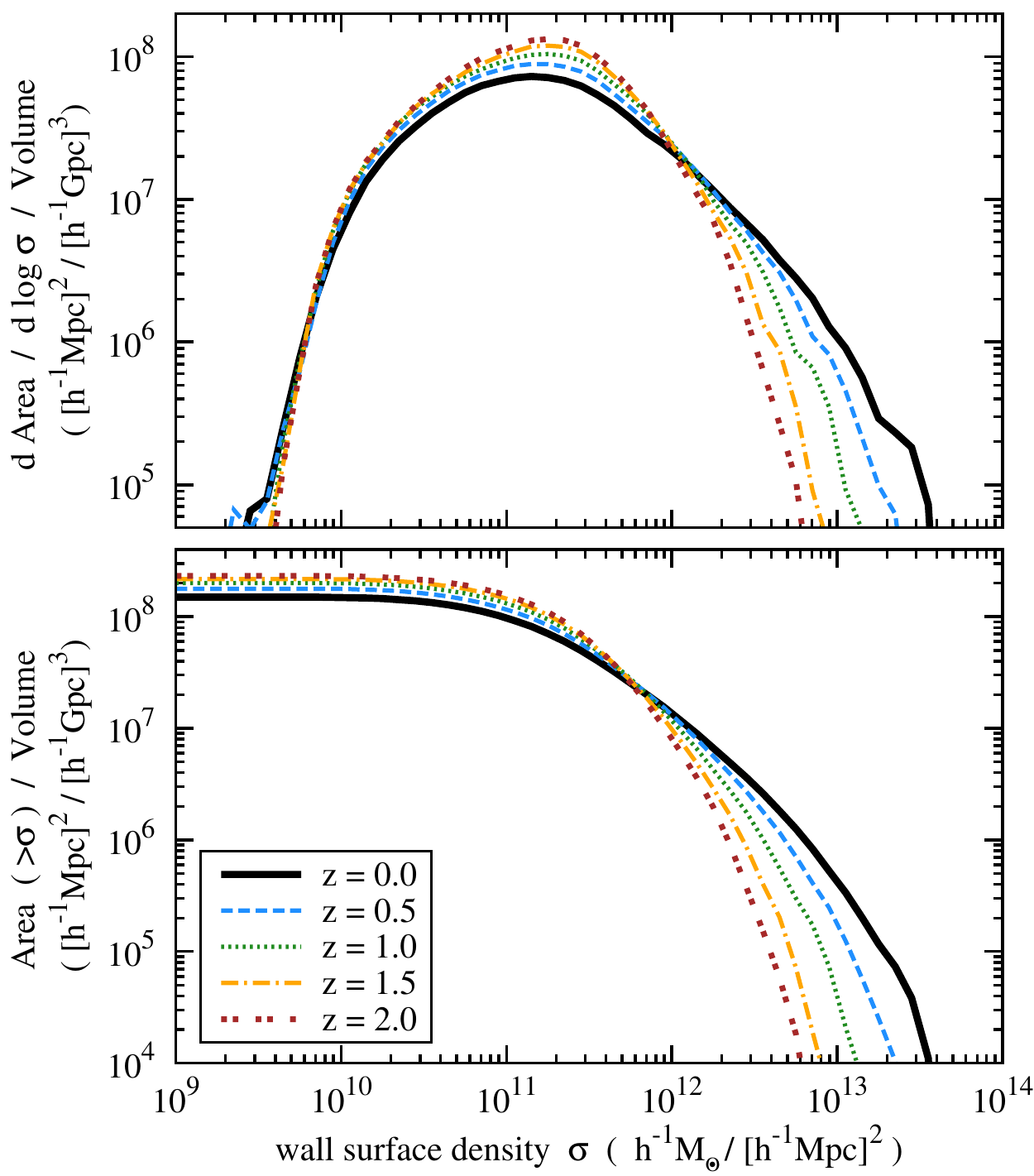}
    \caption{ The area of walls per unit volume as a function of wall mass surface density and redshift. It gives both the differential length (top panel) as well as the cumulative one (bottom panel). }
    \label{fig:evolution_env_wall_mass}
\end{figure}


In contrast to filaments, the typical sheet regions become less massive at present time, as shown in \reffig{fig:evolution_env_wall_mass}. The decrease in wall surface density $\sigma_\rmn{wall}$ is seen as the shift in the peak of the distribution towards lower $\sigma_\rmn{wall}$ values at later times. As we already saw in \refsec{subsec:evolution:evolution_mass_fraction}, the mass fraction of walls decreases towards the present time. \Reffig{fig:evolution_env_wall_mass} shows that this takes place via two processes. First, as we just argued, typical sheet stretches become less massive. And secondly, the extent of the wall network reduces at later times, as seen in decreasing peak values of the $\sigma_\rmn{wall}$ distribution.

Of considerable interest is the variation in the tails of the $\sigma_\rmn{wall}$ distribution. Similarly to filaments, the extent of the massive sheet regions increases substantially since early times. So while most walls decrease in mass, there are a few structures that do become more massive. On the other hand, the low $\sigma_\rmn{wall}$ tail shows very little time variation. It suggest a freeze-out in the mass distribution of tenuous sheets, which may be indicative of the fact that these structures are succumbing to the accelerated expansion of the universe. This is the case since most walls are predominantly in underdense regions (see \reffig{fig:evolution_density_pdf}), which, due to the reduced matter content, experience a faster expansion than overdense regions.

\begin{figure}
    \centering
    \mbox{\hskip -0.0truecm\includegraphics[width=\linewidth]{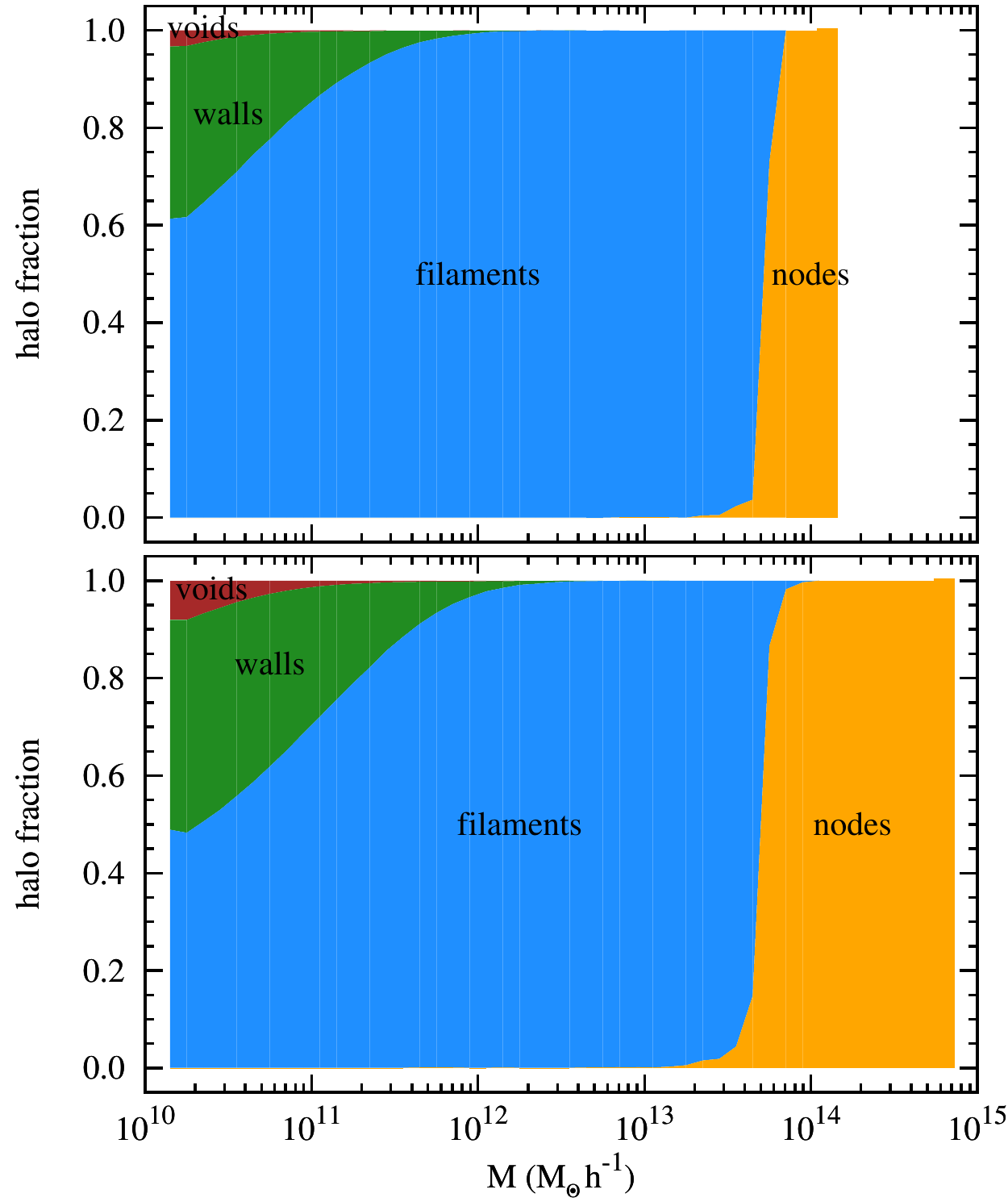}}
    \caption{ The fraction of haloes in the different components of the cosmic web as a function of halo mass. The two panels show the results for two different redshifts: $z=2$ (top frame) and $z=1$ (bottom frame). Compare with the bottom panel of \reffig{fig:evolution:halo_mass_function} which gives the same results for $z=0$.} 
    \label{fig:evolution:halo_mass_fraction_redshift}
    \centering
    \includegraphics[width=\linewidth]{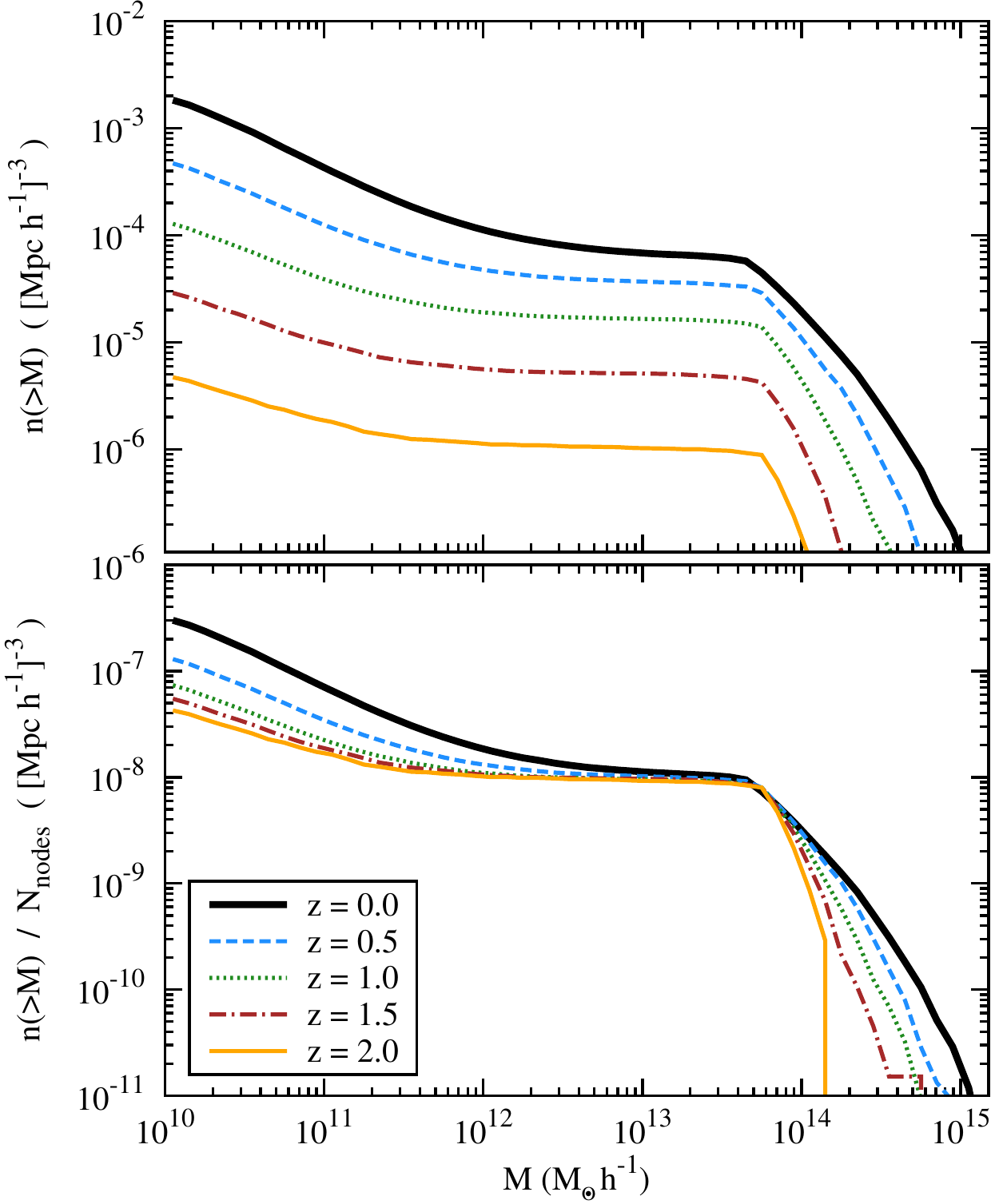}
    \caption{ The evolution of the halo mass function in node environments. The top panel gives the mass function normalized to the volume of the entire simulation. The bottom frame gives the same result further divided by the number of cluster regions. }
    \label{fig:evolution:halo_mass_cluster}
\end{figure}

\subsection{Halo distribution}
\label{subsec:evolution:evolution_halo_population}

The evolution of the halo population across the cosmic web can be easily argued to be of considerable importance, due to the close connection between haloes and galaxies. Motivated by this, we explore in \reffig{fig:evolution:halo_mass_fraction_redshift} the evolution of the halo population segregated according to the morphological component in which the halo resides. It clearly shows the variation with time of the distribution of fixed mass haloes across environments.

\begin{figure}
    \centering
    \includegraphics[width=\linewidth]{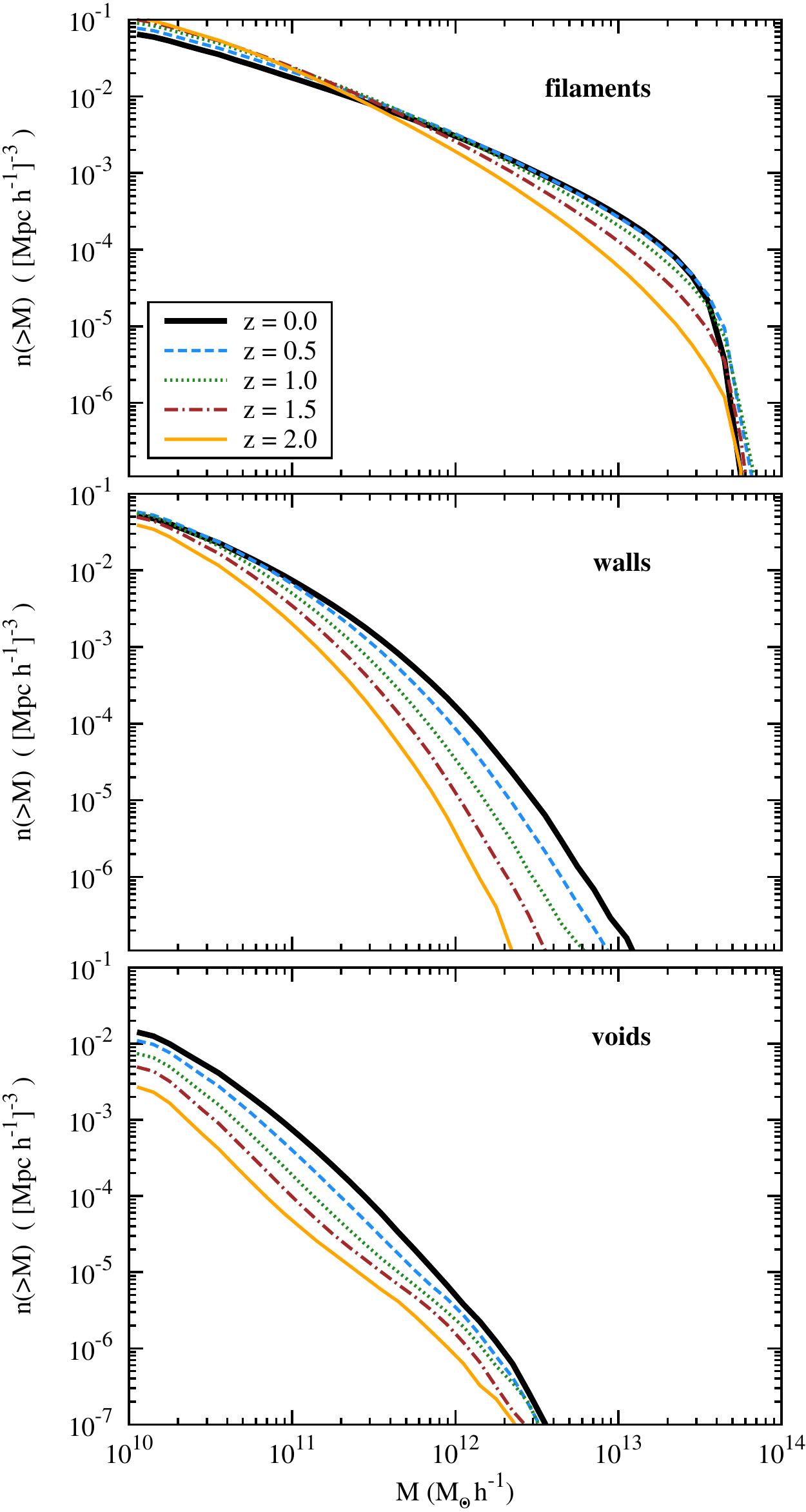}
    \caption{ The evolution of the mass function for haloes residing in filament (top), wall (centre) and void (bottom) environments. The mass function is normalized to the entire volume of the simulation. }
    \label{fig:evolution_halo_mass_boxVolume}
\end{figure}

\Reffig{fig:evolution:halo_mass_cluster} shows that the halo mass function of cosmic web nodes shows a considerable evolution, with significantly lower values at earlier times. Most of this variation is given by the rapid increase in the number density of node environments since high redshift till today. In fact, the change in the mass function of node haloes almost vanishes when normalizing it by the total number of cluster regions. In such a case, the only significant evolution is at the high mass tail of the distribution. It implies that at later times not only that there are more cluster regions, but that the cosmic web nodes contain many more massive haloes.

In contrast to node regions, the halo population across filaments shows a much slower evolution, as can be assessed from \reffig{fig:evolution_halo_mass_boxVolume}. The largest changes take place at high halo masses, with the population of such objects showing a considerable increase in number. The trend is reversed at lower masses suggesting that many small haloes are accreted by their more massive counterparts and therefore feeding the growth of these massive haloes. The slow evolution of the filaments halo population is even more interesting given that the volume of  filament halves since $z=2$ till present time (see \reffig{fig:evolution_mass_volume_fraction}). It implies that the same number of haloes has a much more compact distribution at later times.

Both wall and void regions show a major increase in their halo numbers since high redshift. In the case of walls, the increase is most pronounced at the high mass tail, while for voids the largest variation is found in the number of low mass haloes. The rapid change in these populations indicate that same mass haloes living in wall and void regions are more likely to be younger than their counterparts found in filaments and clusters \citep{Aragon07,Hahn2007a,Hahn2007b}. It also suggest that galaxies living in wall and void regions evolve more slowly that their higher density analogues and therefore probably correspond to earlier stages of the galaxy formation process. On account of this, a comparison between void galaxies and their higher density counterparts can offer insights into the dominant galaxy evolution mechanisms acting at different times, without the need of high redshift observations \citep{Kreckel2011,Kreckel2012}.


\section{Mass transport across the cosmic web}
\label{sec:evolution:mass_trasport}

\begin{figure*}
    \centering
    \includegraphics[width=0.99\linewidth]{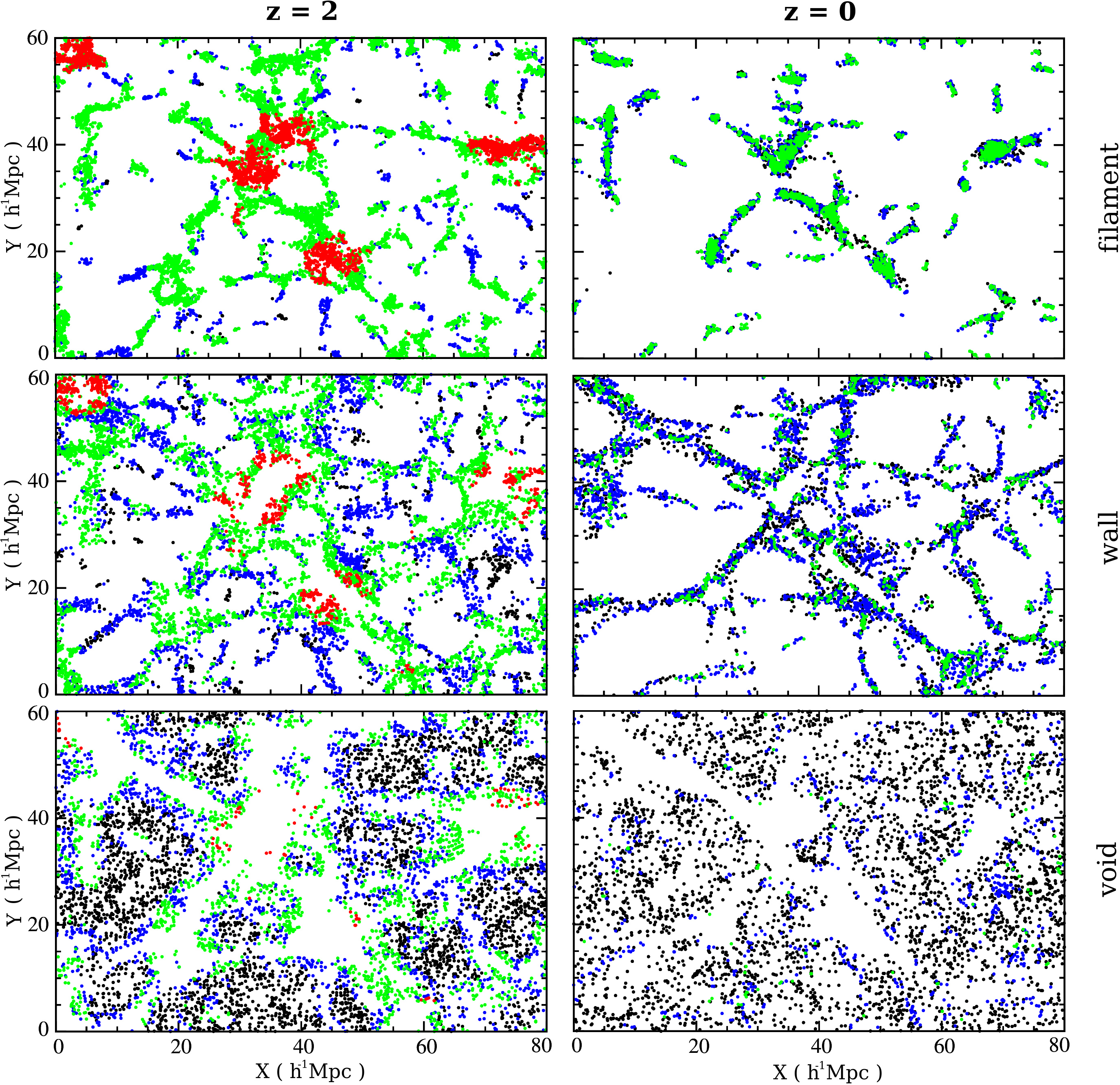}
    \caption{ The DM particles segregated according to their environment at two different redshifts: $z=2$ (left column) and $z=0$ (right column). Each row shows only the particles located in: filaments (top row), walls (centre row) and voids (bottom row). The particles plotted at $z=2$ (left column) are coloured according to their environment tag at $z=0$: cluster (red), filament (green), wall (blue) and void (black). Similarly, the particles shown at $z=0$ are coloured according to their environment at $z=2$, using the same colour scheme. The graph shows a small fraction, selected randomly, of the DM particles found in a $2\Mpch$ thick slice. }
    \label{fig:evolution:env_particle_transport}
\end{figure*}

\begin{figure}
    \centering
    \includegraphics[width=\linewidth]{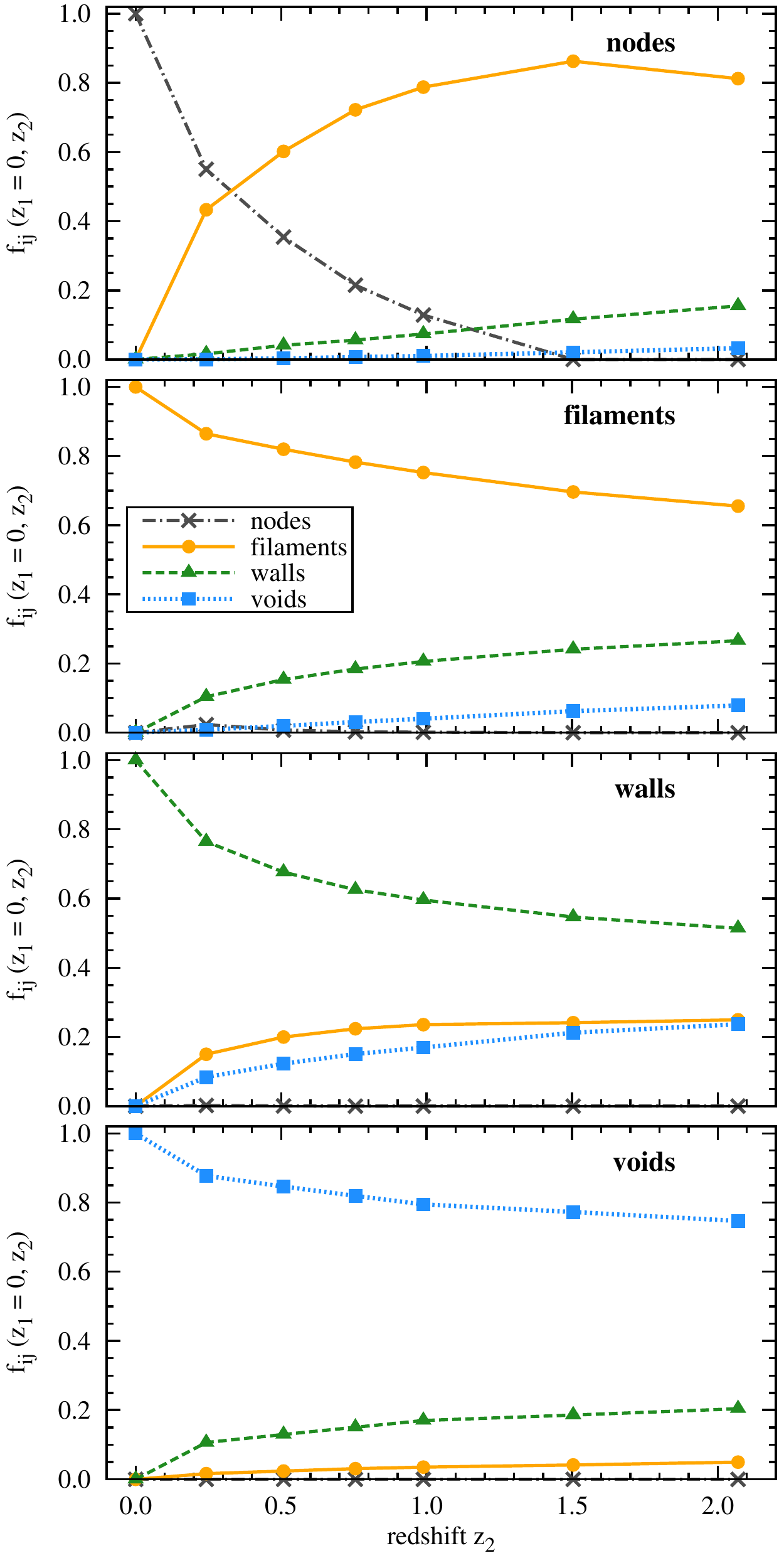}
    \caption{ Tracing back in time which environments contributed to the current mass in cosmic web nodes (top row), filaments (second row), walls (third row) and voids (bottom row). The y-axis gives how the mass of a given environment at $z_1=0$ was split among the cosmic web components at a higher redshift $z_2$.  }
    \label{fig:evolution_env_mass_present}
\end{figure}

\begin{figure}
    \centering
    \includegraphics[width=\linewidth]{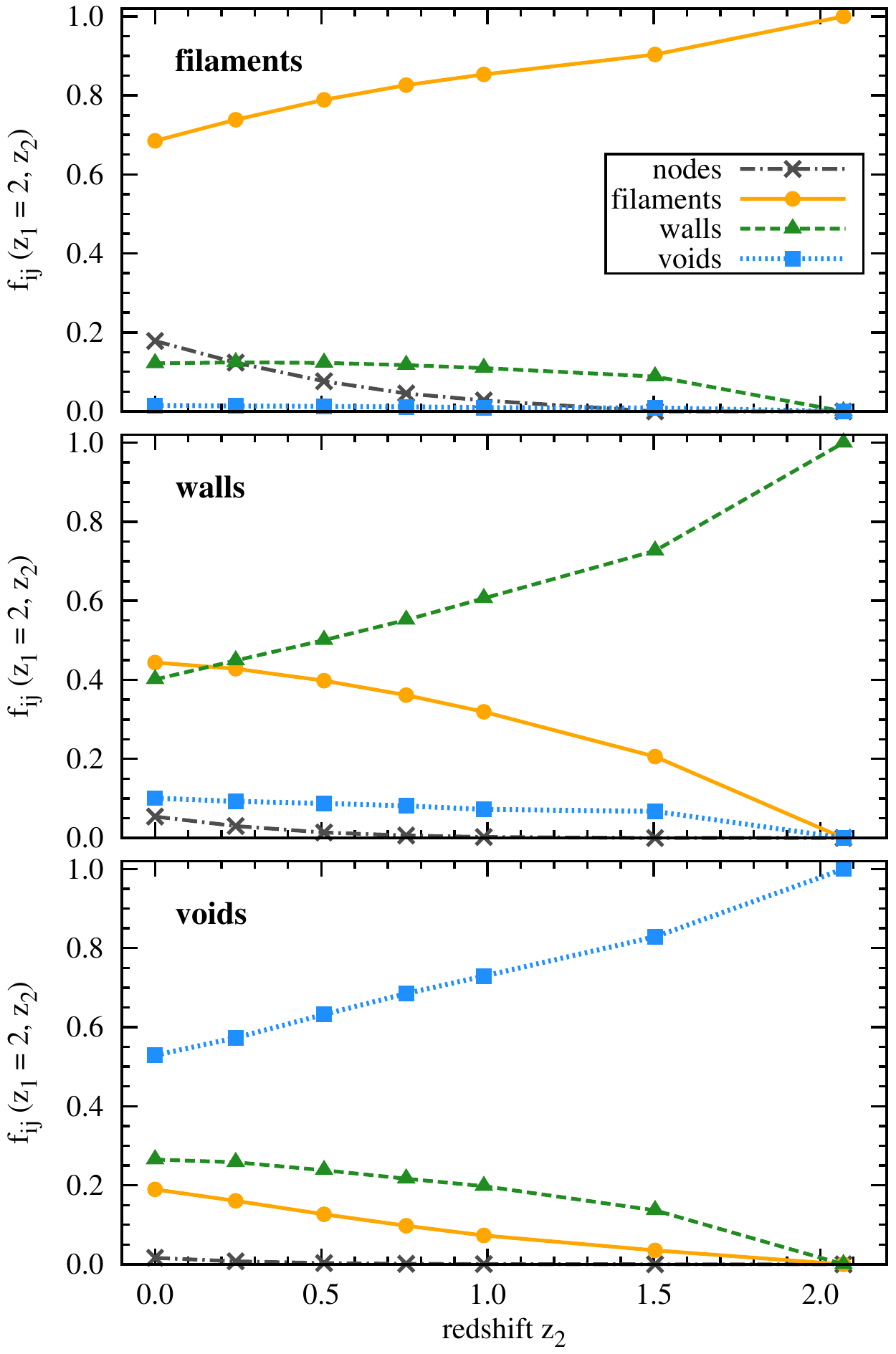}
    \caption{ Tracking the final destination of the $z=2$ mass found in cosmic web filaments (top), walls (centre) and voids (bottom). The y-axis gives how the mass in a given environment at $z_1=2$ was split among the cosmic web components at lower redshift $z_2$. Note that at $z=2$ we do not find any cosmic nodes. }
    \label{fig:evolution_env_mass_past}
\end{figure}

Following the time evolution of the cosmic web raises an important question: what is the path that matter follows before arriving into its present environment? This question is related to how gravitational collapse takes place for an anisotropic distribution of matter. According to gravitational instability theory \citep[][]{Zeldovich70,Icke73,White1979,Sheth2001,Shen06}, an overdense region first collapses along the direction with the largest positive eigenvalue of the deformation tensor to give rise to a pancake-like matter distribution. If the second largest eigenvalue is positive too, than a collapse along this second axis takes place resulting in filament-like regions. And last, regions with a third positive eigenvalues contract along the third direction to give rise to fully collapse objects. This sequence of events predicts a well defined evolution of the matter distribution, with mass flowing from voids into walls, than into filaments and only in the last step into cosmic web nodes. The predictions of this standard view should be easily testable, given the identification of the cosmic web components at different redshifts.

We find that the majority of mass elements flow according to the predictions of the gravitational instability theory, from less dense to more dense environments. For example, most of the DM particles located in filaments at $z=2$ are found at the present time in cluster and filament regions. Similarly, most wall particles either remain in their current environment or are accreted to filaments and clusters. This is easily seen with the help of \reffig{fig:evolution:env_particle_transport}, which shows the DM particles in a thin slice segregated according to their environment at $z=2$ and $z=0$. To illustrate the changing cosmic web environment, the particles at $z=2$ are coloured according to their environment at present time, while the $z=0$ particles are painted according to their environment at $z=2$. 

While the transport of most mass elements between cosmic web components is in accordance with the predictions of the \MCn{anisotropic collapse theory}, there are a few that show an opposite flow, from more dense to least dense morphological components. For example, several of the $z=2$ filament particles in the top-left panel of \reffig{fig:evolution:env_particle_transport} are classified as wall particles at $z=0$. Similarly, a small fraction of $z=2$ wall particles are found to reside in void regions at the present time. The common link between these outliers is that they populate tenuous filaments and walls. Therefore, rather than presenting a challenge to the standard theory of cosmic web evolution, such results reveal the difficulty in identifying filaments and walls in underdense regions. This challenge is clearly visible when comparing the results of the five identification methods shown in \reffig{fig:evolution:env_filaments}, since most differences arise in the detection of tenuous structures in void-like regions. Therefore, we suspect that the puzzling results are due to an incomplete or incorrect identification of environments in underdense regions.

\begin{figure}
    \centering
   \mbox{\hskip -0.05\linewidth \includegraphics[width=1.1\linewidth]{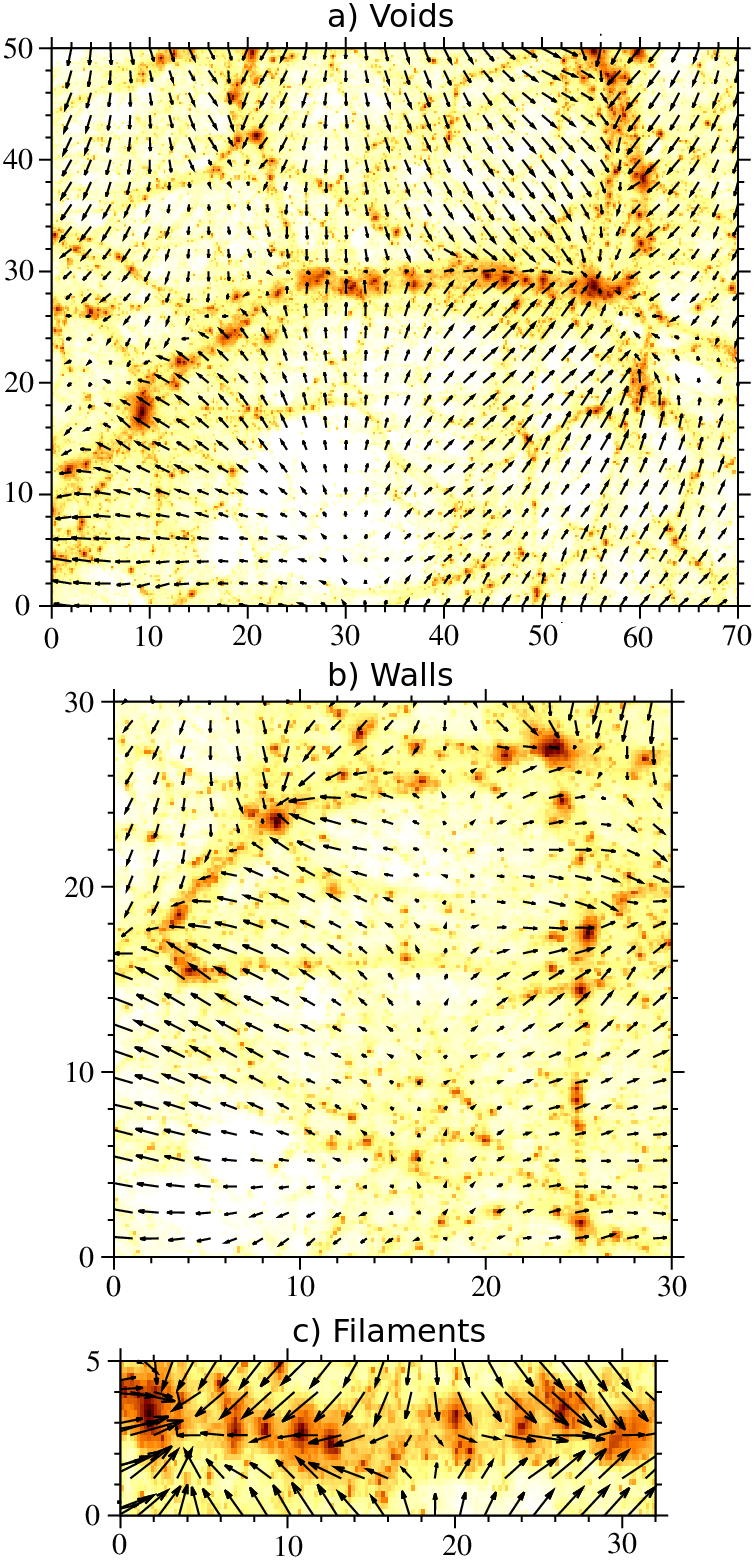}} 
    \caption{ The velocity field along a few typical void (top), wall (centre) and filament (bottom-right) stretches. The background image shows the density field, with light and dark patches corresponding to underdense and overdense regions. The arrows show the direction of the matter flow in the plane of the image, with the size of the arrow proportional to the velocity magnitude. Note that each panel has a different physical size as indicated by the coordinate ticks. }
    \label{fig:evolution:env_velocity_flow}
\end{figure}

\Reffig{fig:evolution:env_particle_transport} offered an intriguing, but only qualitative view on mass transport across morphological components. To undergo a more quantitative analysis, we define the common mass fraction between two cosmic web components at different times, i.e. between component $i$ at redshift $z_1$ and component $j$ at redshift $z_2$, as
\begin{equation}
	f_{ij}(z_1,z_2) = \frac{ M_{(i;z_1)\cap (j;z_2)} } {M_{i;z_1}} \;.
	\label{eq:common_mass_fraction}
\end{equation}
With $M_{i;z_1}$ we denoted the mass in environment $i$ at redshift $z_1$, while $M_{(i;z_1)\cap (j;z_2)}$ denotes the mass overlap between cosmic web components $i$ and $j$, with the first at redshift $z_1$ and the second at $z_2$. To compute the mass overlap we use the id tag of the DM particles to find all the common mass tracers between the two environments at different time steps of the simulation. The quantity $f_{ij}(z_1,z_2)$ has two common interpretations depending on the relation between $z_1$ and $z_2$:
\begin{itemize}
	\item If $z_1<z_2$, then the overlap fraction $f_{ij}(z_1,z_2)$ reveals what percentage of the $z_1$ mass in environment $i$ originated from mass found in environment $j$ at the higher redshift $z_2$.
	\item If $z_1>z_2$, then $f_{ij}(z_1,z_2)$ represents the fraction of $z_1$ mass in environment $i$ that is found at a later time $z_2$ to be contained in environment $j$.
\end{itemize}

In \reffig{fig:evolution_env_mass_present} we investigate the morphological origin of the mass found in the present day cosmic web environments. In the case of cosmic web nodes, most of their mass originated in filaments and only a small fraction of it in walls. This agrees very well with the standard picture according to which clusters accumulate mass from the filaments that have the cluster as one of their end-points \MCn{\citep[e.g.][]{Shandarin1984,Shandzeld89,vanHaarlem1993,Aubert2007,WeyBond08a}}. Most of the mass in filaments seems to have been part of filamentary environments since early redshift, with only a small fraction coming from walls and voids. It has important implications for the population of filament haloes and galaxies since it implies that the majority of such objects have been in filament environments since at least $z=2$. 

\Reffig{fig:evolution_env_mass_present}  also characterizes the fraction of mass that changed environment in opposite way than predicted by the gravitational instability theory. As we already argued, this is indicative of the limitations of our method in the identification of tenuous structures. Quantifying this artefact is important in understanding if this drawback represents a serious problem for our analysis. We find that up to $20\%$ of present day mass content of sheets has been identified as part of filaments at an earlier time, with a similar mislabelling of void mass content too. Therefore, the artefacts arising from the difficulty of detecting tenuous environments, though not dominant, cannot be neglected.

\Reffig{fig:evolution_env_mass_past} shows another way of looking at the evolution of matter in the cosmic web. It plots the successive destinations of the matter that is initially, at $z=2$, identified as being part of filaments, walls and voids. Compared to the previous figure, it illustrates the rapid outflow of mass from walls and voids. Less than $40\%$ of the walls $z=2$ mass is still part of present days sheets, with most of the mass flowing into filaments. For voids, around half of their high redshift mass has streamed out into sheets and filaments, showing the significant outflow of mass from underdense regions. These results show that even though individual structures extended on tens of megaparsecs scales, the cosmic web components are evolving rapidly and are far from being static structures.

Fully understanding the mass transport across morphological components can only be done by investigating the large scale velocity field, since this is the main driver behind megaparsec-scale mass flows. To that end, we present in \reffig{fig:evolution:env_velocity_flow} the peculiar velocity field across a few representative void, wall and filament stretches. The void regions are characterized by a strong outflow, with the velocity clearly pointing out towards the sheets that act as void boundaries. One such sheet is visible in the centre of panel a), and it shows that the walls accrete mass from both of the two voids separated by the sheet. Moreover, the direction of the inflow is close to the normal to the wall, which resides along the horizontal line in the plane of the figure. This is true for most parts of the sheet, except close to large agglomerations of mass. Once in sheets, the matter outflows towards the filaments that border the wall, as clearly seen in panel b) of \reffig{fig:evolution:env_velocity_flow}. In filaments, the flow points towards the two massive clusters which act as the filament's endpoints.

According to the above velocity field, matter outflows from voids into walls, while the mass content of sheets streams towards filaments. In turn, the filaments act as matter transport highways towards the clusters bounding them. This is in very good agreement with both the predictions of anisotropic collapse theory and the mass transport results we obtained in this section. Moreover, it offers conclusive proof that the artificial transport of mass from filaments into walls or from walls into voids is an artefact of the cosmic web identification methods and does not pose a challenge to current cosmic web evolution theories.


\section{Size and distribution of filamentary and wall networks}
\label{sec:evolution:morphology_fractal}

In the previous sections we characterized the evolution of the cosmic web in terms of both its mass and halo content, finding a marked time variation in these quantities. But these are not the only way of describing morphological components, since such structures also have certain spatial extents, which, according to \reffigS{fig:env_fila_evolution} and \ref{fig:env_wall_evolution}, show a significant evolution too. For this reason, this section is focused on characterizing the evolution in spatial extent of the filamentary and wall networks. We investigate the total length of the filament network as well as the typical diameter of these objects. Of particular interest is the density profile perpendicular to the filament spine and how this correlates to the filament width. The wall network undergoes a similar analysis, with studies of the total area of sheets and their typical thickness. At the end of this section we examine the spatial distribution of filaments and walls by performing a fractal dimensional analysis of these structures.

\subsection{Total extent of filaments and walls}
\label{subsec:evolution:evolution_filament_length}
 
\begin{figure}
    \centering
    \includegraphics[width=\linewidth]{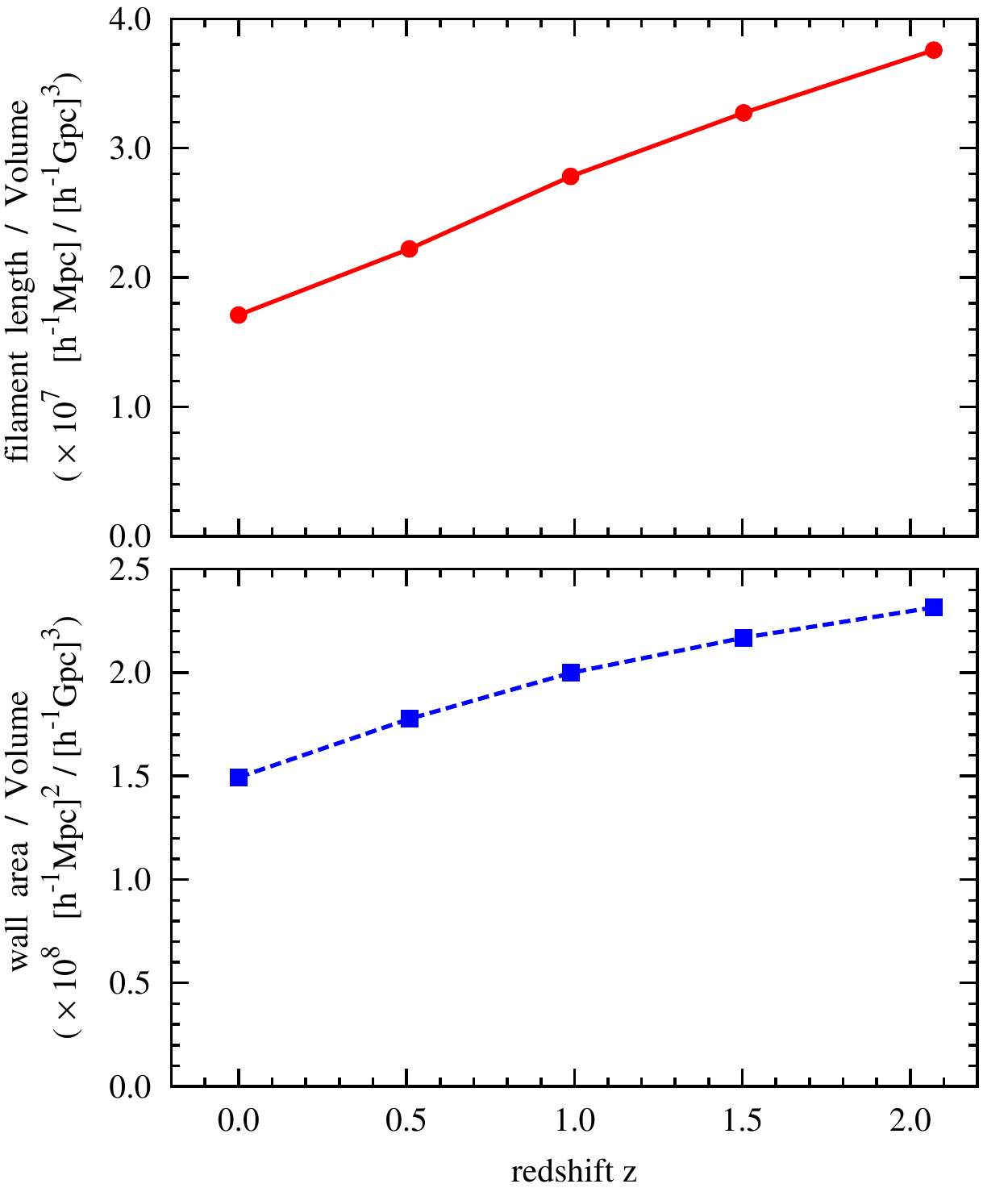}
    \caption{ \textit{Top panel:} The total length of the filamentary network as a function of redshift. \textit{Bottom panel:} The total area of the wall network as a function of redshift. We show volume normalized quantities, such that at present time a $1\GpchVolume$ volume contains filaments with a total length of ${\sim}1.7\times 10^7\Mpch$ and walls with a total area of ${\sim}1.5\times 10^8\MpchArea$. }
    \label{fig:evolution_env_size_on_redshift}
\end{figure}

The most basic way of characterizing the spatial extent of morphological components is in terms of the length of filaments and area of sheets \citep[e.g.][]{Sousbie08a,Pogosyan2009a,Gay2012a}. To that end,  \reffig{fig:evolution_env_size_on_redshift} shows the time variation of these quantities, which were computed using the procedures described in \refsec{subsec:evolution:fila_length_computation} and \refsec{subsec:evolution:wall_area_computation}. As expected, the overall length of filaments has decreased dramatically since high redshift. Nowadays, the filamentary network has only one third of its extension compared to $z=2$. Similarly, the total area of sheets has also decreased since high redshift, but only to a lesser extent. These findings are in very good agreement with the visual impression given by \reffigS{fig:env_fila_evolution} and \ref{fig:env_wall_evolution}, and reinforce the view that at later time both the filament and wall networks have a smaller extension. 

More interestingly, the change in size of both filaments and walls seems to be almost independent of redshift. This is a puzzling result, given that qualitatively we find only a minor evolution of the cosmic web after $z=0.5$ (see \reffigS{fig:env_fila_evolution} and \ref{fig:env_wall_evolution}). The answer to this may lie in the main limitation of the investigated quantities, since the total extension of the filament and wall networks is most sensitive to the tenuous structures, and not to the prominent ones. The tenuous environments are the ones that contribute the most to the length of filaments and area of walls. A more telling analysis involves exploring the change in prominent versus tenuous environments, as characterized by their mass or width distributions. The former has been done in \refsec{subsec:evolution:evolution_linear_density}, while the latter is carried out in the next subsection.

\subsection{Width of filament and wall networks}
\label{subsec:evolution:evolution_filament_diameter}

\begin{figure}
    \centering
    \mbox{\hskip -.1\linewidth \includegraphics[width=1.1\linewidth]{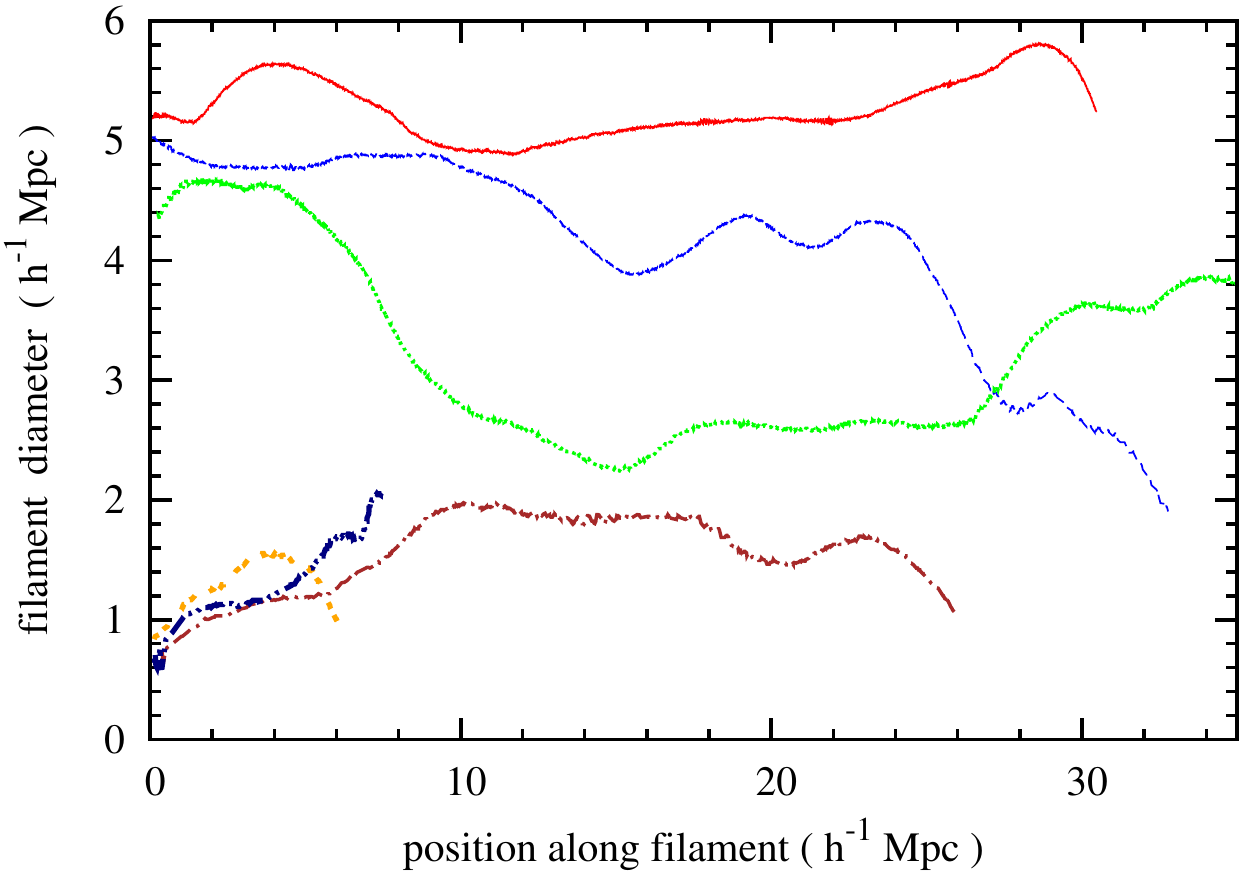}}
    \caption{ The variation of the local filament diameter across a few representative filamentary branches. We exemplify this for six structures, with the three thick ones corresponding to filaments between clusters. The remaining three examples, with smaller thickness, are objects found in underdense regions. }
    \label{fig:evolution:test_fila_diameter_2}
\end{figure}
\begin{figure}
    \centering
    \mbox{\hskip -0.0truecm\includegraphics[width=1.00\linewidth]{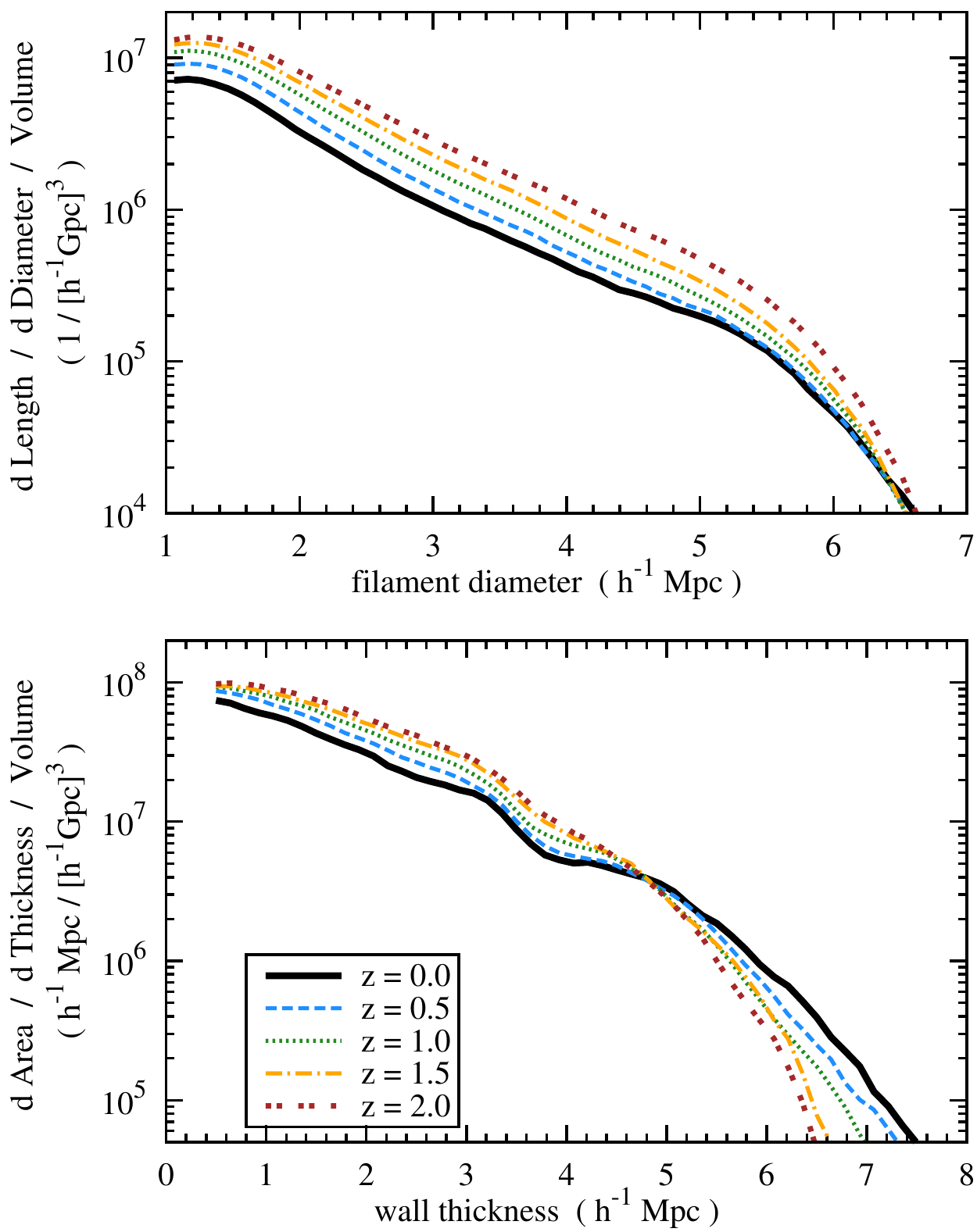}}
    \caption{ \textit{Top panel:} The length of filaments per unit volume as a function of the filament diameter. \textit{Bottom panel:} The area of walls per unit volume as a function of wall thickness. Both frames show the time evolution of those two quantities. The results for filament diameter and wall thickness below ${\sim}1\Mpch$ are affected by finite resolution effects which give rise to the artificial cut-off present at small values. }
    \label{fig:evolution_env_size}
\end{figure}

The cosmic web components are complex structures which show a large degree of variation not only in their mass content, but also in their width distribution. It is easily appreciated from \reffigS{fig:env_fila_evolution} and \ref{fig:env_wall_evolution} that there is a large variation in the width of filaments and walls, not only between different structures, but also along the same object. These observations raise important questions like what is the typical width of morphological components and how does this quantity change in time?

To explore these questions, we proceed by computing the local filament diameter and wall thickness via the procedures described in \refsec{subsec:evolution:fila_diameter_computation} and \refsec{subsec:evolution:wall_thickness_computation}. The resulting quantities describe the width of locally representative stretches of filaments and walls, allowing us to compare not only the sizes of different objects, but also how the width of a structure varies at different points along its spine. This is illustrated in \reffig{fig:evolution:test_fila_diameter_2} where we show the local diameter of a few typical filamentary branches. As expected, we observe a large degree of variation both between different structures as well as between different points along the same object. In general, the thicker filament segments correspond to a large enclosed mass within that stretch, while the thin structures have low masses and are typically found in underdense regions. This is easily seen when comparing with \reffig{fig:evolution:test_fila_density}, which gives the linear mass density for the filaments shown in \reffig{fig:evolution:test_fila_diameter_2}. The two figures demonstrate the close connection between the local mass density and the local width of filaments, relation which is further investigated in \refsec{subsec:evolution:evolution_width_density}.

The distribution of filament and sheet widths is shown in \reffig{fig:evolution_env_size}. The quantitative results support the visual impression of filament and wall networks that have few thick structures and many more thin ones. Moreover, it underlines the limitation of using global quantities, like the total length of filaments, to characterize the evolution of morphological components. Such quantities are most sensitive to the tenuous structures and cannot describe the time variation of the prominent objects, which contain both most of the mass and most of the haloes.

Both the filaments and sheets have a wide range of widths, with a sharp cut-off at high values. The distribution of thin objects, with widths smaller than $\lsim 1\Mpch$, is affected by resolution effects, since in those cases the thickness of filaments and walls is comparable to the grid spacing used to identify these structures. We find that \nexus{} filaments have typically diameters below $5\Mpch$, though there are some rare structures which locally have even higher widths. These results are in good agreement with the findings of \cite{2010MNRAS.408.2163A} which suggest that the prominent filaments have a radius of ${\sim}2\Mpch$ \citep[see also][]{Colberg05,Gonzalez09,Bond10}. In the case of sheets, we find very few structures of $5-8\Mpch$ thickness as reported by \cite{2010MNRAS.408.2163A}, with most of our walls being thinner than this. A visual comparison of the \nexus{} sheets and those identified by the MMF method, which was used in \cite{2010MNRAS.408.2163A}, suggests that the latter one has problems in identifying coherent planar structures in the matter distribution (for details see \nexusPaper{}). The discrepancy is indicative of the difficulties arising in the detection of cosmic sheets, given that these morphologies typically correspond to tenuous structures.

The time variation of the filament and wall thickness offers another important insight into the evolution of the cosmic web. For example, we find a consistent decrease in the length of the filamentary network at fixed filament diameter. For thin filaments this can be easily understood, given that since $z=2$ many such structures merge with thicker filaments and therefore results in fewer thin objects. In contrast, this process cannot explain the evolution of thick filaments given that we find the same network of prominent filaments at all times. This suggests another process at work, the contraction of filaments to become more concentrated and therefore having smaller diameters at later times. Most probably this effect plays a role in the evolution of both thin and thick filaments, but it is more obvious for prominent structures since in this case it does not compete with other processes.

The change in sheet thickness is governed by the same processes as for filaments, since walls show a very similar evolution of their widths. The only exception is for very thick sheets, whose number seem to increase at later times. We suspect that this is an artefact arising from the difficulty of identifying sheets. Typical walls are emptier of matter at later times (see \refsec{subsec:evolution:evolution_density_distribution}), which means that \nexus{} needs to use a lower morphological signature threshold. This leads to identifying the massive sheets as being slightly thicker, since the density profile decreases only slowly at the boundary of such walls (e.g. see \reffig{fig:evolution:test_fila_density_profile}). Anyway, such an effect does not seem to play an important role for most sheets.

\subsection{The width - density relation}
\label{subsec:evolution:evolution_width_density}

\begin{figure}
    \centering
    \mbox{\hskip -0.5truecm\includegraphics[width=1.05\linewidth]{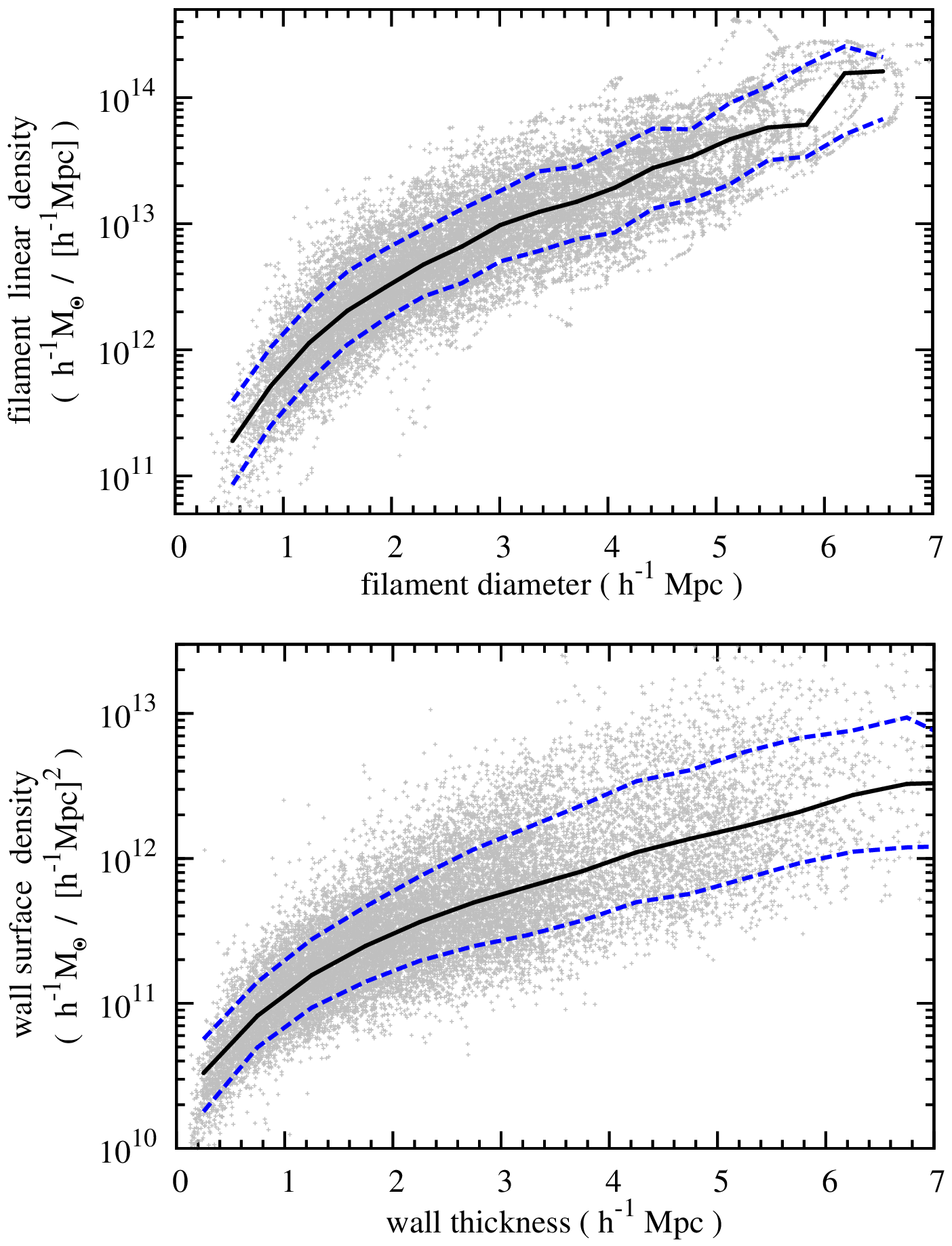}}
    \caption{ The correlation between the width and the mass distribution of filaments and walls. The points show a small subset of filament and wall stretches. The solid lines show the median relation and the dashed curves show the $16$ and $84$ percentiles. \textit{Top panel:} The filament linear density as a function of filament diameter. \textit{Bottom panel:} The wall surface density as a function of wall thickness. }
    \label{fig:evolution:test_fila_diameter_density_relation}
\end{figure}
\begin{figure}
    \centering
    \includegraphics[width=\linewidth]{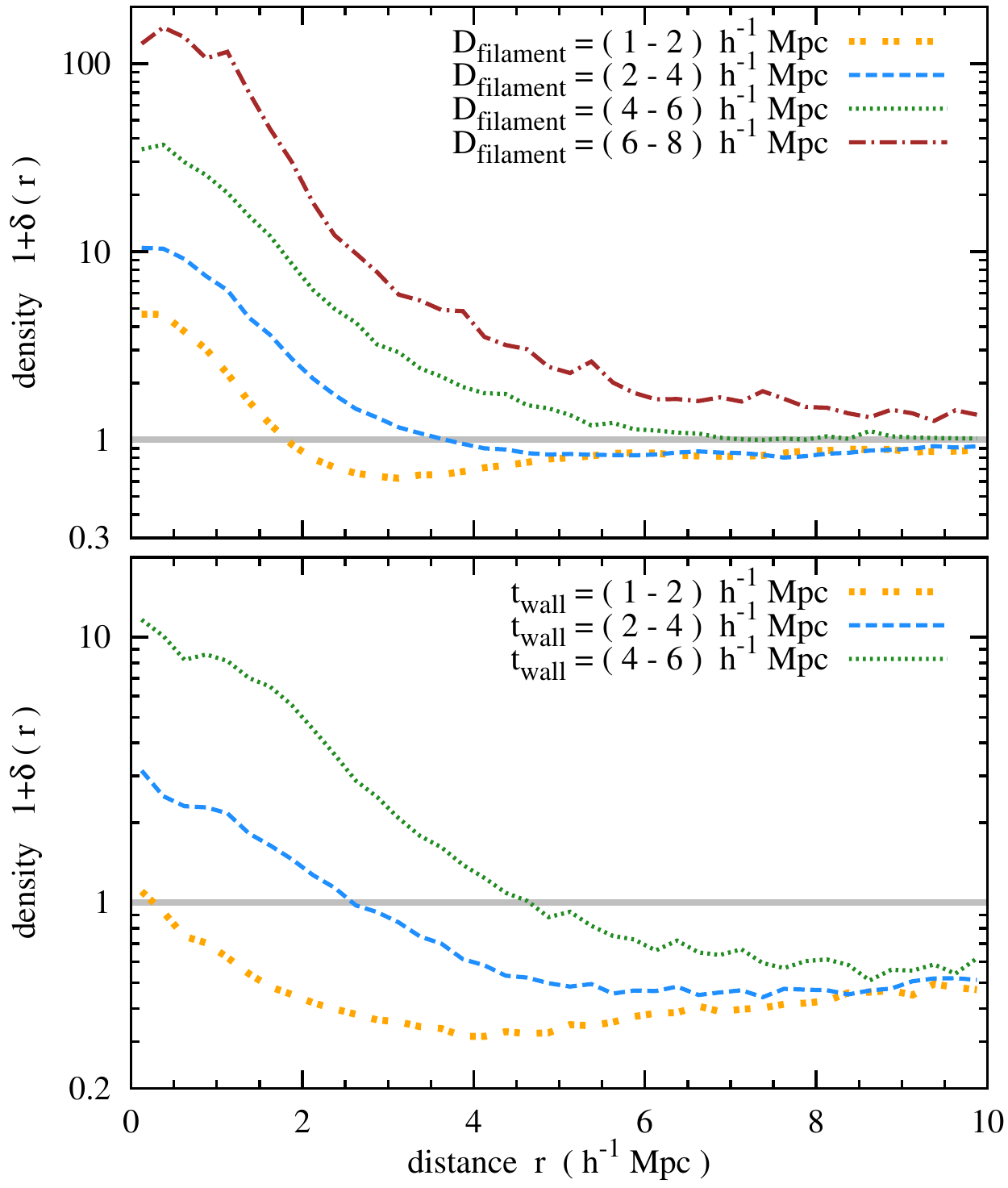}
    \caption{ The mass distribution around filaments (top) and walls (bottom) as a function of the distance $r$ from the centre of these structures. The quantity $1+\delta(r)$ denotes the mean density at that distance. We show separate profiles for structures of different widths, as indicated by the filament diameter $D_\rmn{filament}$ and by the wall thickness $t_\rmn{wall}$. Each profile corresponds to an average over all the objects with a given width. }
    \label{fig:evolution:test_fila_density_profile}
\end{figure}

As was already hinted in the previous section, there seems to be a correlation between the width and density of filaments, and possibly for walls too. Such a relation can shed light on whether the thickness of a structure is only dictated by its mass or if there are other factors at play. If the former holds true, then a filament's width, which is easier to assess observationally, can be used to infer the mass enclosed within the structure. 

\Reffig{fig:evolution:test_fila_diameter_density_relation} highlights the correlation between the width and mass distribution of filament and wall segments, by showing a scatter plot of these quantities. We do observe a general trend, with thicker structures characterized by larger linear or surface mass densities. Our data give a good measurement of this relation for objects wider than $\gsim2\Mpch$, while the results for objects thinner than this are subject to resolution effects. This is clearly visible as a change in the mean trend below $2\Mpch$, given that the width estimate is more prone to effects arising from the coarse grid used in our study.

While we do find a clear correlation between the local diameter and the local linear mass density of filaments, we also see a large object-to-object scatter. Given the width of a filament, its mass can only be estimated to within a factor of 5 (see the $1\sigma$ curves in \reffig{fig:evolution:test_fila_diameter_density_relation}). Therefore, while a filament's diameter cannot be used to reliably assess its mass, a statistical approach can be employed to estimate the typical mass of a filament population. This can be further used to indirectly search for the missing baryons, a large fraction of which are thought to be inside filaments in the form of the warm-hot intergalactic medium \citep[e.g.][]{Cen1999,Nicastro2005}. Such an analysis can be applied only to thick filaments, since these are the only ones with a high enough number density of galaxies to allow for a good determination of their width.

A visual inspection of the density field suggests that both filaments and walls are characterized by a dense inner region surrounded by more diffuse matter \citep{Colberg05,Dolag06,2010MNRAS.408.2163A}. This suggests a natural way of defining the edge of these structures as the point where the density field drops to the local background value. Given this alternate way of describing the width of filaments and sheets, how does it compare with our previous width estimates presented in \refsec{subsec:evolution:evolution_filament_diameter}. In general, the two thickness measurements do not have to agree, since the latter method uses the physical extent of the morphological components as identified by \nexus{}, without any knowledge on the position of the density drop. 

\Reffig{fig:evolution:test_fila_density_profile} shows the density profile as a function of the distance $r$ to the filament spine and to the central plane of walls.  As expected, we see a high density peak for small values of $r$ which corresponds to the dense inner regions, followed by a sharp drop at larger distances. The distance where the density profile becomes approximatively flat indicates the edge of the filament or sheet. To connect the two width estimates described in the previous paragraph, we compute the density profile separately for filaments and walls of different morphological widths. Moreover, given the noisy nature of individual density profile \citep{2010MNRAS.408.2163A}, we compute generic profiles by averaging over all objects with a given width. The results show a very good consistency between the edge of the density profile and the \nexus{} width estimates. Therefore, our environment identification technique is also very successful in correctly identifying the edges of filaments and walls.

\Reffig{fig:evolution:test_fila_density_profile} illustrates two more interesting facts. On average, the width of a structure and its core density are related, which has already been shown in a slightly different form in \reffig{fig:evolution:test_fila_diameter_density_relation}. And more importantly, its shows that thin filaments are found in underdense regions while thick ones live in overdense areas. This is easily assessed from the density profile, since the local nature of the region, i.e. underdense versus overdense, can be estimated using the value of the density at large distance $r$. The thinnest filaments, though overdense in their cores, are located in very underdense domains outside their edges, with $\delta(r)\sim-0.5$. In contrast, the very thick structures are found in areas which are overdense even at distances as high as $10\Mpch$. On the other hand, all sheets, indifferent of their thickness, are located on average in very underdense regions \citep{Pogosyan98}. The thinnest of these objects, whose core density barely reaches the mean background density, are found deep inside voids of very low densities.

\subsection{Fractal dimension}
\label{subsec:evolution:fractal_dimension}

The visual impression given by \reffigS{fig:env_fila_evolution} and \ref{fig:env_wall_evolution} is that of a cosmic web that evolves strongly with redshift, and which, at later times, becomes dominated by fewer, but more massive structures. These changes affect the spatial distribution of filaments and sheets, with both the branching characteristics and space filling capacity varying in time. The fractal dimension \citep{Mandelbrot1983} represents one possible way of obtaining a more quantitative description of the time variation in the spatial distribution of filaments and walls. It measures how details in the pattern change with the observed scale and also describes the space filling capacity of the system. Such an analysis has been used in various fields to get an understanding of complex patterns, from the shape of neurons \citep{Smith1989fractal,Jelinek1998} to the intricate structures seen in galactic gas and star forming regions \citep{Feitzinger1987,Elmegreen2001} and to the large scale distribution of galaxies \citep{Jones1988,Martinez1990,Martinez90}.

One simple way to measure the fractal dimension involves the use of the box counting method \citep{Mandelbrot1983}. For our application, it involves overlying on to the simulation box of a regular grid with spacing $l$ and counting how many of the grid cells intersect the pattern that we measure, i.e. the filamentary and wall networks. The number of intersecting grid cells gives the box count $N$ at scale $l$. The method works by measuring the box counts at different scales and then investigating the dependence of $N$ on $l$. In the case of a fractal there is a well defined relation 
\begin{equation}
   N\propto l^{-d}
   \; ,
\end{equation}
with $d$ being the fractal dimension of the pattern. For example, the fractal dimension of an infinitely thin line is $d=1$, of a zero thickness plane is $d=2$ and that of a filled box is $d=3$. In general, a fractal pattern has a non integer fractal dimension showing an intermediate behaviour between the ideal cases just described.

To obtain the fractal dimension, we proceeded by first measuring the box count $N$ for the largest possible box, which is the simulation box. After which, in each successive step, we reduced the box length $l$ by a factor of $2$ and measured $N$ again. This process was stopped when $l$ was equal to the grid spacing used to obtain the filamentary and wall networks. We applied this procedure to the \MII{} data since it allowed us to obtain a larger dynamical range at small $l$, which shows the most interesting behaviour. When comparing between \MII{} and \MI{} results we could not find any important differences, suggesting that the \MII{} findings that we discuss below are not significantly affected by cosmic variance.

\begin{figure}
    \centering
    \includegraphics[width=\linewidth]{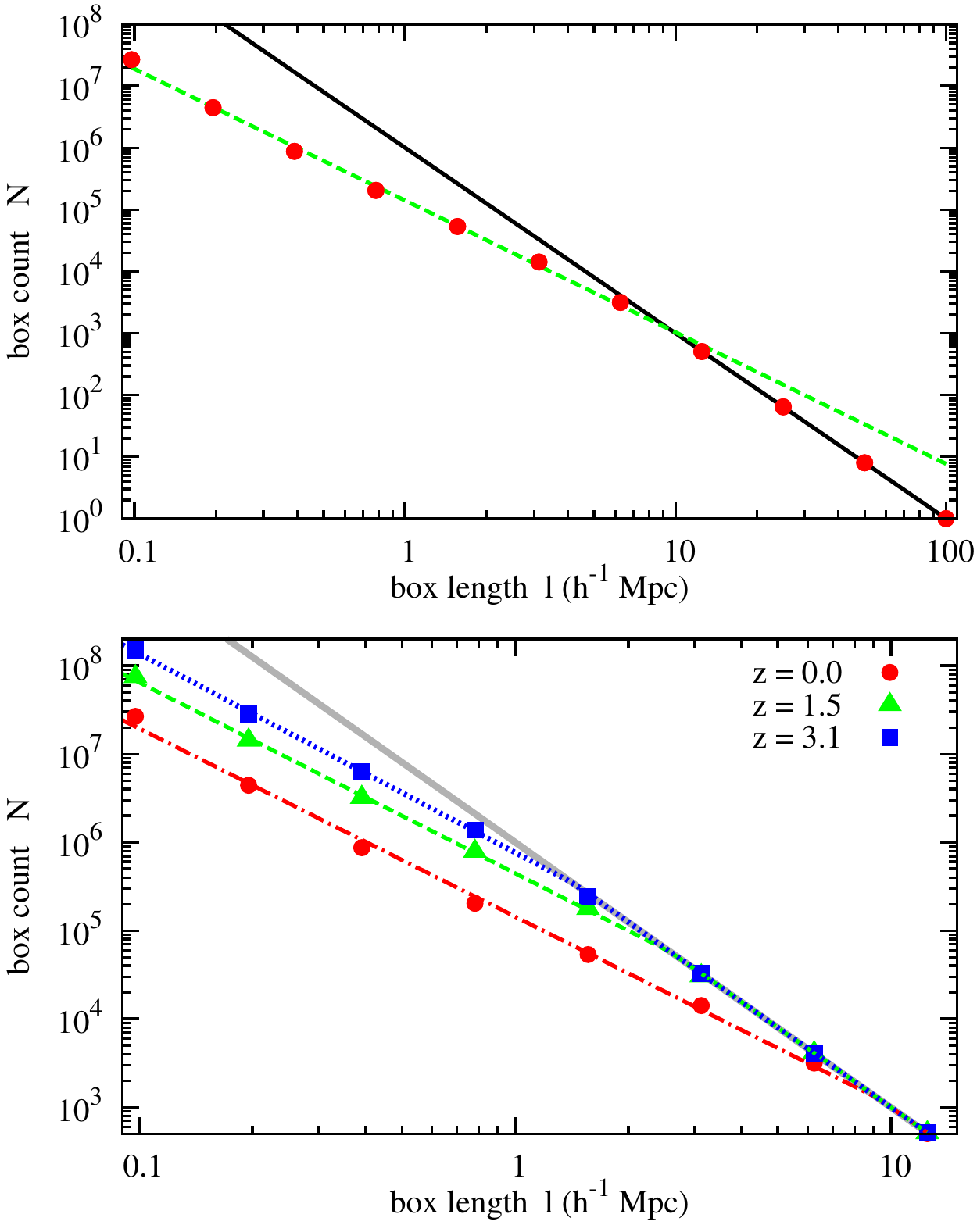}
    \caption{ The output of the box-counting method for the determination of the fractal dimension. \textit{Top panel:} The number of box counts that contain filaments as a function of box length. We distinguish two regimes: for small boxes we see the fractal behaviour of filaments (dashed line), while at larger lengths we find $N\propto l^{-3}$ (solid line) which suggests that all boxes at that scale intersect at least one filamentary region. \textit{Bottom panel:} Zoom-in on the left frame showing the box counting results for filaments at three redshifts: 0.0, 1.5 and 3.1. The solid line shows an $N\propto l^{-3}$ behaviour while the remaining lines show the best fitting results for each data set. }
    \label{fig:fractal_dim_test}
\end{figure}

\begin{figure}
    \centering
    \includegraphics[width=\linewidth]{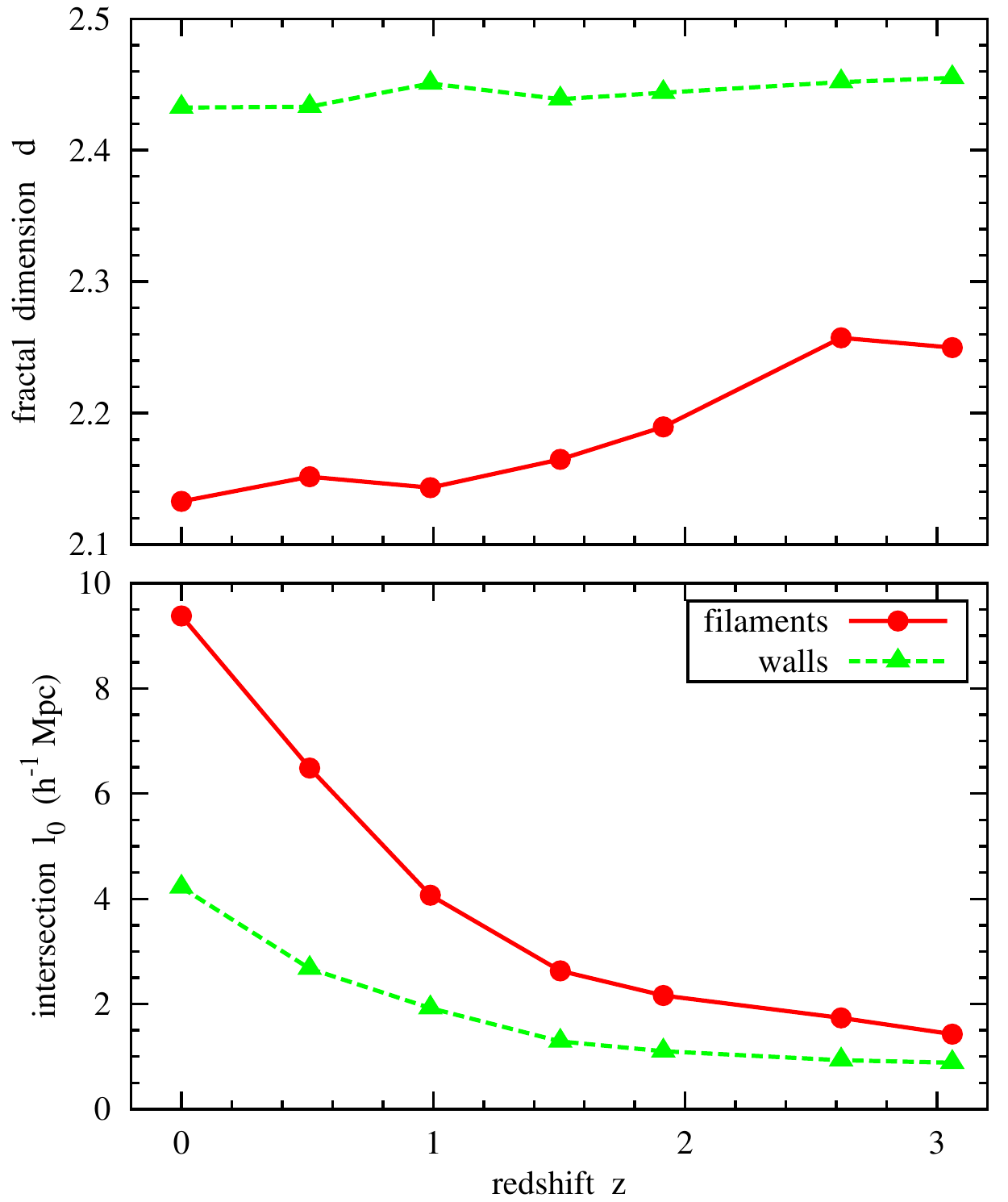}
    \caption{ \textit{Top panel:} The variation with redshift $z$ of the fractal dimension $d$ for filaments (solid curve) and walls (dashed curve). \textit{Bottom panel:} The variation with redshift of the breaking scale $l_0$ for both filaments and walls. The breaking scale gives the intersection between the solid and dashed curves from the left panel of \reffig{fig:fractal_dim_test}. }
    \label{fig:fractal_dim}
\end{figure}

The box count measurements for the present day filaments are shown in the top frame of \reffig{fig:fractal_dim_test}. The figure shows an interesting two regime dependence on the value of the box length $l$. For large boxes, i.e. coarse grids, every box intersects at least one filament segment. It means that filaments fully fill the simulation volume at these scales resulting in an $N\propto l^{-3}$ behaviour shown by the solid curve. For smaller $l$ values, only a fraction of the measuring boxes intersect the filamentary regions, which gives rise to the dependence illustrated via the dashed line. The fact that all the data points for $l\lsim10\Mpch$ lie along the dashed line shows the fractal-like behaviour of the filamentary network below this threshold value.

According to the points we just argued above, the box count $N$ can be expressed as a two component dependence on scale $l$ via:
\begin{equation}
	N = 
	 \begin{cases}
        c_1\; l^{-d} & \rmn{for}\; l\le l_0  \\
        c_2\;l^{-3} & \rmn{for}\; l> l_0
    \end{cases} \;,
    \label{eq:evolution_fractal_dimension}
\end{equation}
where the breaking scale $l_0$ denotes the threshold that differentiates between the two behaviours. The quantities $c_1$ and $c_2$ denote two normalization constants that can be easily computed noticing that: the two components need to be equal at $l=l_0$ and that $N(l=L_\rmn{box})=1$, where $L_\rmn{box}=100\Mpch$ is the side length of the \MII{} volume. Therefore, the two constants can be expressed as:
\begin{equation}
	c_1 = \left( \frac{l_0}{L_\rmn{box}} \right)^{d-3} 
	\quad\quad \rmn{ and } \quad\quad
	c_2 = L_\rmn{box}^{3} \;.
\end{equation}
To find the fractal dimension for filaments and walls, we fit the two component form presented in \eq{eq:evolution_fractal_dimension}, using two free parameters: $d$ and $l_0$. We find that this simple function gives a very good description of the box count data not only for $z=0$, but also at high redshift. This is presented in the bottom panel of \reffig{fig:fractal_dim_test}, which shows the data and the fit function for filaments detected at three different redshifts. For each snapshot we find the same qualitative behaviour, but quantitatively both the fractal dimension $d$ and the breaking scale $l_0$ depend on redshift. A similar result, though not shown, holds true for the wall network too.

\Reffig{fig:fractal_dim} shows the time evolution of the fractal dimension $d$ and that of the breaking length $l_0$ for both filaments and walls. Filaments are characterized by $d\sim2.2$ which shows that they have a fractal dimension higher than that of a thin plane. This is puzzling at a first sight, given that the filamentary network is made of many line-like objects and therefore we would expect $d<2$. This is not the case since filaments are not infinitely thin lines, but they do have an intrinsic width and hence we can have $d>2$. The fractal dimension of filamentary environments shows a strong time evolution, with larger values at high redshift. It shows the decrease in complexity of filaments, with lower values suggesting a simpler network with fewer branches, in agreement with the visual impression given by \reffig{fig:env_fila_evolution}. In the case of walls, we find only a very weak time evolution of the fractal dimension, with slightly lower values at present day. While the variation is small, it also shows the decrease in complexity of the wall network at later times. 

The variation with time of the breaking length can offer some interesting insights too. The length $l_0$ gives the scale beyond which all counting cells are occupied by filaments or sheets. The point where this behaviour is reached gives a characteristic scale that is related to the properties of the pattern under study \citep{Jones1988}. The breaking length is presumably related to the typical separation between filaments and walls or to the clustering scale of these structures. The breaking length shows a rapid increase since high redshift, suggesting a larger separation between present day filaments and sheets. Compared to high redshift when we hardly find large regions without filaments or walls, at present there are many large contiguous volumes empty of such structures.

\section{Segmenting the filamentary network}
\label{sec:evolution:filament_segmentation}

Up to now, our analysis focused on the evolution of the filamentary and wall networks as a whole. While this approach gives numerous insights into the global evolution of environments, it cannot fully characterize the growth of individual objects that, through their connectivity, form the global networks. To study individual structures, we introduce a method that splits the filamentary network into distinct branches. The segmentation takes place at the points where two or more filaments intersect and therefore offers a natural way of dividing the web into individual objects. While the segmentation method can be applied to sheets too, we restrict our analysis to filamentary environments since, together with clusters, they are the more prominent features of the cosmic web.

Following the application of \nexus{} and other Hessian based methods \citep[e.g.][]{Aragon07b,Hahn2007a}, the cosmic web components are identified as a set of points, distributed on a regular grid. Given the large scale coherence of the web, such voxels connect to each other to give rise to intricate patterns that pervade most of the cosmic volume. The resulting filamentary network has a complex structure, connecting many objects of various shapes and sizes. While identifying the distinct branches of the network comes naturally to the human brain, the same task is very challenging to implement via the use of computer algorithms. The problem is made even more difficult due to the hierarchical and multiscale character of the cosmic web, since the branches have a large variety of widths and lengths and intersect together at a wide spectrum of angles.

We are not the first to deal with the segmentation of the filamentary network into individual objects. Several previous studies had tackled this challenging issue using different methods, with various degrees of success. The most popular method uses cluster mass haloes to naturally split the filamentary network into individual objects. In practice, this approach has been implemented slightly different by identifying only filaments found between pairs of clusters \citep{Pimbblet04,Colberg05,Gonzalez09,Noh2011}. By design, such a procedure is biased towards the detection of the most prominent filaments and does not allow for a complete description of both thin and thick structures. Another technique makes use of the percolation of filaments as a function of density threshold to split the network into individual objects using the percolation threshold \citep{Shandarin04,2010MNRAS.408.2163A}. While it produces distinct structures, there is no clear one-to-one connection between these objects and the branches of the network. A different approach involves the use of minimal spanning tree algorithms for the segmentation of the filament network into main and secondary branches \citep{Colberg07,Park2009}. It gives a very good separation of the global network into distinct objects, but it can also introduce unwanted artefacts \citep[for details see][]{Colberg07}. And last, the change in orientation between two close points along the filament can be used to probe the intersection of distinct filaments \citep{Bond10}. This formalism is motivated by the picture of filaments as 1D continuous strands that end abruptly at the intersection of two or more such objects \citep{Bond1996}. We choose this last approach motivated by the clear and intuitive way of splitting the filamentary network into individual branches.

In the remaining of this section, we use a simple test configuration to introduce the filament segmentation method and discuss its characteristics. We further test the splitting technique by applying it to Voronoi clustering models, which represent a simple procedure of generating large-scale structures similar to those found in the cosmic web. We end by applying the method to the filamentary network identified in the \MI{} and \MII{} data, which, compared to the previous test cases, involves additional layers of complexity due to the hierarchical and multiscale nature of the cosmic web.

\subsection{The segmentation procedure}
\label{subsec:evolution:filament_segmentation_toy_model}

\begin{figure}
    \centering
    \includegraphics[width=1.05\linewidth]{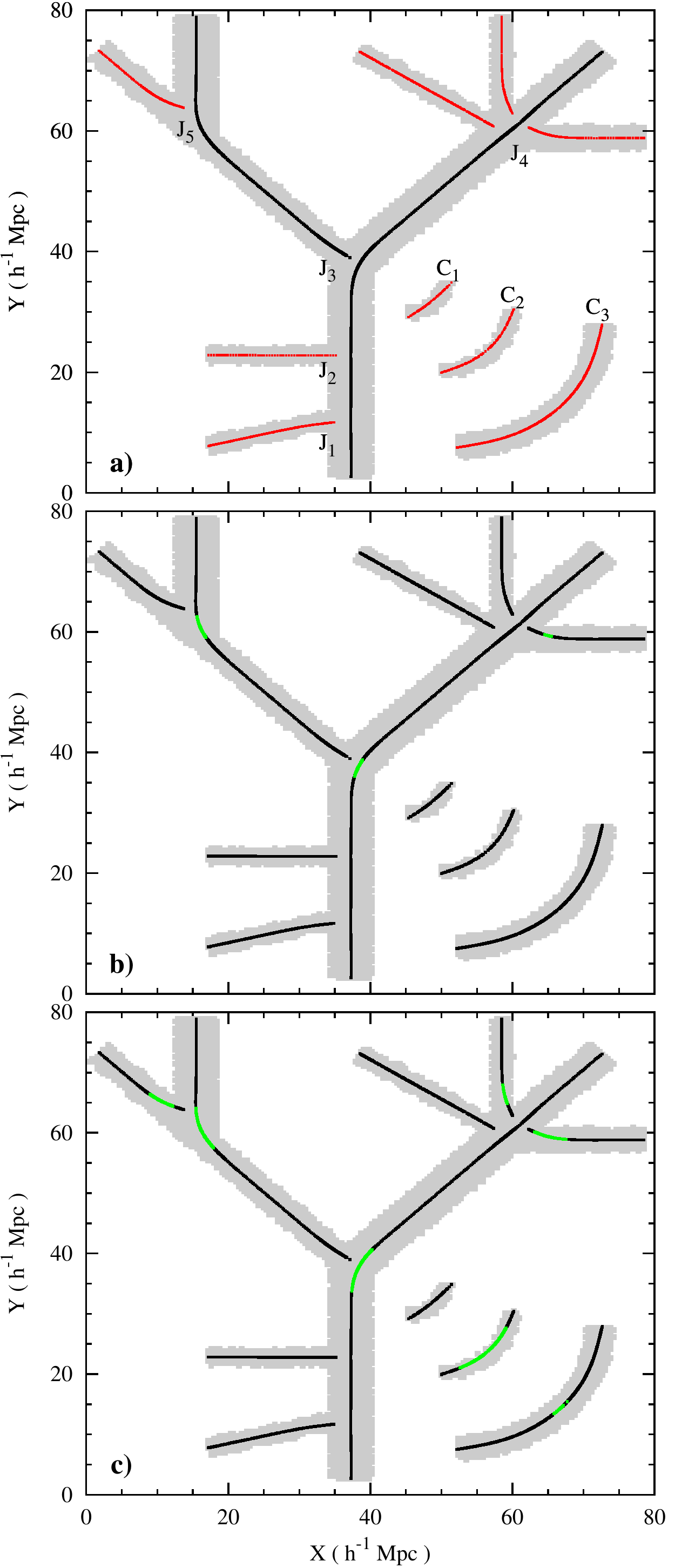}
    \caption{ An illustration, using a simple test configuration, of the filament network segmentation into individual branches. The light grey areas show the filaments, with the black and red curves showing the spine of the objects. The green curves in the bottom frames show the regions with $\meanTheta>2.5^\circ\MpchInverse$ (panel b) and  $\meanTheta>1.5^\circ\MpchInverse$ (panel c).  }
    \label{fig:evolution:split_toy_example}
\end{figure}

We illustrate the segmentation method using the simple filamentary configuration shown in the top frame of \reffig{fig:evolution:split_toy_example}. The filamentary network is shown via the light grey colour and, for visualization purposes, is fully confined to the plane of the figure. The configuration is composed of a few representative types of filamentary intersections. The $J_1$ and $J_2$ junctions are the most common and show the intersection of thin structures with more prominent filaments. The $J_3$ and $J_4$ points show the bifurcation of filaments into two or more branches, with the latter case especially common at the nodes of the cosmic web. The fifth junction, $J_5$ shows the intersection of a thick object with the middle of a much thinner structure. Though such a crossing is very rare or even completely absent in the cosmic web, we included it here to illustrate some of the limitations of the method. Additionally, the test configuration has three curved filaments $C_i$, with constant curvature radii of $5$, $10$ and $20\Mpch$. These curved structures are used to exemplify that the splitting procedure can also deal with non-straight filaments.

The first step in the filament segmentation process involves the compression of the filaments to their spine, via the technique described in \refsec{subsec:evolution:fila_contraction_description} and \refappendix{appendix:evolution:filament_contraction}. This contracts the objects to a single curve that corresponds to their central axis, as clearly seen in \reffig{fig:evolution:split_toy_example}. The compression procedure not only identifies the filament spine, but it also splits the network at its bifurcation points. Every junction is divided into a main continuous branch (black curves) and one or more secondary ones (red curves). In some of the cases (i.e. $J_3$ and $J_5$), the main branch shows a change of orientation around the region of the junction suggesting that this single object corresponds in fact to two or more branches. Therefore, the filament compression step gives only a partial segmentation of the filamentary network.

\begin{figure}
    \centering
    \includegraphics[width=\linewidth]{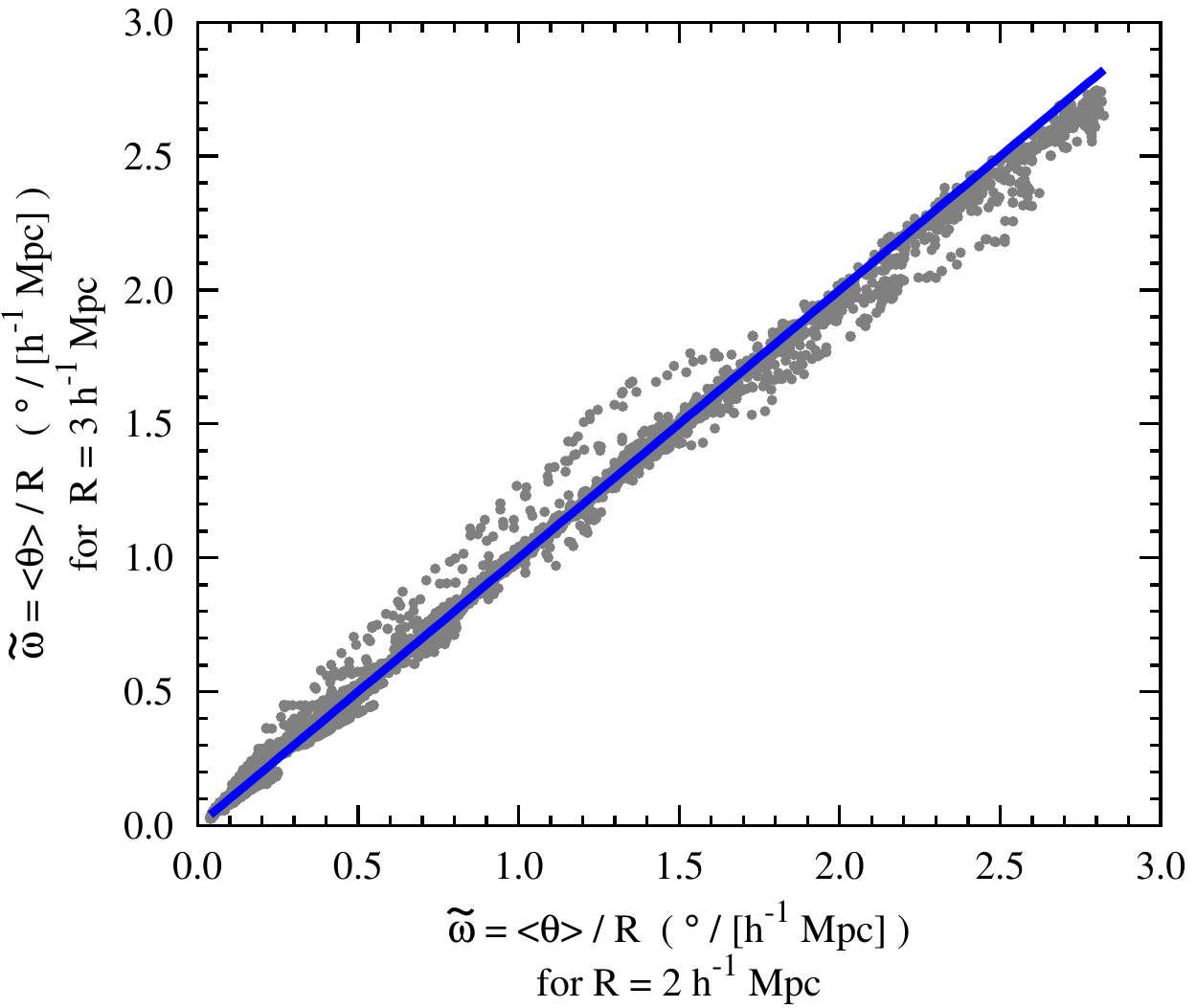}
    \caption{ The mean change in filament orientation $\mean{\theta}$ computed at two different radii $R$. The scaled ratio $\meanTheta$ is largely insensitive to the value of $R$. The solid diagonal line shows a one-to-one relation. }
    \label{fig:evolution:split_toy_example_R1_R2}
\end{figure}

\begin{figure*}
    \centering
    \includegraphics[width=0.65\linewidth]{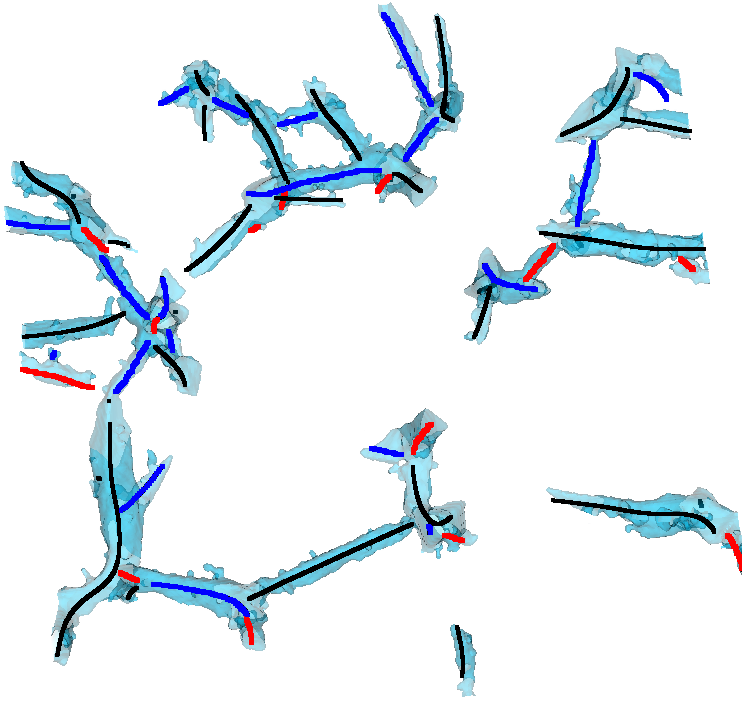}
     \caption{The filaments and their central spine in a $10\Mpch$ slice of the Voronoi clustering model. The spine is shown in various colours to better emphasize the different individual filaments. }
    \label{fig:evolution:split_voronoi}
\end{figure*} 

The second step of the segmentation procedure involves further dividing the main branches that show signs of being two or more distinct objects. Such branches have a rapid change in orientation around the junction points where the branch needs to be further split (i.e. $J_3$ and $J_5$). To characterize this, we take two points $i$ and $j$ along the branch and compute the change in filament orientation between those points as
\begin{equation}
    \theta_{ij} = \arccos ( \Vector{u_i} \cdot \Vector{u_j} )
    \label{eq:evolution:change_in_angle} \;,
\end{equation}
where $\Vector{u_{i,j}}$ denotes the filament orientation at those two points (see \refappendix{appendix:evolution:filament_direction} for details on computing the local filament orientation). We are interested in the mean change of orientation $\mean{\theta_i}$ around point $i$, which is given by
\begin{equation}
    \mean{\theta_i} = \sum_{i=1}^{N_i} \frac{ \theta_{ij} } {N_i}
    \label{eq:evolution:mean_angle} \;.
\end{equation}
The sum is over all the $N_i$ filament points found within distance $R$ from point $i$. While the value of $\mean{\theta}$ depends on the radius $R$ used to find neighbours, the scaled quantity 
\begin{equation}
    \meanTheta = \frac{ \mean{\theta} } {R}
    \label{eq:evolution:scaled_angle}
\end{equation}
is largely independent of $R$. This is shown in \reffig{fig:evolution:split_toy_example_R1_R2} which shows the $\meanTheta$ values obtained for two different distances $R$. Therefore, the segmentation algorithm is rather insensitive to the actual value of the $R$ parameter. For the rest of this study we take $R=2\Mpch$.

The bottom panels of \reffig{fig:evolution:split_toy_example} evaluates how effective is a $\meanTheta$ threshold in detecting the additional segmentation points along the main branches. The two graphs show the spine segments which have $\meanTheta>2.5^\circ\MpchInverse$ (panel b) and  $\meanTheta>1.5^\circ\MpchInverse$ (panel c). As expected, the highest values of $\meanTheta$ correspond to the junction points where the main branches change direction. Therefore, a branching threshold value $\meanTheta_\rmn{T}$ is an effective way of fully segmenting the filamentary network into distinct branches.

Care needs to be taken when choosing the branching threshold value, since a too low value leads to the artificial segmentation of curved branches. This is the case for the bottom-right panel, in which the low threshold value divides the $C_2$ and $C_3$ structures as well as some of the other secondary branches. While the $\meanTheta$ threshold values used in \reffig{fig:evolution:split_toy_example} are heuristically determined, \refsec{subsec:evolution:filament_segmentation_voronoi} and \refsec{subsec:evolution:filament_segmentation_filaments} present a quantitative way of identifying $\meanTheta_\rmn{T}$. It gives very good results for realistic distributions of filaments, like those found in the cosmic web.

Before proceeding to the analysis of more realistic filament distributions, we discuss some of the limitations of the segmentation technique.  Following filament compression, some of the secondary branches have curved ends at the junction points (e.g. $J_4$, $J_5$). These curved ends can have large $\meanTheta$ values and can be interpreted as indicating the junction point of two branches. Such an example is visible in the top-right region of panel c) where it gives rise to a spurious short branch. Such issues are rare and bring small artefacts only when measuring the properties of very short filaments. The properties of such short objects are anyway unreliable since their length is the same order of magnitude as the spacing of the underlying grid used to identify these structures.


\begin{figure*}
    \centering
    \includegraphics[width=0.65\linewidth]{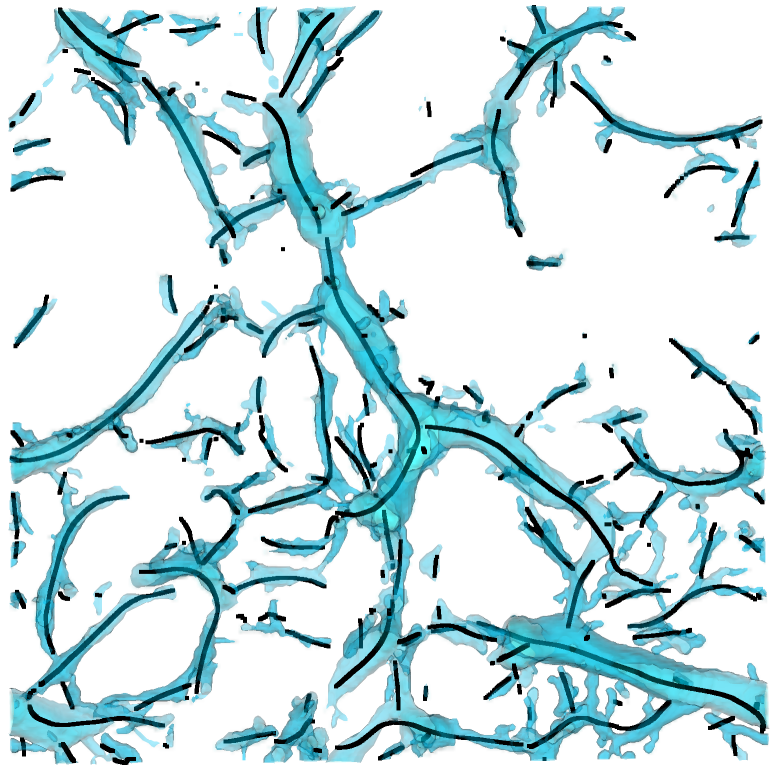}
    \caption{ A $5\Mpch$ slice showing the filaments and their central spine as identified in \MII{}. }
    \label{fig:evolution:m2_fila_split_Nbody}
\end{figure*}

Another limitation of the procedure is illustrated at junction $J_5$ in panel a) of \reffig{fig:evolution:split_toy_example}. In this case, a thick filament intersects the middle of a thinner one. The filament compression procedure fails, since it connects the thick structure with half of the thin filament, while the correct outcome is to connect the two halves of the thin filament. This leads to the segmentation of the thin structure into two objects. The limitation arises since the filament compression algorithm always selects as the main branch the two thickest filaments that enter the bifurcation, without regard to their orientation. In reality, configurations as those shown at junction $J_5$ are very rare or non existent in the distribution of cosmic web filaments, and therefore this drawback does not play a prominent role in our study.

The most significant limitation of the method affects the short and highly curved filaments, which are typically found in void regions. We illustrated this via the curved filament $C_1$ in \reffig{fig:evolution:split_toy_example}. Following the compression procedure, the computed spine is considerably straighter than the input object. It is a consequence of the $1\Mpch$ smoothing radius used by the compression algorithm, whose value is similar to the curvature radius of filament $C_1$. As a result of this, all highly curved structures are being significantly straightened. In contrast, filaments with curvature radii ${\gsim}10\Mpch$ are hardly affected, as shown by filament $C_2$. 

\subsection{Segmenting Voronoi clustering models}
\label{subsec:evolution:filament_segmentation_voronoi}

\begin{figure}
    \centering
   \includegraphics[width=\linewidth]{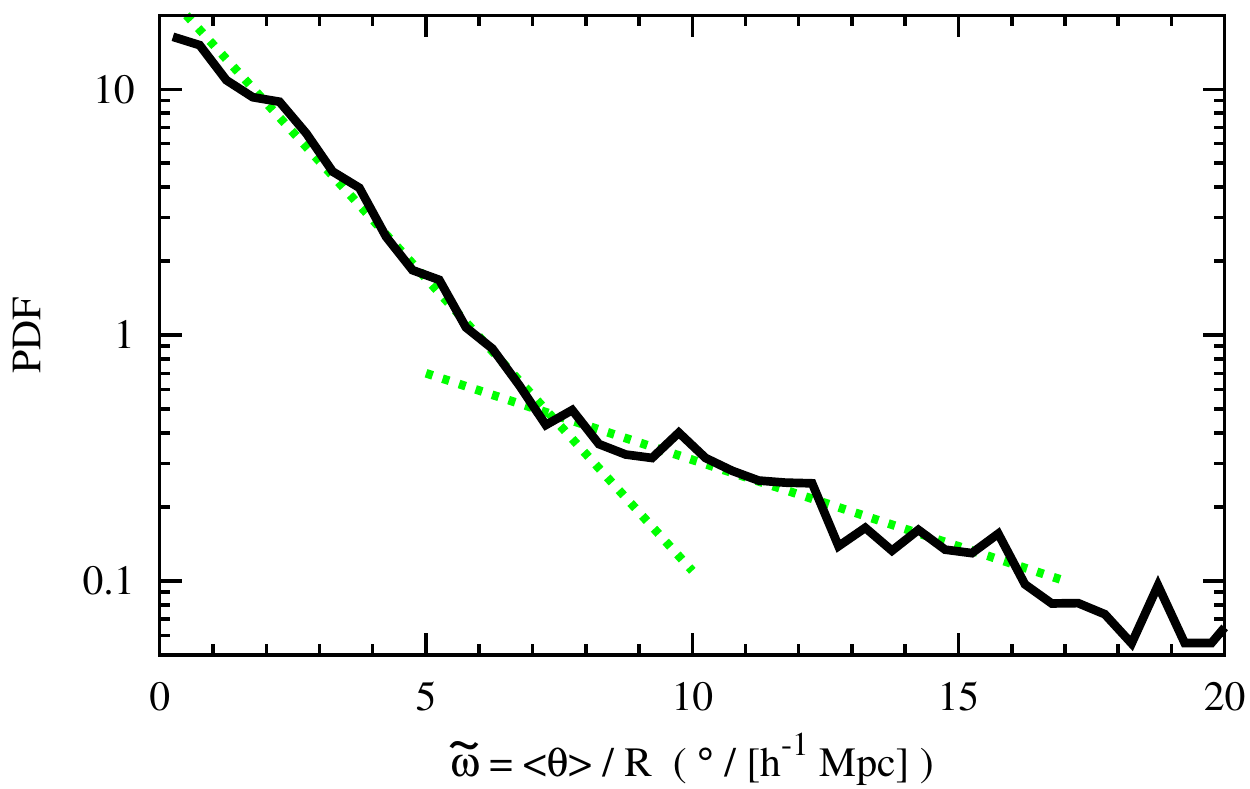}
    \caption{ The PDF of $\meanTheta$ for the Voronoi clustering model (solid curve). The dotted curves are only for illustrative purposes to show that the PDF changes its behaviour at $\meanTheta \sim 7.5^\circ\MpchInverse$. Values above this threshold correspond to filamentary intersections, as confirmed by a visual inspection of the filamentary network. }
   \label{fig:evolution:split_voronoi_pdf}
    \vskip 0.5truecm
    \includegraphics[width=\linewidth]{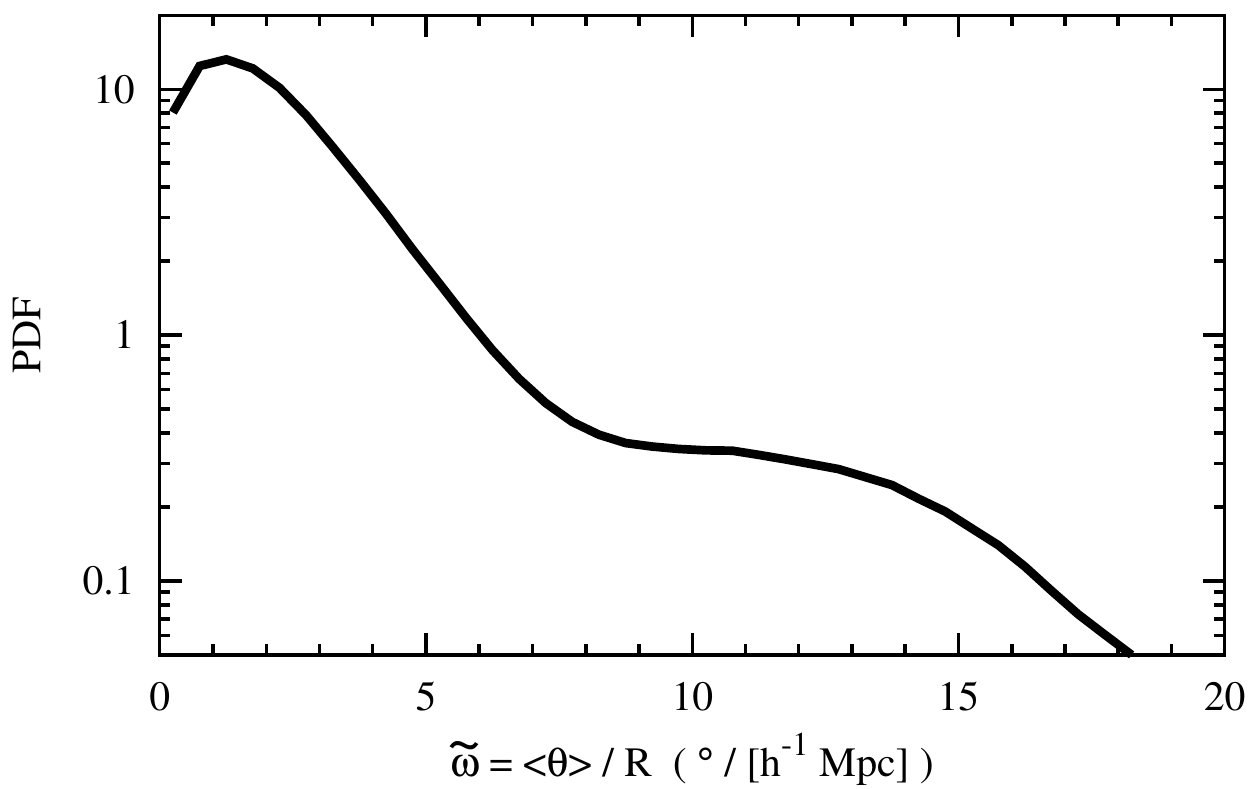}
    \caption{ The PDF of $\meanTheta$ for filamentary structures detected in the \MI{} simulation. }
    \label{fig:evolution:split_angle_pdf}
\end{figure}

While the previous test configuration was very useful in exemplifying the inner workings of the segmentation procedure, it offers at best a poor description of the real filamentary network. Consequently, we need to test the segmentation approach on a more realistic distribution. The filaments obtained within the framework of Voronoi clustering models represent one such network \citep{vandeWeygaert1991a,vandeWeygaert2002,2009LNP...665..291V}. Such distributions share many of the characteristics of cosmic web filaments, since they describe complex 3D networks composed from objects of various sizes and with multiple branches intersecting at the same location. On the other hand, filaments in these heuristic models do not follow a hierarchical and multiscale distribution and also do not contain any curved structures. Therefore, Voronoi models represent a configuration of intermediate complexity, much more elaborate than the simple test configuration used before, but still without exhibiting all the challenges of realistic filamentary networks.

The Voronoi clustering models are a class of heuristic models used to describe the cellular-like pattern of the large scale distribution of galaxies \citep{vandeWeygaert1991a,vandeWeygaert2002,2009LNP...665..291V}. Such an approach uses the Voronoi tessellation as the skeleton of the large scale matter distribution, with Voronoi cells corresponding to cosmic voids, while the cell walls represent cosmic sheets. Moreover, the edges of the Voronoi walls exemplify filaments, with the intersection of these edges corresponding to the cosmic web nodes. 

The simple Voronoi models, which we employ here, confine the distribution of galaxies or DM particles to one of the four components of the tessellation discussed above. This proceeds by first classifying the particles into node, filament, sheet and void types according to a heuristic assignment scheme which is of no importance to our study. Starting from a spatially random distribution, each particle is localized inside a Voronoi cell and moved to its final position according to the morphological tag of the particle. Void objects are kept at their initial random positions, while sheet, filament and node particles are moved to the closest tessellation wall, edge or vertex, respectively \citep[for a more detailed description see][]{vandeWeygaert1991a}. It results in a matter distribution following a cellular-like pattern, not very different to the large-scale structure of the Universe.

For the purpose of our study, we apply the \nexus{} technique to the resulting Voronoi matter distribution. While the Voronoi models can be used to identify cosmic web components, we choose to detect filaments using \nexus{}. This way, we make sure that any artefacts due to the identification procedure, if present, are included in the Voronoi filament detection, giving a closer representation of the real cosmic web. The results discussed below are obtained using a Voronoi model employing $128^3$ particles distributed among $64$ Voronoi cells generated in a $100\Mpch$ box.

\Reffig{fig:evolution:split_voronoi} shows the Voronoi filaments in a thin slice through the simulation box, along with their central spine. To better emphasize individual branches, we used different colours for the spine of various objects. A visual inspection finds that the $\meanTheta_\rmn{T}$ threshold employed in the figure performs very well in dividing the filamentary network into individual branches. Most of the branches, which are fully contained in figure, start and end at the intersection points of several filaments and are approximatively straight, as expected in the case of Voronoi model filaments.

\begin{figure}
    \centering
    \includegraphics[width=\linewidth]{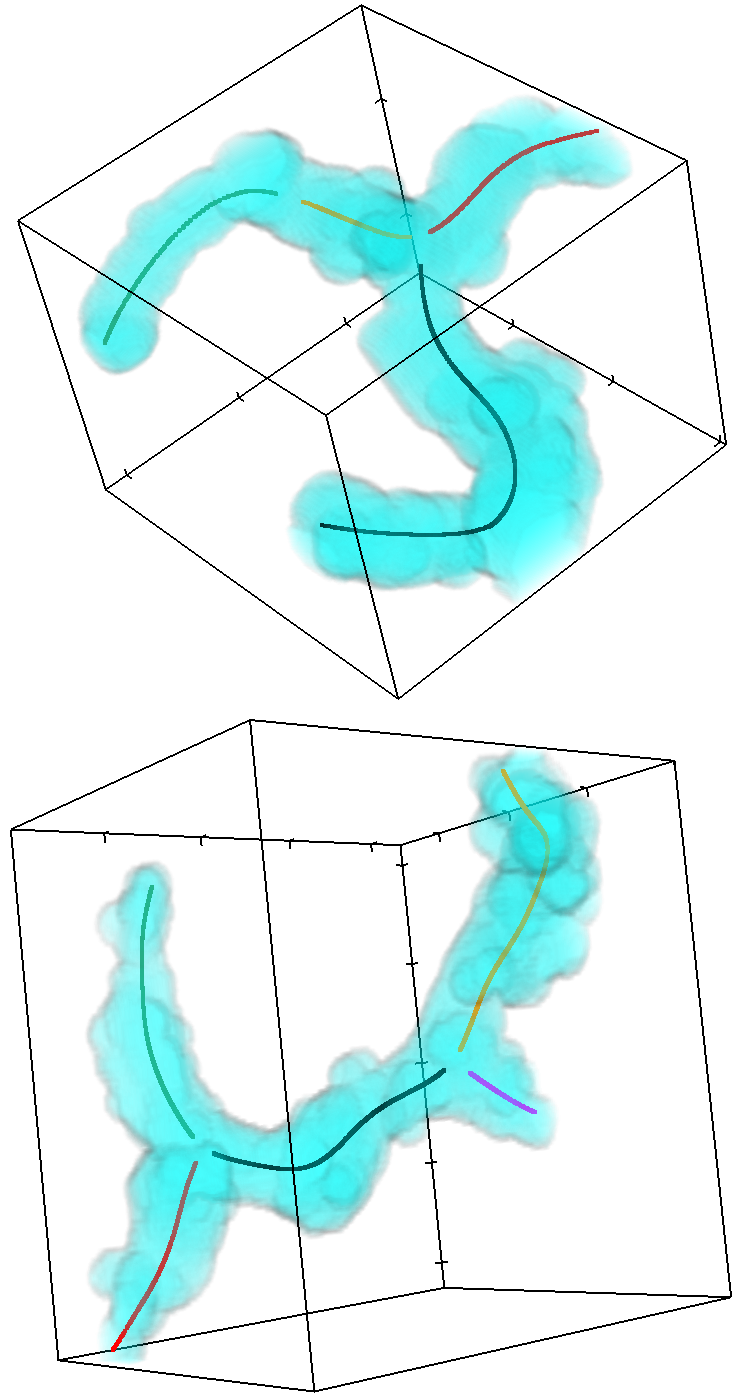}
    \caption{ Examples of filamentary structures that were identified as single objects following the compression process. These structures were further segmented into individual branches using the segmentation procedure described in \refsec{subsec:evolution:filament_segmentation_toy_model}. We show the spine of these branches, using a different colour for each distinct object. }
    \label{fig:evolution:m2_fila_split}
\end{figure}

To identify the optimal value for the branching threshold $\meanTheta_\rmn{T}$, we examined in \reffig{fig:evolution:split_voronoi_pdf} the PDF of $\meanTheta$ values. As expected, the PDF peaks at small $\meanTheta$ values indicating that most filamentary stretches are close to being perfectly straight. At higher $\meanTheta$ values, the PDF shows a two regime behaviour as highlighted via the two dotted lines, with a rapid decrease till $\meanTheta \sim 7.5^\circ\MpchInverse$, after which it changes to a slower decline. While within Voronoi models the filaments are perfectly straight, sampling such objects on a grid introduces discreteness effects which in some cases lead to slightly curved filaments. Such effects are typically small and unlikely to give rise to large curvatures, resulting in the rapidly decreasing regime seen in the $\meanTheta$ PDF. In contrast, the slowly declining region of the PDF corresponds to branching points and indicate the points where filaments need to be further divided. Therefore, the $\meanTheta$ value where the PDF changes behaviour represents a natural choice for the branching threshold $\meanTheta_\rmn{T}$. Such a threshold value leads to the individual filaments shown in \reffig{fig:evolution:split_voronoi}, providing a reliable way of segmenting the filament network into its distinct branches.  

\subsection{Segmenting the cosmic web filaments}
\label{subsec:evolution:filament_segmentation_filaments}

\begin{figure*}
   \centering
   \includegraphics[width=.95\linewidth]{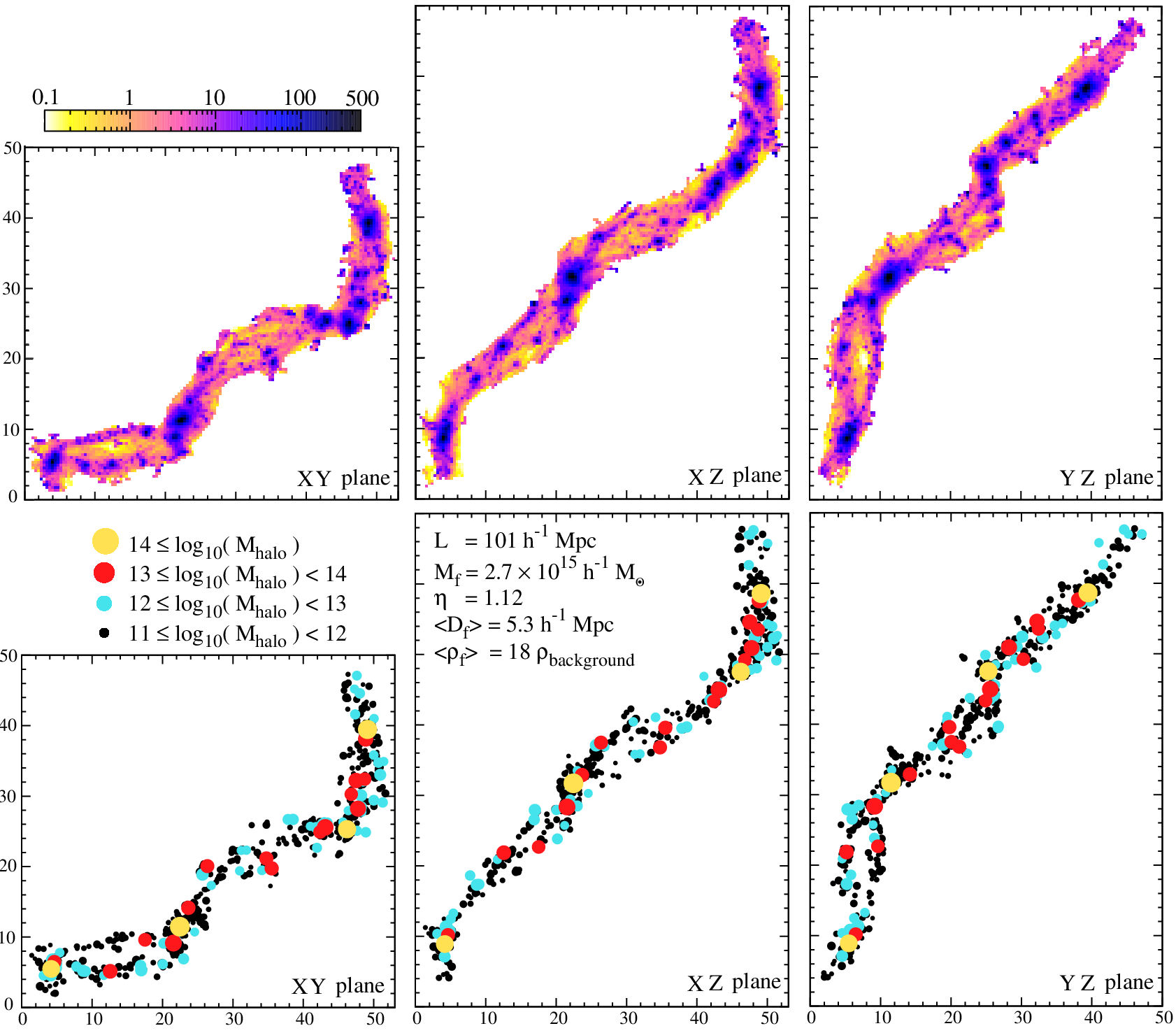}
   \caption{ Three orthogonal projections of a long and approximatively straight filament. It shows both the density field along the structure (top panels) as well as the spatial distribution of haloes more massive than $10^{11}\Msolar$. The points representing the haloes are coloured and sized according to the halo mass, as shown in the legend. The figure also gives some basic properties of the object: its length $L$, its mass $M_\rmn{f}$, the shape proxy $\eta$, the mean diameter $\mean{D_\rmn{f}}$ and the mean density $\mean{\rho_\rmn{f}}$.  }
   \label{fig:evolution:fila_example_1}
\end{figure*}

The cosmic web filaments contain several layers of additional complexity compared to the two examples discussed previously. Due to its hierarchical and multiscale character, the filamentary network is composed of structures of various lengths and widths that come together at a multitude of angles. Moreover, the naturally curved filaments found in the cosmic web pose an additional challenge since their curvature can be incorrectly interpreted as a sign of filament bifurcation. These aspects motivate us to further investigate the behaviour of the segmentation procedure when applied to realistic filamentary distributions.

We start with an illustration of the filament compression outcome in \reffig{fig:evolution:m2_fila_split_Nbody}. It exemplifies that the method performs very well in finding the spine of most objects, even in the case of realistic filament distributions. A closer visual inspection reveals that there are some structures which show differences between their contours and the spine, with such cases typically limited to thin filaments that are highly curved or in densely packed regions. While present, such artefacts hardly affect the population of thick and long objects, which represent the main focus of our study.

To find the branching threshold, we investigate in \reffig{fig:evolution:split_angle_pdf} the PDF of $\meanTheta$. Similarly to \reffig{fig:evolution:split_voronoi_pdf}, the PDF shows a two regime behaviour, with the transition between the two regions taking place at $\meanTheta = 7.5^\circ\MpchInverse$. This gives the value of the branching threshold $\meanTheta_\rmn{T}$. We further explore in \reffig{fig:evolution:m2_fila_split} how effective such a threshold value is, by examining structures that were identified as single objects following the compression process. A visual inspection shows that the chosen $\meanTheta_\rmn{T}$ threshold performs extremely well in dividing the network into individual branches.


\section{The evolution of individual filaments}
\label{sec:evolution:filaments}

\begin{figure*}
   \centering
   \includegraphics[width=.95\linewidth]{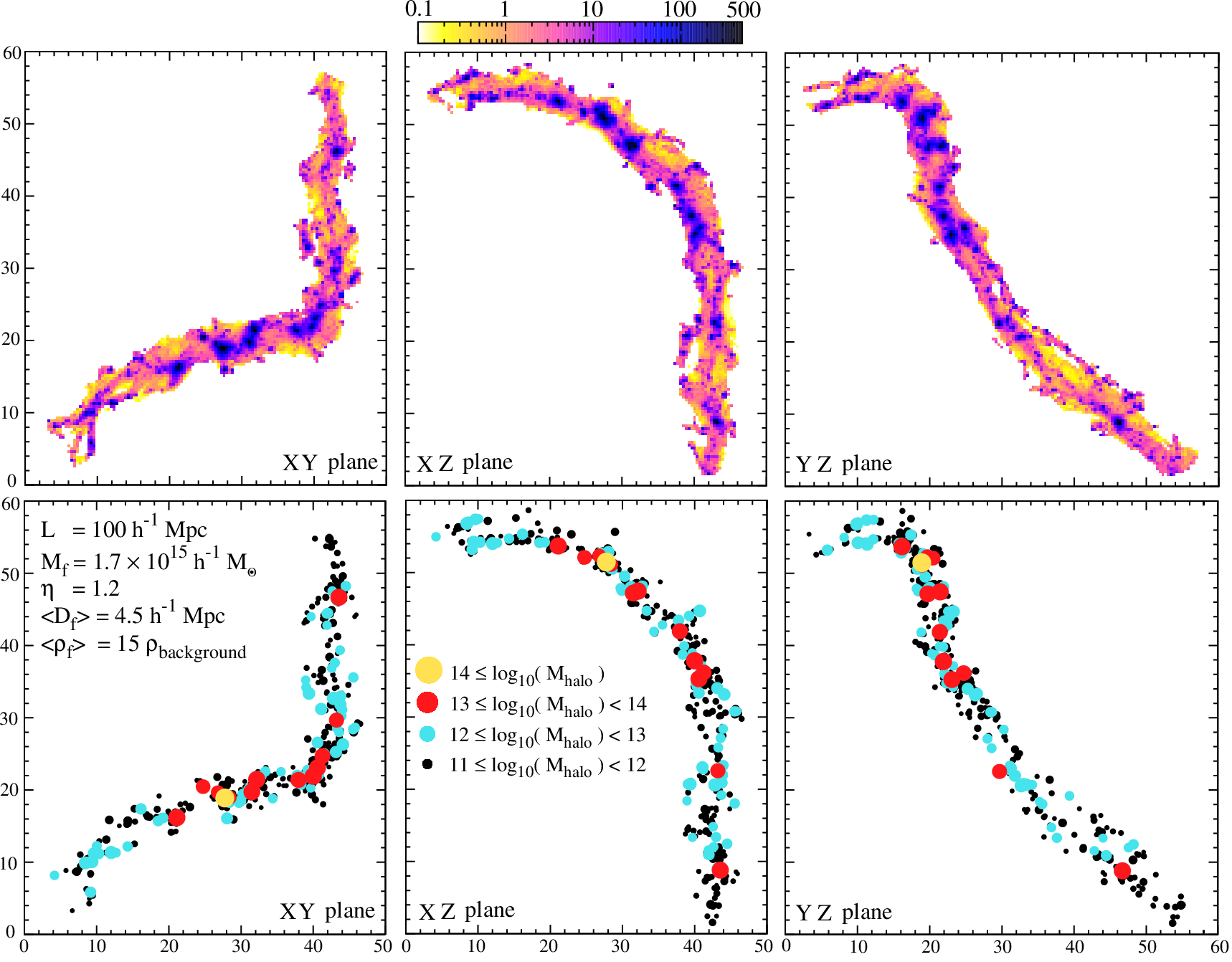}
   \caption{ Same as \reffig{fig:evolution:fila_example_1}, but showing a more curved filament. }
   \label{fig:evolution:fila_example_2}
\end{figure*}
\begin{figure*}
   \centering
   \includegraphics[width=1.\linewidth]{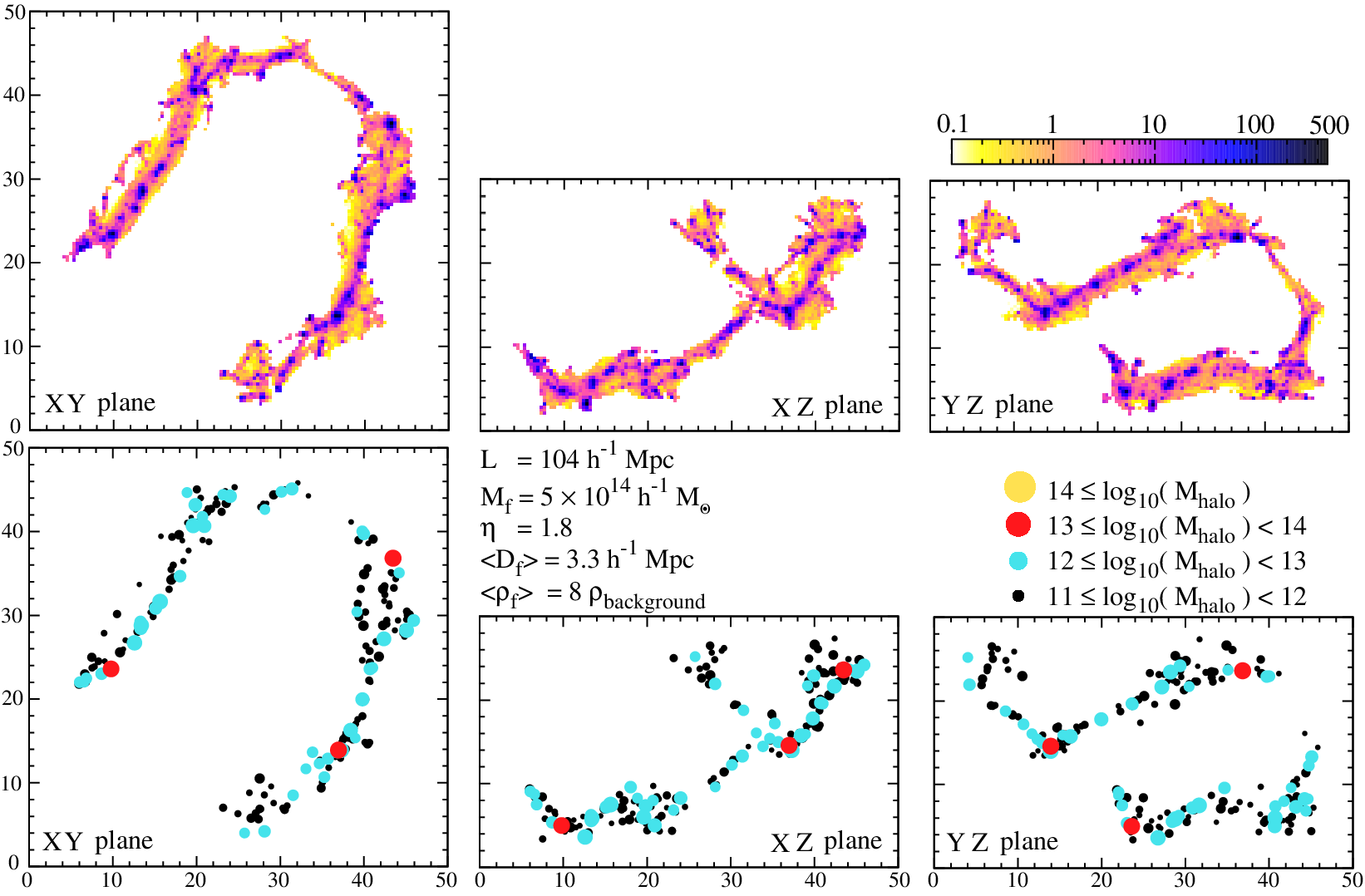}
   \caption{ Same as \reffig{fig:evolution:fila_example_1}, but showing the most curved filament with length above $100\Mpch$. In fact, this structure is composed of at least two branches that are connected by a thin and tenuous bridge. The segmentation algorithm wrongly labels these as a single filamentary branch. }
   \label{fig:evolution:fila_example_3}
\end{figure*}

Without a doubt, filaments represent the most salient features of the cosmic web, extending over tens of megaparsecs and incorporating the largest share of the matter content of the universe. Given the predominance of filamentary features, this section is focused on investigating basic properties of individual filaments such as the length and mass distribution, shapes and sizes. 

Before analysing the properties of the filament population, we present in \reffigs{fig:evolution:fila_example_1}{fig:evolution:fila_example_3} the structure and halo population of a few representative filaments. We focus these examples on the properties of very extended objects, since we have already shown in the previous sections plenty of typical, shorter, filaments. To fully appreciate the intricate structure of these objects, we show them from three different angles. These vantage points hint at the large variation in shape and size of these structures when seen in a 2D projection. It also illustrates that to fully identify such objects in observations, one needs the full 3D information, and especially accurate spectroscopic redshifts to obtain a good handle on the depth of these structures. 

The filamentary examples shown in \reffigs{fig:evolution:fila_example_1}{fig:evolution:fila_example_3} were chosen to have a large variety of shapes, starting from approximatively straight objects to very curved ones, to illustrate the close connection between the shape and the matter content of these objects. The structure shown in \reffig{fig:evolution:fila_example_1} is the quintessential filament which forms a prominent bridge between one or more pairs of clusters \MCn{\citep{Shandarin1984,Shandzeld89,Bond1996,Pimbblet04,Colberg05,WeyBond08a,2010MNRAS.408.2163A}}. In this case, it connects four massive clusters which are spatially distributed in a near straight line. In general, the filaments connecting pairs of clusters are the most prominent ones, having high densities and being tightly packed with haloes, especially massive ones. The closer the two clusters are, the more dense and packed with haloes the filament is. This can be easily appreciated in \reffigS{fig:evolution:env_example_haloes} and \ref{fig:evolution:fila_example_1}.

A second widely common category of filaments are those that have as one of their end points a cluster mass halo, while at the other end they branch out into underdense regions. Such an example is shown in \reffig{fig:evolution:fila_example_2}, where two such structures connect at the cluster endpoint to form a much longer filament. Compared to the previous example, such filaments have a lower mass and density and host smaller mass haloes. More importantly, these structures tend to be more curved, exhibiting more intricate shapes. 

The third class of filaments are those that do not connect directly to clusters, since they extend only between smaller mass haloes, similar to the structure shown in \reffig{fig:evolution:fila_example_3}. Such objects are typically found in lower density regions and are highly meandering, thin and only loosely populated with haloes. In fact, the filament shown in \reffig{fig:evolution:fila_example_3} has at least two main parts, which are connected by a very tenuous bridge. It suggests that this structure should probably be divided into two or more objects, and that it was misclassified by the segmentation procedure as being a single branch. Comparing the properties of this filament with the other two examples shows that this object is a clear outlier, having for its length a significantly lower mass, density and diameter. It suggests that simple criteria can be successfully used to reduce and even eliminate any misclassified filaments. This study does not employ such criteria since it would introduce a subjective bias on what is considered a proper filament. Moreover, such artefacts affect only a small fraction of the overall population of haloes and do not significantly change our results.  

The above discussion on the various types of filaments is very illustrative in the context of the cosmic web theory, which describes the formation and connectivity of large-scale structures \citep{Bond1996}. Filaments arise from quadrupolar mass distributions in the primordial fluctuation field, which evolve into the canonical cluster-filament-cluster configuration \MCn{\citep{Shandarin1984,Shandzeld89,Bond1996,WeyBond08a}}. This naturally explains the very close connection between filaments and clusters, with massive clusters indicating the presence of prominent filaments and vice versa.

\subsection{Spatial extent of filaments}
\label{subsec:evolution:filament_individual_extent}

The simplest way of characterizing individual filaments is by determining their properties, with length being one of the most basic such measurements. This plays an important role in the light of galaxy redshift results, which show that our Universe contains coherent linear structures on large spatial extents. Analyses of observational data have consistently found a wide range of filament lengths, from short structures with length of only ${\sim}5\Mpch$ to very long objects that extend above $100\Mpch$ \citep{Bharadwaj04,Pimbblet04,Pandey2011,Smith2012,Tempel2013a}.

\begin{figure}
   \centering
   \mbox{\hskip -0.65truecm\includegraphics[width=1.10\linewidth]{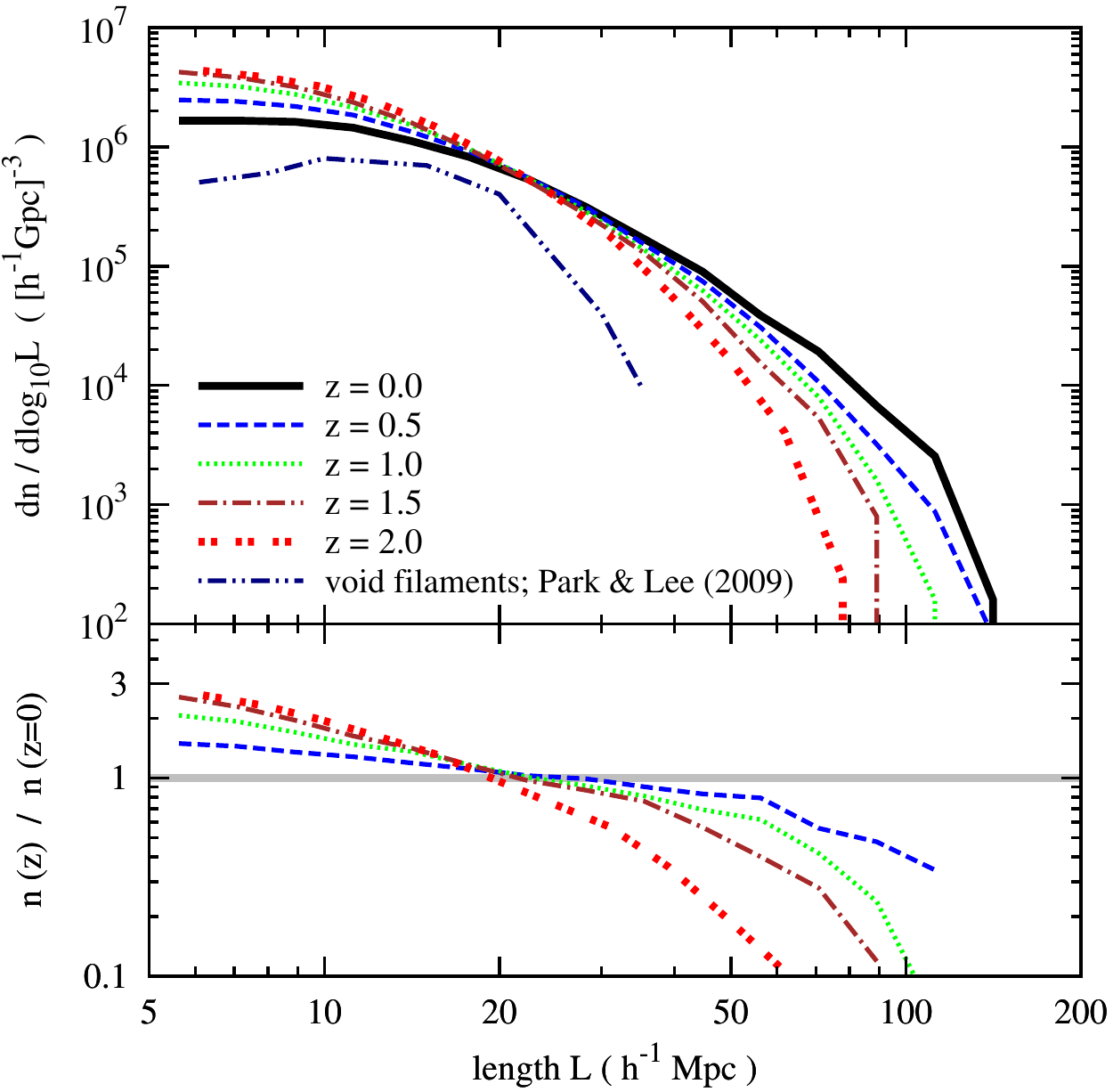}}
   \caption{ The number density of filaments $n$ as a function of filament extent $L$ and redshift $z$. For comparison, the figure also shows the previous results of \citet{Park2009} for the number density of filaments in voids at $z=0$. The lower panel shows the ratio of the filament number density at different times with respect to the results at $z=0$. }
   \label{fig:evolution_fila_length_distribution}
   \vskip 1.0truecm
   \centering
   \mbox{\hskip -0.65truecm\includegraphics[width=1.10\linewidth]{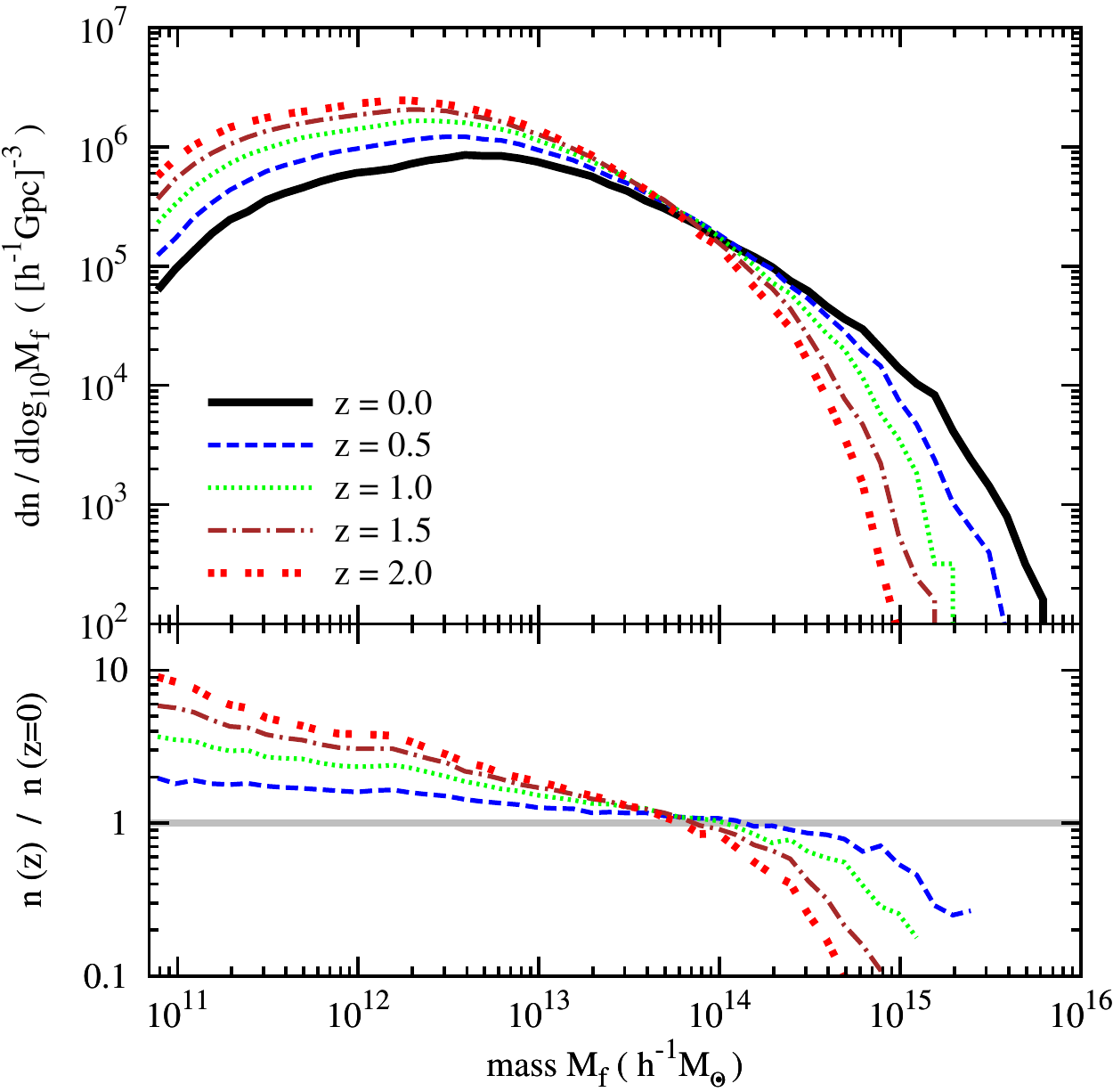}}
   \caption[]{ The filament mass function at different cosmic epochs. The quantity $M_\rmn{f}$ denotes the mass of an individual filament branch. The mass function is complete for $M_\rmn{f}>10^{13}\Msolar$. The lower panel shows the ratio of the filament mass function at different times with respect to the results at $z=0$. }
   \label{fig:evolution:fila_mass_function}
\end{figure}

Given the above findings, we explore the distribution of filament lengths in \reffig{fig:evolution_fila_length_distribution}, where we show the number of filaments of a given spatial extent. We find a large variation in filament size, from very short objects to structures extending above $100\Mpch$, in very good agreement with the observational results we just discussed. Short filaments are clearly more abundant than long ones, as expected in a hierarchical evolution scenario. The very extended objects, while very prominent in the distribution of haloes and galaxies, are few in number and account for only a very small fraction of the filament population \citep{Pimbblet04,Colberg07,2010MNRAS.408.2163A,Bond10,Gonzalez09,Tempel2013a}.

The \citet{Park2009} study offers an interesting comparison point since it analysed the number and properties of filaments found in void regions. As expected, given that they focused on void regions, \citet{Park2009} found a smaller number of filaments at all $L$ values. But more interestingly, the void filament number shows a much sharper decline for longer objects, which suggests, unsurprisingly, that extended filaments are mostly found in overdense regions and not in voids.

\begin{figure}
   \centering
   \includegraphics[width=\linewidth]{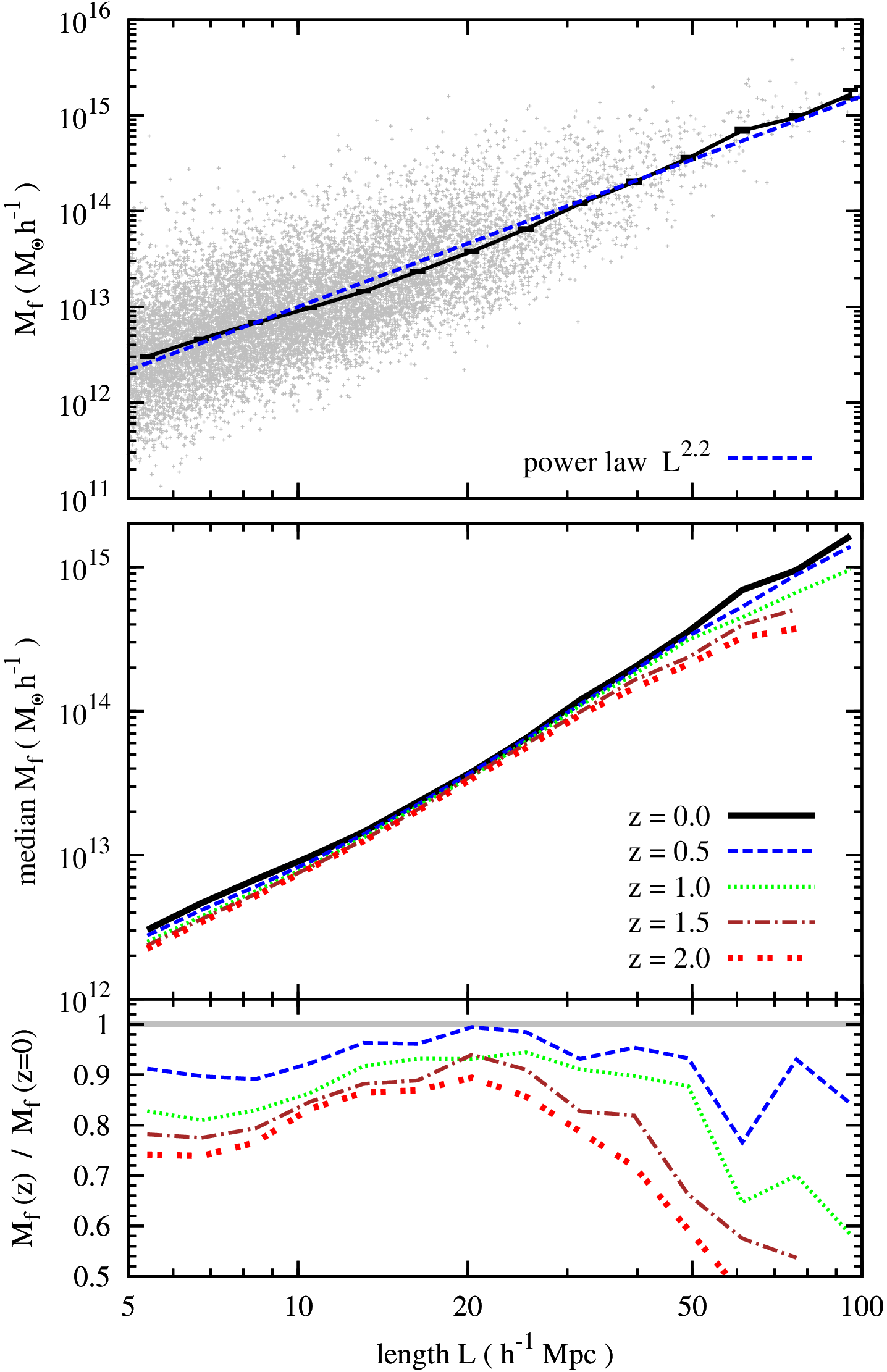}
   \caption{ \textit{Top panel:} The relation between the mass $M_\rmn{f}$ and the length $L$ of filaments at $z=0$. The solid curve shows the median mass at fixed filament length. The dashed curve shows a ${\sim}L^{2.2}$ power law dependence. The scatter plot shows only $10\%$ of the filaments, selected randomly.  \textit{Bottom panel:} The median value of the filament mass $M_\rmn{f}$ as a function of object length at different redshifts. The lower insert shows the ratio with respect to the $z=0$ results. }
   \label{fig:evolution:fila_length_mass}
\end{figure}

\begin{figure*}
   \centering
   \includegraphics[width=.96\linewidth]{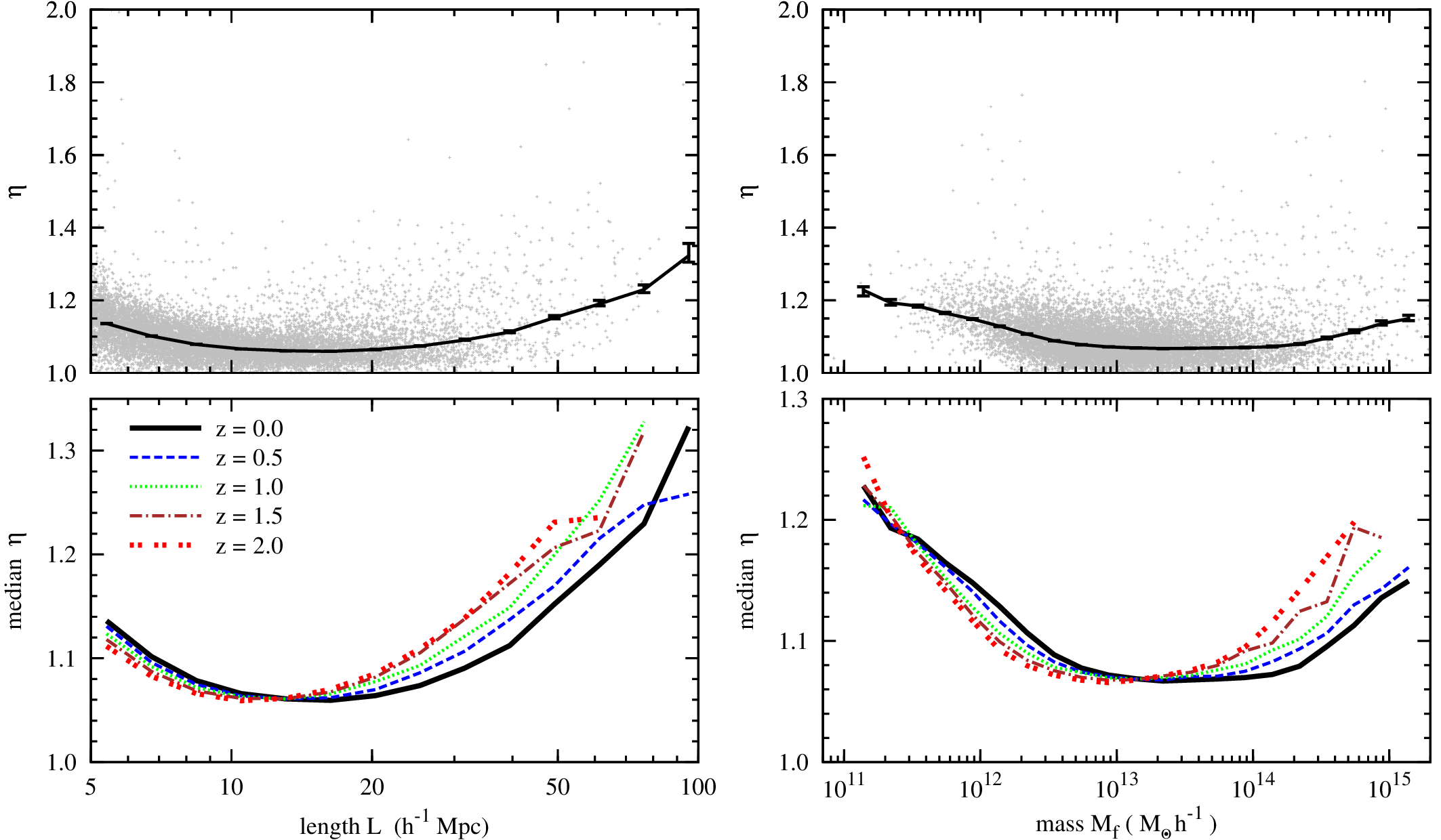}
   \caption{ \textit{Top panels:} The dependence of the filament shape, as characterized by $\eta$, on the length $L$ (left frame) and mass $M_\rmn{f}$ (right frame) of the filament. The solid curves indicate the median relation at fixed length or mass. The scatter plots show only a random $10\%$ subset of the filaments.  \textit{Bottom panels:} The variation of the median filament shape with redshift. } 
   \label{fig:evolution:fila_individual_shape}
\end{figure*}

Compared to present time, at high redshift we find significantly more short filaments and fewer extended ones. Most of the short filaments disappear by merging with other objects, to give rise to new, longer structures. This has been exemplified by \citet{Aragon07}, who showed how present day filaments started as a set of mini-filaments, orientated parallel to the final filament, which collapsed hierarchically to form the present day structure. It indicates the hierarchical evolution of the filament population towards ever more extended structures to the detriment of short objects. On the other hand, \citet{Park2009} found that the number and size of void filaments barely changes since $z=2$ up to present. Our results and that of \citet{Park2009}, when taken together, suggest that various filament populations evolve differently, with a small to no change in underdense regions and a much bigger evolution in overdense regions.

\begin{figure*}
   \centering
   \includegraphics[width=.97\linewidth]{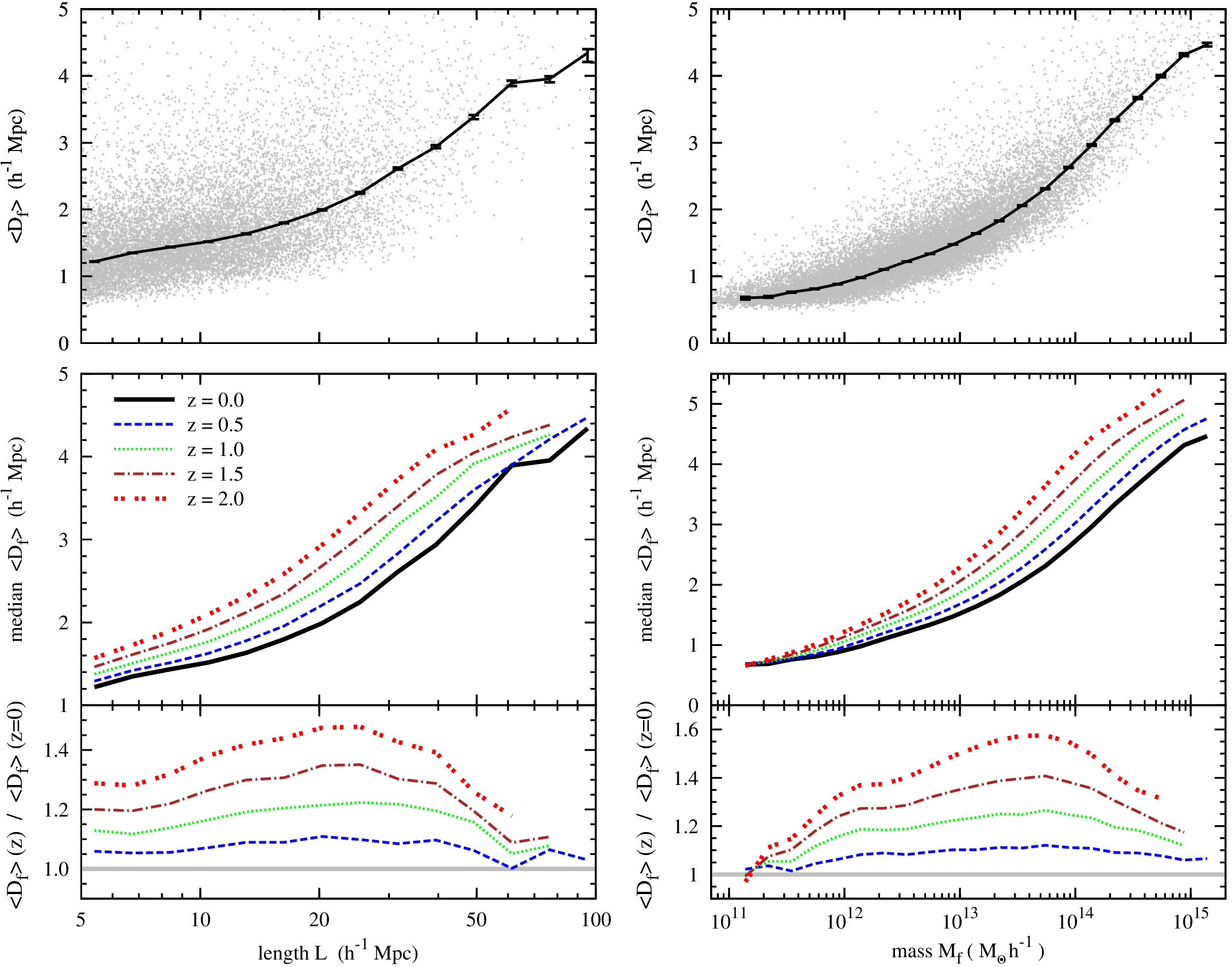}
   \caption{ \textit{Top panels:} The dependence of the mean filament diameter $\mean{D_\rmn{f}}$ on the length $L$ (left frame) and mass $M_\rmn{f}$ (right frame) of the filament. The solid curves indicate the median relation at fixed length or mass. The scatter plots show only a random $10\%$ subset of the filaments. \textit{Bottom panels:} The variation with redshift of the median relation found in the top panels. The bottom-most graphs show the ratio with respect to the $z=0$ result.  }
   \label{fig:evolution:fila_individual_diameter}
\end{figure*}
\begin{figure*}
   \centering
   \includegraphics[width=.98\linewidth]{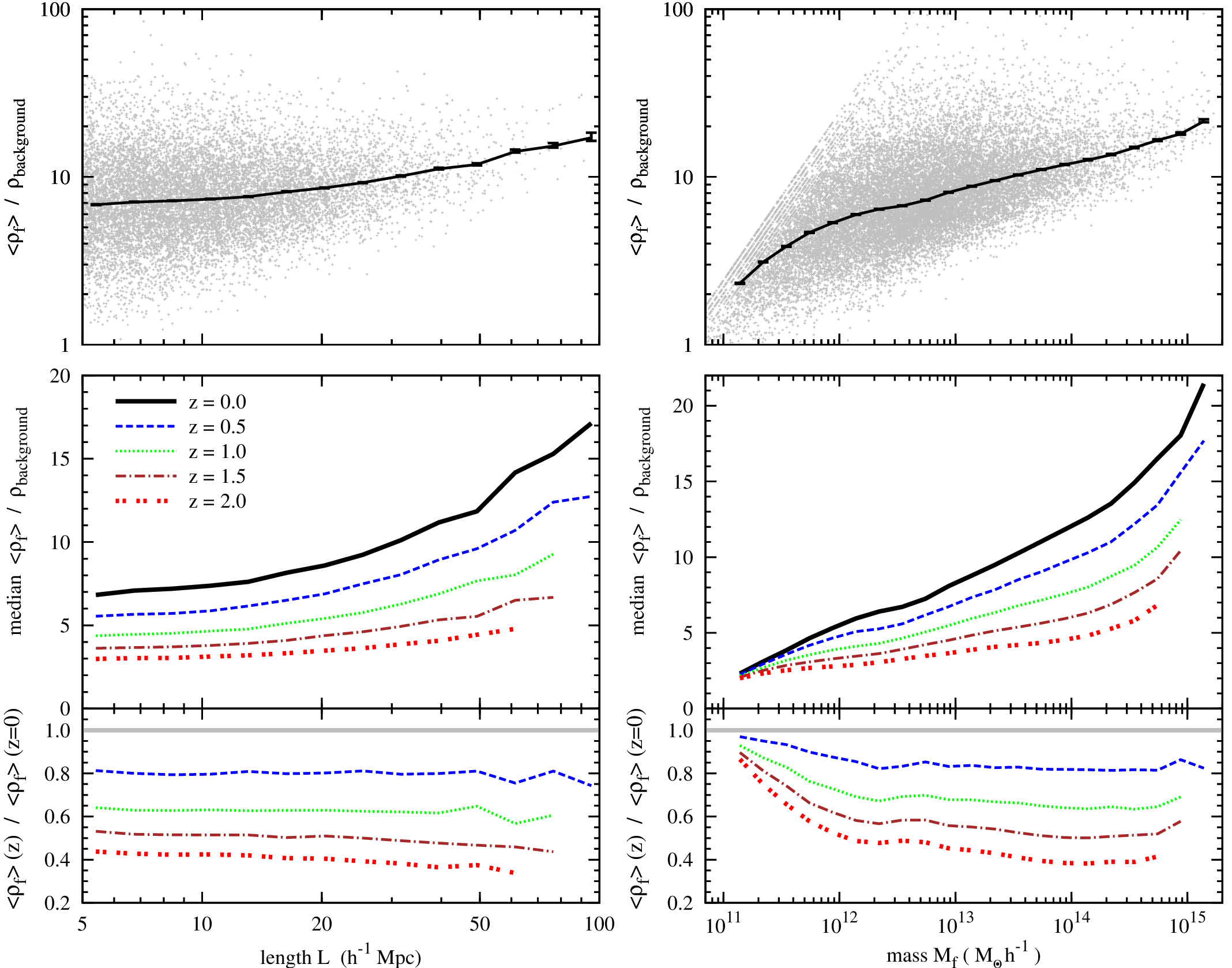}
   \caption{ \textit{Top panels:} The dependence of the mean filament density $\mean{\rho_\rmn{f}}$ on the length $L$ (left frame) and mass $M_\rmn{f}$ (right frame) of the filament. The solid curves indicate the median relation at fixed length or mass. The scatter plots show only a random $10\%$ subset of the filaments. \textit{Bottom panels:} The variation with redshift of the median relation found in the top panels. The bottom-most graphs show the ratio with respect to the $z=0$ result. }
   \label{fig:evolution:fila_individual_density}
\end{figure*}

\subsection{Filament mass function}
\label{subsec:evolution:filament_individual_mass_function}

The distribution of filamentary masses is investigated in \reffig{fig:evolution:fila_mass_function}, with the mass function complete for $M_\rmn{f}\gsim10^{13}\Msolar$. For lower masses, we only have a partial population of objects, since we miss out the short and dense filaments. This limitation comes from the filament segmentation algorithm, which due to its intrinsic smoothing scale, cannot recover structures below a certain length. Nonetheless, the filament mass function clearly shows the broad range of masses that these objects have, which, similar to wide distribution of filament lengths, is a consequence of the hierarchical processes that shape these objects. The time variation of the mass function highlights again the conclusions of \refsec{sec:evolution:present_properties}, that filament mergers as well as mass transport from sheet environments leads to the accumulation of mass in a few massive filaments.

The large fraction of filaments with cluster or higher mass suggests the importance of these structures for the dynamics of the universe. It implies that together with clusters, filaments are an important source of tidal fields, which in turn shapes the formation and evolution of large-scale structures \citep{WeyBond08a}.

\Reffig{fig:evolution:fila_length_mass} analyses the relation between the length and the mass of filaments. While there is a large amount of object-to-object scatter, we do find a correlation between the two quantities, with longer filaments being also more massive. The interesting fact about this is that the mean length-mass relation scales as $L^{2.2}$. It indicates that long filaments are distinct well defined objects and not simply two or more shorter structures connected end to end. 

\subsection{Filament shape}
\label{subsec:evolution:filament_individual_shape}

Filamentary branches come in a variety of shapes, from straight to highly curved objects, as already exemplified at the beginning of this section, in \reffigs{fig:evolution:fila_example_1}{fig:evolution:fila_example_3}. It raises questions on what is the typical shape of a filament and if this depends on the physical properties of the object, like mass or length. Quantifying visually the shape of these filaments is not feasible, given that we find a very large number of objects. Moreover, popular shape measurements like the shape ellipsoid used for haloes \citep[e.g.][]{Cole1996,Hopkins2005,Bett2007} have their limitation when it comes to characterizing the morphology of extended, highly anisotropic, structures. We choose to proceed via a simple shape proxy
\begin{equation}
    \eta = \frac{L}{L_\rmn{diagonal}}
    \;,
\end{equation}
which is the ratio between the filament length and the diagonal of the box that fully encompasses the object. This latter quantity is easily computed as
\begin{equation}
    L_\rmn{diagonal} = \sqrt{ (x_\rmn{max}-x_\rmn{min})^2 + (y_\rmn{max}-y_\rmn{min})^2 + (z_\rmn{max}-z_\rmn{min})^2 } 
    \label{eq:evolution:spatial_extent} \;,
\end{equation}
where $(x_\rmn{min},y_\rmn{min},z_\rmn{min})$ and $(x_\rmn{max},y_\rmn{max},z_\rmn{max})$ are the minimum and maximum coordinates of the box that fully contains the filament. In the case of a perfectly straight filament, $L_\rmn{diagonal}=L$ and $\eta=1$, which is also the lowest value that $\eta$ can take. Curved filaments are characterized by $\eta>1$, with small departures from unity indicating only slightly bent objects. A better intuitive idea on the connection between $\eta$ and filament curvature can be obtained by inspecting \reffigs{fig:evolution:fila_example_1}{fig:evolution:fila_example_3} which show objects with $\eta = 1.1$, 1.2 and 1.8. Therefore, the $\eta$ shape proxy represents a quick and simple way to discriminate between straight and curved structures. 

The top panels of \reffig{fig:evolution:fila_individual_shape} show the distribution of filament shapes as a function of the length and mass of these objects. In general, most of the branches have $\eta$ values close to unity indicating that they are approximatively straight. It implies that the filamentary network is built of many branches that slowly bend to create a very elaborate and entangled pattern. The shape proxy shows a clear trend with filament length, with both very short and very long branches more likely to be curved. For the small $L$ region, the higher curvature signal comes from the population of void filaments which are typically short and curved. At high $L$ the $\eta$ trend is indicating that longer filaments are more likely to be curved, which has already been seen in several previous studies \citep{Pimbblet04,Colberg05,Gonzalez09}. A similar trend in $\eta$ is present when binning the filaments according to their mass, though the increase in curvature at the high mass end is not as pronounced. 

The bottom panel of \reffig{fig:evolution:fila_individual_shape} shows that filament shapes do not show a strong evolution with redshift. Any variations are consistent with a shift of the $\eta$ distribution to longer and more massive objects, due to the tendency of filaments to evolve into more extended and massive structures (see \refsec{subsec:evolution:filament_individual_extent} and \refsec{subsec:evolution:filament_individual_mass_function}). 

\subsection{Filament diameter}
\label{subsec:evolution:filament_individual_diameter}

\Reffig{fig:evolution:fila_individual_diameter} shows the dependence of the mean filament diameter $\mean{D_\rmn{f}}$ on the length and mass of the object. On average, $\mean{D_\rmn{f}}$ shows a clear trend with both the extent and mass of filaments, with longer or more massive branches more likely to be thicker too. Compared to the huge object-to-object scatter in the $L\mean{D_\rmn{f}}$ plane \citep[see also][]{Colberg05,Bond10}, a closer analysis reveals a a tighter relation between $\mean{D_\rmn{f}}$ and $M_\rmn{f}$ suggesting that the mass of filaments is the main factor that determines their width. This is in good agreement with the findings of \refsec{subsec:evolution:evolution_width_density} which demonstrate the correlation between the local width and local mass content of filaments.  

The evolutionary processes shaping filaments also have an impact on the width of these structures, with later time objects thinner than their progenitors. This change is not present in the case of the least massive filaments, which are typically found in underdense regions (see \refsec{subsec:evolution:evolution_width_density}). It implies that the evolution of thin and tenuous filaments stops at an earlier time. This freeze-out could be due to void filaments being too far apart to merge with neighbouring structures or due to their feeble mass which does not exert a strong enough local influence to accrete a significant amount of matter.

\subsection{Filament density}
\label{subsec:evolution:filament_individual_density}

In \reffig{fig:evolution:fila_individual_density} we investigate the mean density $\mean{\rho_\rmn{f}}$ distribution for the population of filaments. We find a very large object-to-object scatter \citep{Colberg05,Gonzalez09}, which is especially significant in the case of short and lower mass structures. Tiny filaments are found in both overdense and underdense regions which lead to the large variation in mean density that we see. Filaments in voids, though locally overdense, can be underdense when compared to the mean background density and therefore contribute to the low $\mean{\rho_\rmn{f}}$ values. In contrast, objects in higher density regions can contain massive haloes which when compared to the small volume of these filaments can easily result in $\mean{\rho_\rmn{f}}\gsim50\rho_\rmn{background}$. On the other hand, long filaments live in mainly overdense regions and, due to their large volumes, sample a significant region of space. These characteristics mitigate the variations given by the highly clustered distribution of matter, resulting in objects with similar mean densities.

The median value of $\mean{\rho_\rmn{f}}$ shows a strong trend with the extent and, not surprisingly, the mass of filaments, with longer and massive structures having higher average densities. It implies that prominent filaments are characterized not only by their extreme length or mass, but also by high mean density values. Therefore, these objects should be visible as prominent linear concentrations of galaxies extending many tens of megaparsecs. Examples of these can be found in the Sloan Great Wall, which contains numerous clusters and superclusters connected by extended filaments \citep{Gott2005,Platen09,Einasto2011d,Park2012}.

The mean density shows a strong variation with time, with the median value of $\mean{\rho_\rmn{f}}$ increasing towards $z=0$ for objects of all lengths. Especially interesting is that the increase of $\mean{\rho_\rmn{f}}$ is more pronounced for massive filaments than for tenuous ones. This is another manifestation of the slow evolution of void filaments that we discussed in the previous subsection. In contrast, massive filaments show a large fractional increase in density, regardless of their mass, indicating that even today these structures are rapidly changing.


\section{Conclusions}
\label{sec:evolution:conclusions}
In this work we investigated the characteristics and the evolution of the large scale matter distribution as traced by the cosmic web. This cosmic pattern raises a lot of interest given that it is the transitional stage between linear primordial fluctuations and fully developed non-linear structures. As a consequence of this, it contains an optimal amount of easily accessible information about the early phases of structure formation processes. To assess this information, we have explored these morphological environments using a multitude of complimentary approaches, from characterizing the mass and volume content, to describing the sizes, density and halo distribution of these structures. Following the extensive analysis performed in this study, we summarize below some of the most interesting findings.

The study contains three main parts that focus on key aspects of cosmic web analysis and evolution. In a first stage, we investigated the properties of present day morphological components, with special emphasis on how such results change when using various tracer fields for environment identification. The goal was to understand which method is best suited to reveal the time evolution of the cosmic web. This is continued with an analysis of how the large-scale structures change across cosmic times, with emphasis on the flow of matter between environments and on the characteristics of the filamentary and wall networks. The last part of the study, after segmenting the filamentary network into individual objects, follows the time evolution of distinct filamentary structures characterizing their size, mass and shape. 

To compare the robustness of cosmic web detection techniques, we used the \Nexus{} algorithm applied individually to the density, tidal, velocity divergence and velocity shear fields, to give rise to four different environmental identification procedures. These were complemented by a fifth method, \nexus{}, which uses the density field. Compared to \Nexus{}, \nexus{} uses a specialized filter to better deal with the orders of magnitude variation in density between overdense and underdense regions, which allows for a better detection of both prominent and tenuous components. We found that while the massive structures are consistently identified by each technique, there are significant variations in the detection of more tenuous components. For example, most of the mass contained in filaments (${\sim}90\%$) is correctly classified as part of this environment by all methods. In contrast, only around ${\sim}60\%$ of the filament volume is identified as such by all the five techniques. The following are some of present day environmental characteristics that are consistent across all methods: 
\begin{enumerate}
	\item[$\bullet$] The filaments contain half of the mass in the universe, followed by walls, voids and clusters with $25\%$,  $15\%$ and $10\%$ respectively. In terms of volume, the voids are dominant with $78\%$ of the volume, followed by walls and filaments with $18\%$ and $6\%$. Cluster environments take an insignificant fraction of the cosmic volume.
	\\[-.3cm]
	\item[$\bullet$] We found a clear segregation of the halo population between morphological components. The most massive haloes reside at the nodes of the cosmic web, while most of the other haloes are found in filaments. In contrast, the halo population of walls and voids becomes significant only at masses below $10^{12}\Msolar$.
	\\[-.3cm]
	\item[$\bullet$] Except the most prominent structures, most filaments and sheets are only sparsely traced by haloes, which suggests that identifying these structures in galaxy surveys is very challenging. This point is especially clear when comparing our results to filamentary networks found in observations \citep[e.g.][]{Gonzalez09}, with the latter having only a tenth of the structures identified in the mass distribution.
\end{enumerate}

The second part of this study was focused on following the evolution of morphological components as a whole, with emphasis on the filamentary and wall networks. We use \nexus{} for environment identification given the sensitivity of the method in detecting both prominent and tenuous structures. The time evolution of the cosmic web environments is characterized by the following:
\begin{enumerate}
	\item[$\bullet$] At early times, the filamentary and wall networks are dominated by small scale structures which disappear after merging with bigger objects, such that at present time these small scale structures are mostly gone. In contrast, the most pronounced filaments and walls are already in place since at least $z=2$ and show little evolution in shape and size. 
	\\[-.3cm]
	\item[$\bullet$] Cluster regions become a significant component of the cosmic web only at late times, for $z\lsim0.5$. Our study shows that the accretion of matter into clusters takes place along filaments, with only an insignificant fraction of mass arriving along walls.
	\\[-.3cm]
	\item[$\bullet$] Filaments dominate the cosmic web in terms of mass, with ${\sim}50\%$ of the total matter contained in filaments since at least $z=2$. While at early times this matter is distributed among many filamentary regions, at present time most of the mass is in a few massive structures. The filaments are overdense on average, with a mean density that evolves from $\delta{\sim}5$ at $z=2$ up to $\delta{\sim}10$ at present time.
	\\[-.3cm]
	\item[$\bullet$] For walls, both their mass and volume fractions decrease in time, but in such a way that the mean density of sheets is always $\delta{\sim}0$. While at later times we find fewer walls, the mass distribution of the remaining sheets has hardly changed since $z=2$, indicating a freeze-out of these structures.
	\\[-.3cm]
	\item[$\bullet$] Voids are the dominant component of the cosmos in terms of volume, with their volume fraction increasing from ${\sim}60\%$ at $z=2$ up to ${\sim}80\%$ at present time. Their matter content shows a slow but steady decrease, such that at $z=0$ voids have a mean density of $\delta{=}-0.8$, in good agreement with the predictions of the excursion set formalism \citep{Shethwey04}.
	\\[-.3cm]
	\item[$\bullet$] We find that most of the matter transport takes place from the less dense to the denser environments. Matter flows from voids into walls, from walls into filaments and finally from filaments into clusters, as predicted by the theory of anisotropic collapse \citep[e.g.][]{Zeldovich70,Shen06}.
	\\[-.3cm]
	\item[$\bullet$] We have also characterized the filamentary and wall networks in terms of their width, finding a large variation in size between different objects and also between different points along the structure itself. The thickness is strongly correlated to the mass of filaments and sheets, though there is a large object-to-object scatter.
	\\[-.3cm]
	\item[$\bullet$] Both filaments and walls shows a fractal-like behaviour with fractal dimensions of ${\sim}2.2$ and ${\sim}2.4$ respectively. This behaviour is valid only on a limited set of scales, with the upper bound of the interval increasing rapidly towards $z=0$. The time variation of the fractal dimension gives a good description of the evolution of filaments and walls towards a simpler and less intricate network.
\end{enumerate}

Having studied the properties of the filamentary network as a whole, we then proceeded in identifying distinct structures along the network. This was motivated by the goal of characterizing the building blocks of the filament network, i.e. the individual filamentary structures. To do so, we implemented a filament segmentation procedure which divides the network into individual branches by identifying the filament intersection points. The branching points are easily detectable since they are characterized by a rapid change in the local orientation of filaments. By studying the evolution of individual filaments, we found the following:
\begin{enumerate}
	\item[$\bullet$] The number of filaments decreases rapidly as a function of object length, with a sharp cut-off in the number of very long structures. Although very rare, we found filaments that extend in excess of $100\Mpch$ by connecting linear configurations of several clusters in a row. At high redshift, there are significantly fewer extended structures, but many more short filaments.
	\\[-.3cm]
	\item[$\bullet$] The mass and length of individual objects is related via the relation $M_\rmn{f}\propto L^{2.2}$. It suggests that long filaments are well defined structures and not simply configurations of two or more shorter objects. 
	\\[-.3cm]
	\item[$\bullet$] Both the mean filament diameter and density show a strong dependence on the mass and length of the object, with extended structures significantly wider and denser than their short counterparts. These characteristics show a strong time evolution, with filaments at earlier times being wider and less dense. 
\end{enumerate}

Having investigated the characteristics and the evolution of the cosmic web, we plan to further this study by analysing the velocity field and the dynamics of individual cluster-filament systems. First, these aspects play a major role in better understanding what drives the transport of mass between various morphological components. And secondly, the environments show characteristic features not only in the mass distribution, but also in the velocity field \citep{Shandarin2012,Tempel2013b}. The velocity flow features are important since they govern the infall of matter into the haloes and galaxies residing in that environment, therefore affecting galaxy formation and evolution \citep[e.g.][]{Codis2012,Libeskind2013a}.

The application of the \Nexus{} and \nexus{} methods to galaxy redshift surveys represents an important next step, which allows us to characterize the cosmic web environments in observational data. The goal is to compare the properties of morphological components between theoretical predictions and observations in the hope that such a study will shed additional light on the large-scale structure of our Universe. Before applying the detection techniques to observational data, a number of challenges need to be addressed and understood. For example, the coarse sampling of the density field by DM haloes and galaxies is an important limiting factor in the detection of the more tenuous structures \citep[for details see][]{Colberg07,Bond10}. Within this context, the recovery of the underlying density distribution play a major role in limiting the success of the environmental detection methods. Recent works have shown great progress in this field, with the DTFE and Kriging methods showing a good recovery of the density field \citep{Schaap2000,2009LNP...665..291V,Platen2011}. Another crucial aspect deals with the effect of redshift space distortions which may introduce artefacts in our filament catalogue. A successful correction for most of the Fingers of God effects would mitigate this issue, though there are other ways around this problem. For example, \citet{Jones10} showed that it is feasible to perform environmental studies using only filaments in the plane of the sky, which are not prone to Fingers of God effects, but only at the price of poorer statistics.

\section*{Acknowledgements}
The authors are grateful to the anonymous referee whose suggestions improved the presentation of this paper. MC would like to thank Johan Hidding and Wojciech Hellwing for valuable discussions. MC and CSF acknowledge the support of the ERC Advanced Investigator grant COSMIWAY [grant number GA 267291] and the Science and Technology Facilities Council [grant number ST/F001166/1, ST/I00162X/1]. RvdW acknowledges support by the John Templeton Foundation, grant number FP05136-O. The simulations used in this study were carried out by the Virgo consortium for cosmological simulations.

This work used the DiRAC Data Centric system at Durham University, operated by the Institute for Computational Cosmology on behalf of the STFC DiRAC HPC Facility (www.dirac.ac.uk). This equipment was funded by BIS National  E-infrastructure capital grant ST/K00042X/1, STFC capital grant ST/H008519/1, and STFC DiRAC Operations grant ST/K003267/1 and Durham University. DiRAC is part of the National E-Infrastructure.


\newcommand{\jcap}{Journal of Cosmology and Astroparticle Physics}
\bibliographystyle{mn2e}
\bibliography{references_bib}

\appendix

\section{Compressing the filament and wall environments}
\label{appendix:evolution:filament_contraction}
The following describes the algorithm that we employ to contract the filaments to their central axis and the walls to their central plane. The outcome of this compression process is used twofold. First, we use it to determine the geometrical direction of filaments and walls. The steps necessary for this are described in \refappendix{appendix:evolution:filament_direction}. And secondly, we used the compressed networks to study the individual characteristics of filament and wall segments, as shown in \refsec{sec:evolution:environment_characterisation}. 

\begin{figure*}
    \centering
    \includegraphics[width=0.8\linewidth]{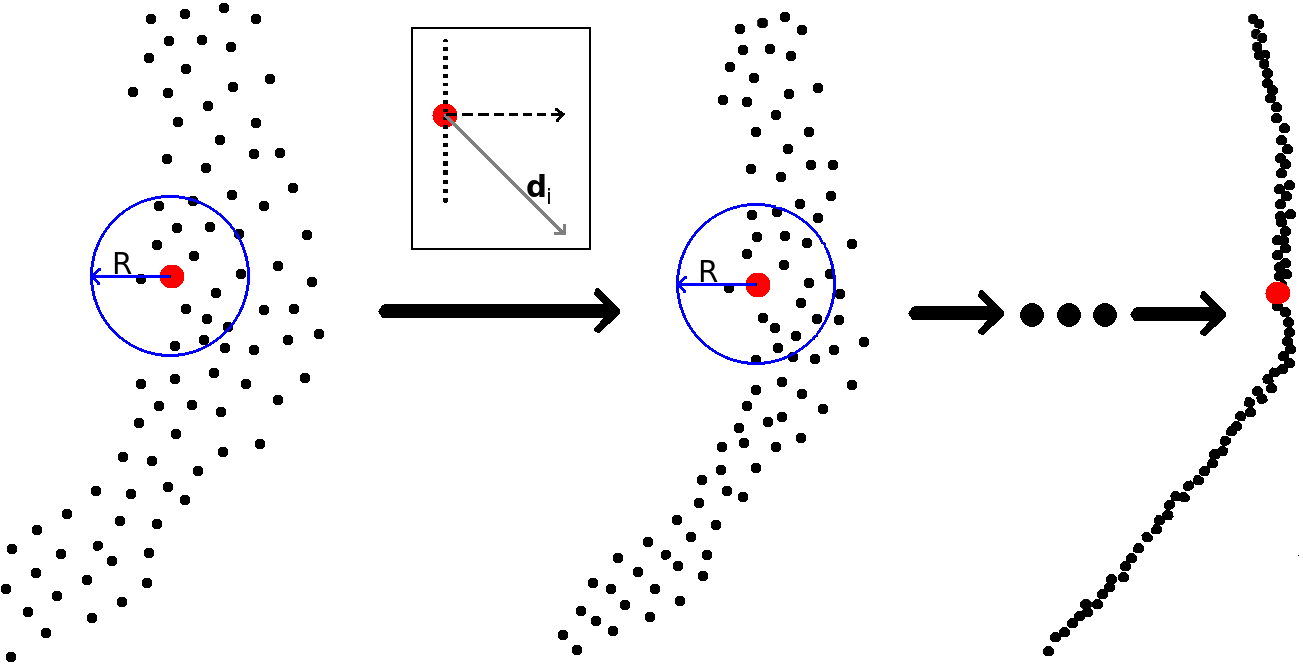}
    \caption{ An illustration of the steps taken for compressing filaments. We start on the left frame, where each point represents a filament voxel. Of those, we select one point, shown in larger size, for which we find all its neighbours within a distance $R$. Following this, the rectangular insert shows the filament orientation (dotted line), the displacement vector  $\mathbf{d_i}$ (solid line) and the component of $\mathbf{d_i}$ perpendicular to the filament orientation (dashed line). The centre panel shows the distribution of points after the first displacement step. We repeat the procedure until all the points lie along a curve which is the central spine of the filament, as shown in the right panel. }
    \label{fig:evolution:example_fila_compression}
\end{figure*}


The compression process is very similar for both filaments and walls, with only minor differences between the two morphological components. Therefore, we first focus on describing the application of the compression algorithm to filaments and only later on we present the few differences that arise in the case of cosmic sheets. For clarity, we illustrate the compression procedure in \reffig{fig:evolution:example_fila_compression}. For filamentary environments, the compression procedure consists of the following steps:
\begin{enumerate}
    \item[I)] Classify as filaments all the grid cells\footnote{Note that the output of the \Nexus{} and \nexus{} methods is a cosmic web pattern described in terms of cells on a regular grid, with each cell identified as being part of a node, filament, wall or void.} that are identified by \Nexus{} or \nexus{} as being part of cluster or filament environments. This is motivated by the empirical finding that cluster regions are fully encompassed inside filaments, so discarding the cluster regions leads to a spurious filamentary fragmentation around the cosmic web nodes.
    \item[II)] Replace each of the cells found at the previous steps with a point located at its centre. Each such point $i$ has associated to it a vector $\Vector{u_i}$ that gives the filament orientation at that point, which was determined according to \refappendix{appendix:evolution:filament_direction}.
    \item[III)] Use a spherical filter of radius $R$ to identify for each point $i$ all its $N_i$ neighbours that are within the sphere. The choice of the filter size is motivated by two constraints. First, $R$ must be significantly larger than the grid spacing used in step \textbf{I)}. Secondly, the value of $R$ should be small such that it does not lead to a considerable smoothing of the filament axis. We use $R=1\Mpch$ which we find that gives a good balance between the two requirements.
    \item[IV)] For each point, find the centre of volume $\Vector{x_{\rmn{CV},i}}$ of the cloud points inside the sphere of radius $R$. The centre of volume is given by:
    	\begin{equation}
    		\Vector{x_{\rmn{CV},i}} = \frac{1}{N_i}\sum_{j=1}^{N_i} \Vector{x_{ij}} \;,
    	\end{equation}
    where $\Vector{x_{ij}}$ denotes the position of the $j$-th neighbour of point $i$. The centre of volume is the same as the centre of mass when all points have equal mass.
    \item[V)] Move each point towards its associated centre of volume, but only along a direction perpendicular to the filament orientation $\Vector{u}_i$. Therefore, the new position of point $i$ is given by:
    	\begin{equation}
    		\mathbf{x_{i,new}} = \mathbf{x_i} +  \left( \mathbf{d_i} - (\mathbf{d_i} \cdot \mathbf{u_i}) {\mathbf{u_i}} \right) \;,\quad \rmn{with}\quad  \mathbf{d_i}=\mathbf{x_{CV,i}} - \mathbf{x_i}
            \label{eq:fila_displacement} \;,
    	\end{equation}
    where $\mathbf{x_i}$ was the initial position of the point and $\mathbf{d_i}$ is the distance of the point with respect to the centre of volume of its neighbours. This step is illustrated in the rectangular insert of \reffig{fig:evolution:example_fila_compression}, where we show the filament orientation ${\mathbf{u_i}}$ (dotted line), the displacement vector $\mathbf{d_i}$ (solid line) and the perpendicular of $\mathbf{d_i}$ on to ${\mathbf{u_i}}$ (dashed line). The point is moved according to the dashed line. The centre frame shows the new point distribution after shifting all the points according to \eq{eq:fila_displacement}.
    \item[VI)] Repeat steps \textbf{III)} to \textbf{V)} until the points have converged to a final position. We consider that each point has converged to its final position, on the spine of the filaments, when the following two criteria are satisfied:
    	\begin{enumerate}
    		\item[a.] The distance $\mathbf{d_i}$ a point is expected to move is smaller than a certain threshold. In practice we select $\mathbf{d_i}<0.01\Mpch$.
    		\item[b.] The point cloud which consists of all the neighbours within radius $R$ shows a very pronounced filamentary morphology. 
    		We check for this in terms of the eigenvalues $e_1\ge e_2\ge e_3$ of the shape of the point cloud (see \eq{eq:evolution:fila_shape} for a definition of the cloud's shape). We require that the ratio $e_2/e_1$ is small, with practical values of $e_2/e_1 < 0.1$.
    	\end{enumerate}
    	The final output of the algorithm is shown in the right panel of \reffig{fig:evolution:example_fila_compression}.
\end{enumerate}

\begin{figure*}
    \centering
    \includegraphics[width=0.8\linewidth]{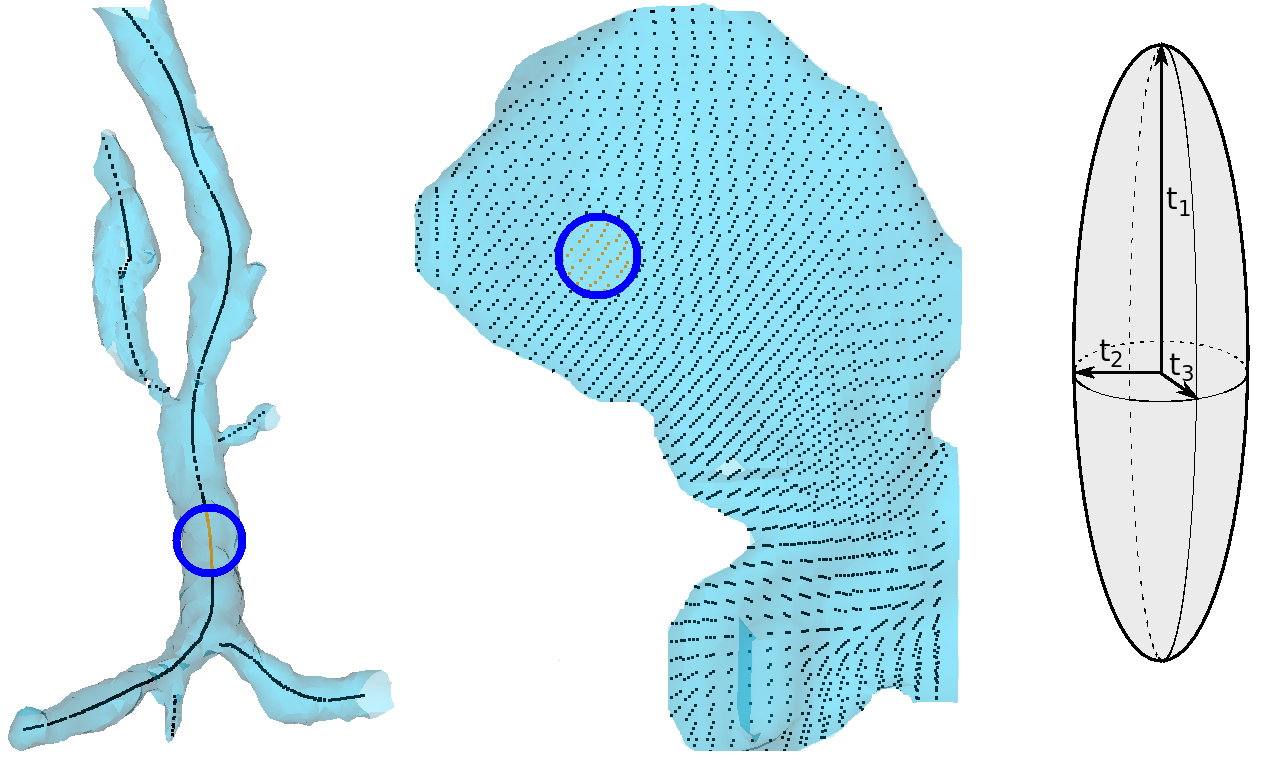} 
    \caption{ Computing the geometric orientation of filaments and walls. The filaments (left panel) and sheets (centre panel) are compressed to their central spine and plane using the procedure described in \refappendix{appendix:evolution:filament_contraction}. The result of this is shown via the black dots. For each point on the spine, we select all the neighbours within a distance $R$ (see the thick circle). The shape of the resulting point distribution is used to determine the local orientation of environments, which for filaments is given by $t_1$ and for walls by $t_3$. The directions $t_i$ sketched in the right panel are the shape eigenvectors corresponding to the eigenvalues $e_1\ge e_2\ge e_3$. }
    \label{fig:evolution:env_test_direction}
\end{figure*}

In the case of wall, the compression algorithm is the same, with only small differences. The following are the modifications that need to be made to apply the technique for sheet environments (note that we only give the steps that are different):
\begin{enumerate}
    \item[I)] Classify as walls all the grid cells that are identified as being part of cluster, filament or wall environments. Cluster and filament regions are encompassed inside sheets, so discarding the regions would lead to significant fragmentation of walls around cosmic web nodes and filaments.
    \item[V)] Move each point towards its associated centre of volume, but only perpendicular to the wall plane. Therefore, the new position of wall point $i$ is given by:
    	\begin{equation}
    		\mathbf{x_{i,new}} = \mathbf{x_i} +  (\mathbf{d_i} \cdot \mathbf{u_i})\mathbf{u_i} \;,\quad \rmn{with}\quad  \mathbf{d_i}=\mathbf{x_{CV,i}} - \mathbf{x_i}
            \label{eq:wall_displacement} \;,
    	\end{equation}
    where $\mathbf{x_i}$ was the initial position of the point and $\mathbf{d_i}$ is the distance of the point with respect to the centre of volume of its neighbours. With $\mathbf{u_i}$ we denoted the direction perpendicular to the plane of the wall at point $i$.
    \item[VI)] Repeat steps \textbf{III)} to \textbf{V)} until the points have converged to a final position. We consider that each point has converged to its final position, on the central plane of walls, when the following two criteria are satisfied:
    	\begin{enumerate}
    		\item[a.] The distance $\mathbf{d_i}$ which a point is expected to move is smaller than a certain threshold. In practice we select $\mathbf{d_i}<0.01\Mpch$.
    		\item[b.] The point cloud which consists of all the neighbours within radius $R$ shows a very prominent wall morphology. We check for this in terms of the shape eigenvalues $e_1\ge e_2\ge e_3$ by requiring that the ratio $e_3/e_2$ is small, with practical values of $e_3/e_2 < 0.1$.
    	\end{enumerate}
\end{enumerate}

Another example of the output of the compression algorithm can be seen in \reffig{fig:evolution_fila_contraction_example}, which shows the initial and contracted distribution of the grid cells that are part of filamentary environments.


\section{Determining the geometrical orientation of filaments and walls}
\label{appendix:evolution:filament_direction}
Several ways have been previously used to characterize the direction of these morphological components. For example, the orientation of filaments has been taken as the direction of constant density \citep{Aragon07a,Aragon2013}, the eigenvector corresponding to the largest eigenvalue of the tidal field tensor \citep{Hahn2007a,Hahn2007b} and largest eigenvalue of the velocity shear field \citep{Libeskind2012,Libeskind2013}. In the case of walls, their orientation was taken as the direction of the largest change in density \citep{Aragon07a,Aragon2013}, the eigenvector corresponding to the smallest eigenvalue of the tidal field tensor \citep{Hahn2007a,Hahn2007b} and smallest eigenvalue of the velocity shear field \citep{Libeskind2012,Libeskind2013}. While in many cases there is a very good agreement between the above directions and the geometrical orientation of filaments and walls, this is not necessarily the case in every region. In fact, many of the above methods fails to characterize the geometrical direction of filaments and walls around massive objects, like clusters, where both the density and tidal fields are heavily distorted by the presence of large mass concentrations. Therefore, applying such methods introduces artefacts when it comes to contracting these morphological components to their central axis or plane.

To overcome the limitations of previous methods, we present here a procedure that computes the geometrical orientation of filaments and walls. The technique is purely a geometric one and does not use information encoded in other cosmic quantities, like density or tidal field\footnote{Of course that density or tidal field data are used for the identification of the morphological components, but these fields do not enter directly into the computation of the geometrical orientation.}. In our approach, the local orientation of filaments is given by the local direction of the filament spine. Similarly, the local orientation of sheets is taken as the normal to the wall's central plane. This is sketched in \reffig{fig:evolution:env_test_direction}. Following the contraction of filaments to a central spine and of sheets to a central plane, we select for each point all the neighbours within a distance $R$. The resulting point distribution is used to compute the local shape of the spine via
\begin{equation}
	I_{ij} = \sum x_i x_j
	\label{eq:evolution:fila_shape} \;,
\end{equation}
with the sum running over all the points within the selection. The shape $I_{ij}$ of this point distribution is characterized by its eigenvectors $t_i$ corresponding to the shape eigenvalues $e_1\ge e_2\ge e_3$, with the orientation of filaments given by $t_1$ and that of walls by $t_3$.

Before describing the algorithm in details, it is important to realize that both the filament spine and the wall central plane depend on the local orientation of these components. Therefore, in our approach, the central spine depends on filament orientation which in turn is determined by the local shape of the spine. We deal with these dependences by following an iterative approach: the initial guess for the orientation of morphological components is improved at each step until we obtained a converged result. The computation proceeds along the following lines:
\begin{enumerate}
    \item[I)] Use an initial guess for the orientation of filaments and walls at each point.
    \item[II)] Compress the filaments to their central axis and the walls to their central plane using the procedure described in \refappendix{appendix:evolution:filament_contraction}. Note that for this we use an estimate of the orientation of these environments, estimate that becomes better after every iteration.
    \item[III)] Using the compressed networks, we compute the geometrical orientation at point $i$ using the distribution of the points in the neighbourhood. We identify all the neighbours within a sphere of radius $R=1\Mpch$ \footnote{Like any of the previous methods, computing the orientation of filaments and walls involves choosing a smoothing scale. For details on why we selected this value for $R$, see \refappendix{appendix:evolution:filament_contraction}.} around point $i$ (see \reffig{fig:evolution:env_test_direction}). Given this point cloud with shape eigenvalues $e_1\ge e_2\ge e_3$ (see \eq{eq:evolution:fila_shape} for the definition of shape), the filament orientation at point $i$ is given by the eigenvector $t_1$ corresponding to eigenvalue $e_1$. The wall orientation is given by the eigenvector $t_3$ corresponding to eigenvalue $e_3$.
    \item[IV)] We repeat steps \textbf{II)} and \textbf{III)} using the updated estimate of the geometrical orientation found in the previous step. We stop the iteration procedure once the change in direction between two successive steps is small enough. In practice we require that for each point the change in direction to be less than $5^\circ$ between two successive steps.
\end{enumerate}

\begin{figure}
    \centering
    \includegraphics[width=\linewidth]{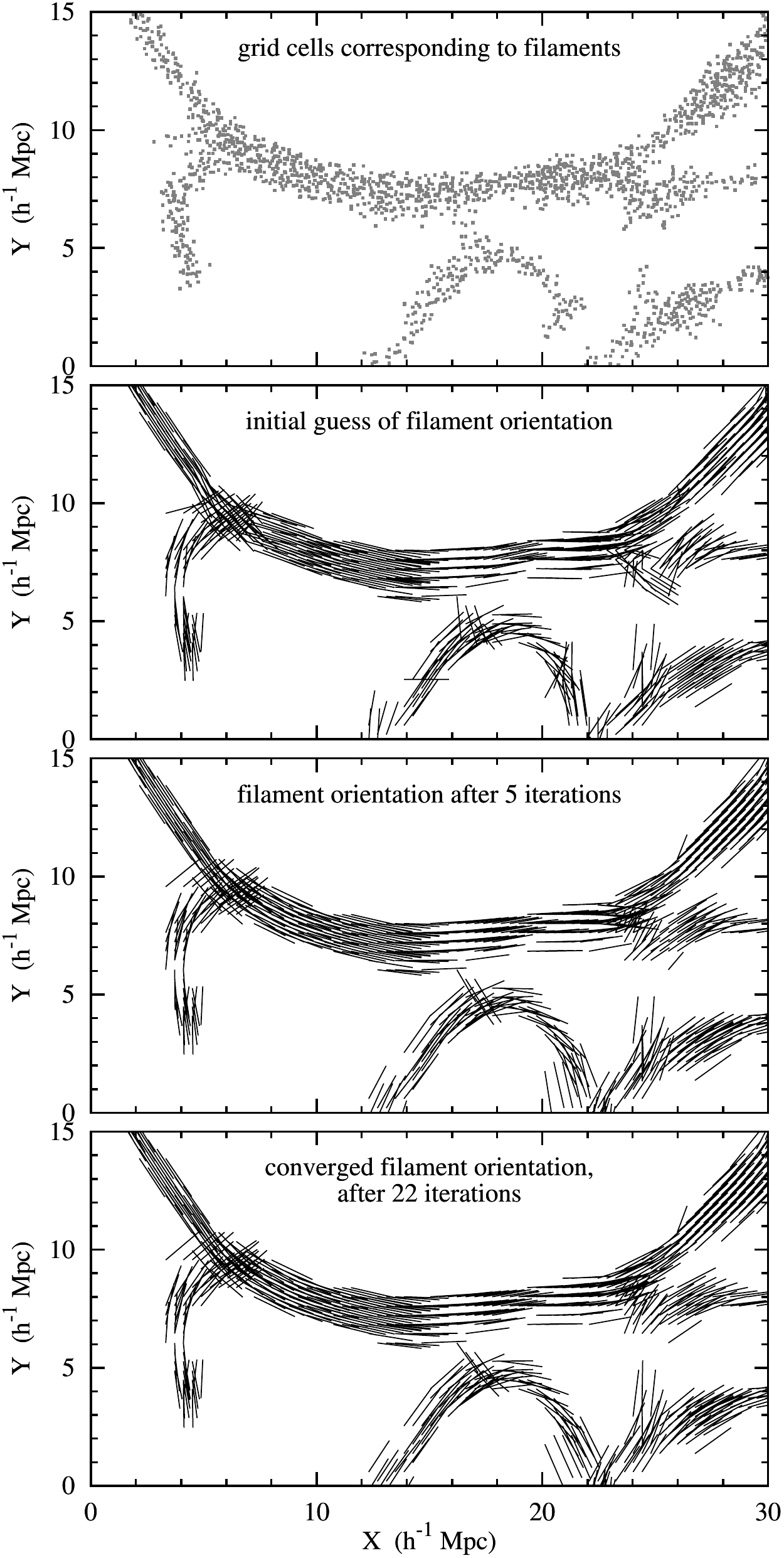}
    \caption{ An example of the geometrical orientation of filaments. The top panel shows the grid cells identified as filaments in a small volume of the \MII{}. The remaining three frames show the filament orientation at every point during a few steps of the iteration procedure: the initial guess (second row), after five iterations (third row) and after the orientation computation has converged orientation, after 22 iterations (fourth row). For clarity we show the orientation of only a third of the filament points. The regions with more than one orientation are due to projection effects and correspond to filaments at different $z$-coordinate values. }
    \label{fig:evolution_fila_direction_example}
\end{figure}

To obtain an initial estimate of the geometrical orientation, we compress the filaments to a central spine and the walls to a central plane without using orientation data. While this does not correspond to the true central axis or central plane, it gives a reasonable approximation that allows us to obtain a rough estimate for the orientation of these components. Contracting the filaments and walls without orientation data necessitates only small changes in the compression algorithm given in \refappendix{appendix:evolution:filament_contraction}. In fact, only \eq{eq:fila_displacement} and \eq{eq:wall_displacement} need to be modified such that both expressions should read:
\begin{equation}
	\mathbf{x_{i,new}} = \mathbf{x_{CV,i}} \;.
\end{equation}  
Once we have computed these compressed networks, we can use step \textbf{III)} of the above algorithm to obtain an initial estimate for the orientation of filaments and walls at each point.

\Reffig{fig:evolution_fila_direction_example} shows the output of the filament orientation computation in a small volume of the \MII{}. The top-most panel shows the spatial distribution of the filamentary network in the selected volume. The remaining panels show the filament orientation at two intermediate steps of the iteration procedure as well as the converged final value (bottom-most frame). From these we conclude that the initial guess for the filament orientations shows a very good match with the final result, with most of the differences limited to the intersection of two or more filamentary branches. Moreover, the iteration procedure converges rapidly, with only small differences between filament orientation after five iterations and the final outcome.

\end{document}